\documentclass[fleqn,usenatbib]{mnras}

\usepackage{newtxtext,newtxmath}
\usepackage[T1]{fontenc}
\usepackage{graphicx}	% Including figure files
\usepackage{amsmath}	% Advanced maths commands
\usepackage{multirow}% http://ctan.org/pkg/multirow
\usepackage{booktabs}
\usepackage{tabu}
\usepackage{breqn}
\usepackage{adjustbox}
\usepackage{pdflscape}
\usepackage{listings}
\usepackage{xcolor}
\usepackage{ifthen}
\usepackage{mathrsfs}
\usepackage{cuted}
\usepackage{rotating}
\usepackage{afterpage}
\newcommand{\bahamas}{\texttt{BAHAMAS}}
\newcommand{\lcdm}{$\Lambda$CDM}
\newcommand{\SN}{SNIa}
\newcommand{\SNe}{SNe}

\newcommand{\FullPars}{\Theta}

\newcommand{\omegam}{\Omega_{M}}
\newcommand{\omegade}{\Omega_{\Lambda}}

\newcommand{\sigmares}{\sigma_{\rm res}}

\newcommand{\be}{\begin{equation}}
\newcommand{\ee}{\end{equation}}
\newcommand{\pars}{{\mathscr{P}}}
\newcommand{\ci}{\hat{c}_i}
\newcommand{\ch}{\hat{c}}
\newcommand{\cobs}{c^{\rm obs}_{sj}}
\newcommand{\sigmaobs}{\sigma^{\rm obs}_{sj}}
\newcommand{\norm}{{\mathcal N}}

\newcommand{\nobs}{n_{\rm obs}}
\newcommand{\zbar}{\bar{z}}
\newcommand{\zCMB}{z_{\rm \tiny{CMB}}}
\newcommand{\zhatCMB}{\hat{z}_{\rm \tiny{CMB}}}
\newcommand{\zhel}{z_{\rm hel}}

\newcommand{\nsn}{{\mathbf n}_{\rm \tiny{SN}}}
\newcommand{\ncmb}{{\mathbf n}_{\rm \tiny{CMB}}}
\newcommand{\ndip}{{\mathbf n}_{\rm dip}}
\newcommand{\zsol}{z_{\rm pec}^\odot}
\newcommand{\vsol}{v_{\odot-{\rm CMB}}}
\newcommand{\zSN}{z_{\rm pec}^{\rm SN}}
\newcommand{\vpecSN}{v_{\rm pec}^{{\rm SN}}}
\newcommand{\Vext}{{\bf V}_{\rm ext}}
\newcommand{\vpec}{\vpecSN}
\newcommand{\vgal}{{\bf v}_{\rm gal}}
\newcommand{\gprime}{{g^\prime}}
\newcommand{\sigvNL}{\sigma_v^\text{\tiny{NL}}}
\newcommand{\Flowpars}{\theta}
\newcommand{\paramspop}{\vartheta}
\newcommand{\paramsBHM}{\Theta}
\newcommand{\dobs}{\hat{d}_{\rm obs}}

\title[Anisotropy constraints from \SN]{New Constraints on Anisotropic Expansion from Supernovae Type Ia}

\author[Rahman {\em et al.}]{
W. Rahman$^{1}$\thanks{E-mail: w.rahman17@imperial.ac.uk}, R. Trotta$^{1,2,3}$, S. S. Boruah$^{4}$, M. J. Hudson$^{5,6,7}$,
 and D. A. van Dyk$^{2,8}$
\\
$^{1}$Astrophysics Group, Physics Department, Blackett Laboratory, Imperial College London, Prince Consort Rd,
London SW7 2AZ\\
$^{2}${Data Science Institute, William Penney Laboratory, Imperial College London, London SW7 2AZ}\\
$^{3}${International School for Advanced Studies (SISSA), Theoretical and Scientific Data Science Group, Physics Department, Via Bonomea 265, 34136 Trieste, Italy}\\
$^{4}$Department of Astronomy and Steward Observatory, University of Arizona, 933 N Cherry Ave, Tucson, AZ 85719, USA \\
$^{5}$Department of Physics and Astronomy, University of Waterloo, Waterloo, ON, N2L 3G1, Canada\\
$^{6}$Waterloo Centre for Astrophysics, Waterloo, ON, N2L 3G1, Canada\\
$^{7}$Perimeter Institute for Theoretical Physics, 31 Caroline St. N., Waterloo, ON, N2L 2Y5, Canada\\
$^{8}$Statistics Section, Mathematics Department, Huxley Building, Imperial College London, Queen's Gate, London SW7 2AZ
}

\date{Accepted 2022 April 22. Received Received 2022 April 22; in original form 2021 August 16}

\pubyear{2021}

\hypersetup{draft=False}

\begin{document}
\label{firstpage}
\pagerange{\pageref{firstpage}--\pageref{lastpage}}
\maketitle

\begin{abstract}
We re-examine the contentious question of constraints on anisotropic expansion from Type Ia supernovae (\SN) in the light of a novel determination of peculiar velocities, which are crucial to test isotropy with \SNe\ out to distances $\sol 200/h$ Mpc. We re-analyze the Joint Light-Curve Analysis (JLA) Supernovae (\SNe) data, improving on previous treatments of peculiar velocity corrections and their uncertainties (both statistical and systematic) by adopting state-of-the-art flow models constrained independently via the 2M$++$ galaxy redshift compilation. We also introduce a novel procedure to account for colour-based selection effects, and adjust the redshift of low-$z$ \SNe\ self-consistently in the light of our improved peculiar velocity model. 

We adopt the Bayesian hierarchical model \bahamas{} to constrain a dipole in the distance modulus in the context of the \lcdm{} model and the deceleration parameter in a phenomenological Cosmographic expansion. We do not find any evidence for anisotropic expansion, and place a tight upper bound on the amplitude of a dipole, $|D_\mu| < 5.93 \times 10^{-4}$ (95\% credible interval) in a \lcdm{} setting, and $|D_{q_0}| < 6.29 \times 10^{-2}$ in the Cosmographic expansion approach. Using Bayesian model comparison, we obtain posterior odds in excess of 900:1 (640:1) against a constant-in-redshift dipole for \lcdm\ (the Cosmographic expansion). In the isotropic case, an accelerating universe is favoured with odds of $\sim 1100:1$ with respect to a decelerating one.

\end{abstract}

\begin{keywords}
supernovae: general, methods: statistical, (cosmology:) dark energy, cosmology: observations, (cosmology:) cosmological parameters
\end{keywords}

%%%%%%%%%%%%%%%%%%%%%%%%%%%%%%%%%%%%%%%%%%%%%%%%%%

%%%%%%%%%%%%%%%%% BODY OF PAPER %%%%%%%%%%%%%%%%%%

\section{Introduction}

A fundamental assumption underpinning the cosmological concordance model is the cosmological principle, namely that the universe is homogeneous and isotropic on sufficiently large scales. Given the ubiquity of the cosmological principle, an observational test of this assumption is an important step towards validating our best description of the large scale universe. Testing of homogeneity is hampered by the need of surveying extremely large scales (see \cite{maartens2011}), although recent studies have found the transition to homogeneity at high ($z\sim 2$) redshift consistent with expectations from the \lcdm{} cosmological concordance model~\citep{2018MNRAS.481.5270G,2020arXiv201006635G}. 

The assumption of isotropy has been tested over a range of redshifts and with many different probes, from the relatively local universe out to the redshift of recombination.
Analyses of Cosmic Microwave Background (CMB) anisotropies data obtained by the Wilkinson Microwave Anisotropy Probe (WMAP) and the Planck satellite found up to $\sim 3\sigma$ evidence of breaches of statistical isotropy in the form of power asymmetry between hemispheres, multipole alignments, anomalous clustering of directions, although the significance of these results is difficult to assess, partially because of issues of {\em a posteriori} testing~\citep{bennett2013,akrami2014,schwarz2016,planck2015_16}. Quasar polarization directions also appear to be aligned along anomalous directions in the CMB \citep{hutsemekers2005} and with coherence scales in excess of 500 Mpc~\citep{refId0}, in potential disagreement with the cosmological principle. Investigating the distribution of galaxies on large scales, \cite{sarkar2019} found however good agreement between the predictions of \lcdm{} and the Sloan Digital Sky Survey data, with a transition to isotropy observed beyond a length scale of $200h^{-1}$ Mpc (where $h$ is the dimensionless Hubble-Lema\^itre parameter). More recently, ~\cite{2020arXiv200914826S} reported a one-sided 4.9-$\sigma$ rejection of the hypothesis that the dipole in a sample of 1.3 million quasars is purely due to our motion with respect to the CMB.

Supernovae Type Ia (\SN) can be used to test the second expression of the cosmological principle, namely that the expansion of the universe is isotropic. \SN\ are a sub-class of supernovae (\SNe), resulting from the thermonuclear explosion of CO white dwarfs accreting mass near the Chandrasekhar limit, whose spectra exhibit no hydrogen lines but strong silicon lines. A series of corrections can be applied to account for correlations of absolute peak magnitude with their light curve decline rate and colour. The correlation with light curve decline rate were first observed by \cite{Rust1974} and further corroborated later by \cite{Pskovskii1977, Pskovskii1984} and \cite{phillips1993} using larger samples of \SNe\ type Ia. Later a correlation with color was also observed and it was noticed that brighter \SN\ had a bluer colour \citep{1995AJ....109....1H,riess1996, Perlmutter1997}. Empirical corrections that account for correlations of absolute peak magnitude with their decline time and colour can then be applied to standardise the \SN\ data ~\citep{Tripp1998,phillips1999}. Light curves of \SNe{} type Ia can be standardised so that the residual scatter of their peak $B$-band magnitude is sufficiently small ($\sim 0.1$ mag) to infer cosmological parameters, as was first demonstrated by \cite{riess1998} and \cite{perlmutter1999}.  \SNe\ type Ia observations can also be used to test the hypothesis of isotropy in the expansion of the universe underpinning the Friedemann-Lema\^itre-Robertson-Walker (FLRW) metric of the concordance cosmological model, which exhibits an isotropic scale factor $a(t)$. To this end, various authors have analyzed increasingly large \SN\ compilations with different statistical approaches, often with sharply discordant results. 

After early works \citep{2001MNRAS.323..859K,2007AA...474..717S,doi:10.1111/j.1365-2966.2008.13377.x}, \cite{cooke2010} analysed a subset of 250 \SN\ from the Union compilation~\citep{kowalski2008} with $z>0.2$ with a maximum likelihood approach to constrain a dipolar modulation to the luminosity distance, finding no significant deviation from isotropy. However, \cite{Cai_2012} claimed that the deceleration parameter shows a preferred direction in the Union2~\citep{amanullah2010} compilation of 557 \SN, a result corroborated by the analysis of~\cite{Antoniou_2010}, who combined \SN\ data with other cosmological probes. \cite{beltran2015} analysed the same Union2 data, additionally including the SNLS3 data, and showed that previous claims of anisotropy disappear if one accounts for correlations among the observations by including the full data covariance matrix in the analysis. Other null results of anisotropic expansion include ~\cite{2014MNRAS.439.1855H} and \cite{lin2016}, who investigated the Joint Light-curve Analysis (JLA) compilation \citep{betoule2014} of 740 \SN. (See also \cite{2019EPJC...79..783S}, who obtain discrepant results from three different compilations of \SN\, namely Union 2.1, JLA and Constitution \citep{Constitution2009}. Similarly \cite{andrade2018}, find the JLA data prefers isotropy, with the results being inconclusive on the Union2.1 data. \cite{Javanmardi2015} however, find that the null hypothesis of isotropy cannot be rejected unless ones specifically takes into account its alignment with the dipole of the Cosmic Microwave Background (CMB), in which case the null hypothesis can be rejected at a level ranging from 95\% to 99\% confidence. An important distinction between these data sets is that Union2/2.1 and Constitution have no corrections for the peculiar velocities of the \SNe\ host galaxies, whereas JLA does. \cite{2017PhLB..765..163B} found from Union2 and LOSS data potential differences between hemispheres in the isotropy of the deceleration parameter.

The situation becomes more confused when considering the largest \SN\ compilation to date, Pantheon, encompassing 1048 objects in the redshift range $0.01 < z < 2.3$~\citep{Scolnic:2017caz}. A major hurdle to any re-analysis that uses Pantheon is the lack of a publicly available full correlation matrix for its \SN\ light curve standardisation coefficients, which hampers a principled statistical approach. Nevertheless, several papers have attempted to use Pantheon-derived measurements of the distance modulus as a function of redshift to investigate potential deviations from an isotropic expansion, finding isotropy is still favoured ~\citep{2018MNRAS.478.5153S,2019PhRvL.122i1301S, 2019MNRAS.486.5679Z, Andrade2018eta}.

Despite the existence of substantially larger compilations like Pantheon, the JLA remains a useful data set for analyses of this kind, because all the necessary statistical and systematics covariance matrixes are publicly available, unlike e.g.~the Pantheon set. Recently, \cite{colin2019}  (henceforth C19) claimed 3.9-$\sigma$ evidence for a dipole in the deceleration parameter from a maximum likelihood analysis of JLA data, leading to a lack of statistical evidence for acceleration in the expansion. The claim was disputed by~\cite{Rubin_2020} (henceforth, RH20), who pointed out the incorrect use of heliocentric redshifts in C19 and other technical assumptions about selection effects which, when corrected, remove the preference for a dipole and restore the high significance for an accelerated expansion. (See also the discussion in~\cite{RH16}, itself a rebuttal of \cite{Nielsen16}.) A further reply by \cite{Nielsen19} appears to concede some technical points, but not the overall conclusion on the actual lack of statistical significance for an accelerated expansion.

The aim of this work is to clarify the status of claims for a statistically significant dipole in the accelerated expansion of the universe, especially in light of the ongoing controversy. In so doing, we also revisit the important question of the level of statistical evidence in favour of the accelerated expansion in an isotropic universe from \SN\ data alone. We address the criticisms of the published JLA data made by C19, who claimed that the peculiar velocity corrections made to the JLA \SN\ data based on local bulk flows are incorrect. In this paper, we introduce a state-of-the-art treatment of peculiar velocities, which are independently constrained using the 2M++ galaxy catalogue~\citep{10.1111/j.1365-2966.2011.19233.x, RN339,RN345}, re-derive correlated peculiar velocity uncertainties (both statistical and systematic) from a fully consistent flow model, and upgrade the Bayesian hierarchical model \bahamas\ \cite{shariff2016} to include an new treatment of residual colour-based selection effects in \SN\ data. 

The remaining of this paper is structured as follows: section \ref{sec:methodology} introduces the cosmological model, the anisotropy model, our Bayesian framework, the data used, our new peculiar velocities treatment and our new colour-based selection effects correction. Section \ref{sec:sim_param_reconstruction} demonstrates the performance of our method on simulated data. Our results from the JLA data, both in terms of parameter inference and Bayesian model comparison, are presented in section \ref{sec:results}. Our conclusions are given in section \ref{sec:conclusions}.

\section{Methodology and Data}
\label{sec:methodology}
\subsection{Cosmological model and dipole modulation}
\label{sec:cosmology}

We investigate the isotropy of the expansion in both a model-specific and a model-independent way: firstly, we consider the \lcdm{} model for the underlying cosmology; secondly, we use the so-called ``Cosmographic expansion''  (i.e., a Taylor expansion of the scale factor as a function in time) as a model-independent description of the underlying matter-energy density of the universe. 

The \lcdm{} model has cosmological parameters $\pars_1 = \{\Omega_m, \Omega_\Lambda, H_0\}$, where $\Omega_m$ and $\Omega_\Lambda$ are the density parameters of matter (both baryonic and dark) and cosmological constant, $\Lambda$, respectively, in units of the critical energy density; $H_0$ is the Hubble-Lema\^itre constant (which we fix to $H_0 = 72$ km/s/Mpc, as it is exactly degenerate with the \SN\ intrinsic magnitude). The curvature parameter $\Omega_\kappa$ is given by 
\be
\Omega_\kappa = 1 - \Omega_m - \Omega_\Lambda
\ee
and we assume a universe with constant dark energy equation of state, $w(z) = -1$. We denote by $\zbar$ the redshift of a comoving galaxy seen by an observer who is also at rest w.r.t. to the CMB restframe (i.e., the ``cosmological'' redshift, with no peculiar velocities from either the source or the observer) and by $\zhel$ the redshift for an observer in the Sun's frame of reference\footnote{We neglect the distinction between geocentric and heliocentric frames of references (the difference due to the $\sim 30$ km/s orbital speed of the Earth is of order $\Delta z \sim 10^{-5}$), since redshift measurements are routinely reported in the heliocentric frame and also already corrected for atmospheric refraction.}. The measured redshift in our heliocentric frame of reference is given by $\hat{z}_{\rm hel}$, and it differs from $\zhel$ by measurement noise. The redshift in our heliocentric frame of reference, $\zhel$, differs from the redshift of a comoving observer, $\zbar$, by virtue of peculiar velocities of the source and the observer, and gravitational red/blueshifts due to the local gravitational potential at the location of the source and observer. In the following, we neglect gravitational effects, which are subdominant (see however~\cite{Calcino_2017}) and focus instead on the impact of peculiar velocities. 

The relationship between heliocentric redshift, $\zhel$, and the redshift of a comoving galaxy as seen by an observer at rest w.r.t. the CMB, $\zbar$ is given by: 
\begin{align}
(1+\zhel) & = (1+\zCMB)(1+\zsol) \label{eq:redshift_relation}\\
(1+\zCMB) & = (1+\zbar)(1+\zSN)\label{eq:CMB_redshift_relation}
\end{align}
where $\zsol$ is the redshift induced by the peculiar velocity of the Solar System w.r.t. the CMB restframe, while $\zSN$ is the redshift caused by the peculiar velocity of the \SN\ w.r.t. the CMB frame. The second equality introduces the redshift in the CMB restframe,  $\zCMB$, i.e., the frame in which our motion w.r.t. the CMB has been removed\footnote{A source of confusion in the literature is the widespread use of the term ``CMB restframe'' to denote what we call $\zbar$ (i.e., the cosmological redshift, with no peculiar motions from either source nor observer). This misleading nomenclature is for example used by  \cite{betoule2014}, as well as in the data products of the JLA data release.}.
With the above definitions, we can write the luminosity distance to redshift $\zbar$, as~\citep{Davis2011}
\begin{equation}
\label{eq:dL_LCDM}
\begin{split}
    d_{L}(\zbar, \zsol,\zSN ,\pars_1) & = \frac{c}{H_{0}}\frac{(1 + \zbar)(1+\zsol)(1+\zSN)^2}{\sqrt{|\Omega_\kappa|}} \times \\ 
    & \mathrm{sinn} \left\{\sqrt{|\Omega_\kappa|}\int_{0}^{\zbar}\frac{dz}{E(z)}\right\}.
\end{split}
\end{equation}
$E(z)$ depends on our choice of cosmology and for the \lcdm{} universe, is given by 
\begin{equation}
E^{2}(z) =\Omega_{M}(1 + z)^{3} + \Omega_{\Lambda} + \Omega_\kappa(1+z)^{2}.
\end{equation}
The sinn$(x)$ function is defined as
\begin{equation}
    \text{sinn}(x) = \begin{cases}
                x & \quad\text{if}\quad \Omega_\kappa =0\\
                \text{sin}(x) & \quad\text{if} \quad\Omega_\kappa < 0\\
                \text{sinh}(x) & \quad\text{if} \quad\Omega_\kappa > 0\\
            \end{cases}.
\end{equation}

In our model-independent approach, we follow \cite{visser2004} and Taylor-expand the scale factor of the FLRW metric up to third order in time around $t_0$ (today), as:
\be
a(t) = a_0 \{1 + H_0(t-t_0) - \frac{1}{2}q_0H_0^2 (t-t_0)^2 + \frac{1}{3!}j_0 H_0^3 (t-t_0)^3 + O([t-t_0]^4)\}
\ee
where $q_0$ is the dimensionless deceleration parameter, defined as 
\begin{equation}
    q_{0} = -\frac{1}{a} \frac{\mathrm{d}^{2} a}{\mathrm{d}t^{2}}\left[\frac{1}{a}\frac{\mathrm{d} a}{\mathrm{d}t}\right]^{-2}_{t=t_{0}}
\end{equation}
and $j_0$ is the so-called ``jerk'',
\begin{equation}
    j_{0} = +\frac{1}{a} \frac{\mathrm{d}^{3} a}{\mathrm{d}t^{3}}\left[\frac{1}{a}\frac{\mathrm{d} a}{\mathrm{d}t}\right]^{-3}_{t=t_{0}},
\end{equation}
which is also dimensionless.
This model-independent expansion only relies on the FLRW metric but makes no assumption about the underlying matter-energy density, and leads to the following form of the luminosity distance at redshift $\zbar$: 
\begin{equation}
    \label{eq:cosmographic_dl}
    \begin{split}
    & d_{L}(\zbar, \zsol,\zSN ,\pars_2) = \frac{(1+\zsol)(1+\zSN)^2}{1+\zbar}\frac{c\zbar}{H_0} \times \\ 
    & \qquad \left[1 + \frac{1}{2} (1 - q_{0})\zbar - \frac{1}{6}(1 - q_{0} - 3q_{0}^{2} + j_{0} - \Omega_\kappa)\zbar^{2} + O(\zbar^3)\right],
\end{split}
\end{equation}
where $c$ is the speed of light and the model-independent parameters are $\pars_2 = \{H_0, q_0, j_0, \Omega_\kappa \}$. 
From expressions \eqref{eq:dL_LCDM} or \eqref{eq:cosmographic_dl}, we obtain the isotropic distance modulus, $\mu_I$, using the standard formula
\begin{equation}
    \label{eq:dist_mod}
    \mu_I(\zbar, \zsol,\zSN,\pars_a) = 25 + 5\log_{10}\frac{d_{L}(\zbar, \zsol,\zSN,\pars_a)}{1 \text{ Mpc}},
\end{equation}
where $a=1,2$ depending on the chosen parameterization.

There are several different ways one can parameterize the possibility of anisotropic expansion, depending on the underlying physical origin for the effect. A spherical harmonics expansion introduces, to lowest order, a dipolar modulation in the direction of a \SN\ situated at redshift $\zbar$ and in direction $\nsn$ in the sky, with $\nsn$ a unit vector pointing from the centre of the coordinate system (the Earth) to the location of the \SN\ on the celestial sphere. Different authors have taken different approaches in the literature, with no consensus as to which quantity should be modulated: one could expand the scale factor $a(t)$, the luminosity distance, the comoving distance, the Hubble parameter, the matter density, the cosmological constant density, or the distance modulus. Each of these possibilities leads to a different anisotropic imprint onto the Hubble-Lema\^itre law. A dipole moment that is constant with distance, $r$, in the peculiar velocity field (i.e., a bulk flow) leads to 
\be
c \zbar \approx H_0 r + D_v (\ndip \cdot \nsn), 
\ee 
By contrast, a constant dipole in either $H_0$ or $r$ leads to 
\be
c \zbar \approx H_0 r + H_0 D_H r  (\ndip \cdot \nsn), 
\ee
which increases linearly with distance. Another possibility is to modulate the distance modulus directly:   
\begin{equation} \label{eq:dipolar_modulation}
\mu = \mu_I(\zbar, \zhel, \pars_a)\left(1 + D_\mu F(\zbar) (\ndip \cdot \nsn)\right), 
\end{equation}
where $F(z)$ is a function of redshift alone which can be used to localize the dipole at a given length scale. Yet another approach, adopted by C19, is to model the dipole on the deceleration parameter, $q_0$, in a Cosmographic expansion:
\be \label{eq:dipolar_modulation_q}
q_0(z) = q_m + D_{q_0} F(\zbar) (\ndip \cdot \nsn).
\ee
In this paper, we add the dipole to either the distance modulus, Eq.~\eqref{eq:dipolar_modulation}, or to the deceleration parameter, Eq.~\eqref{eq:dipolar_modulation_q}, and consider both $F(z)=1$ and, following C19, an exponentially decaying function of redshift with characteristic scale given by the free parameter $S$, namely  $F(z) = \exp{(-z/S)}$.  These two forms have the effect of either creating a dipole that is constant in redshift or constrained to a local scale which could arise for reasons such as existing within a cosmic void.

As noted in previous works, a phenomenological approach as the one taken here that perturbs an underlying FLRW metric may not be entirely consistent. An alternative route would require specifying a physical model for the anisotropy, and then derive the ensuing predictions for the distance modulus and compare those with observations, as done for example in the context of an ellipsoidal universe from Bianchi type I models \citep{campanelli2011}. However, the advantage of a purely phenomenological approach is that it remains agnostic about the underlying cause of any anisotropy, and it provides constraints on the level of anisotropy that can then be applied to other models.

\subsection{Bayesian Hierarchical Model}
\label{sec:BAHAMAS}

In this paper, we improve on previous works constraining anisotropy from \SNe\ data by adopting a fully Bayesian hierarchical model (BHM) for the statistical analysis of \SN\ data, called \bahamas{}. ``Hierarchical'' here refers to the existence, within the model, of a layer of unobserved (so-called ``latent'') variables for each \SN, corresponding to the true value of their light-curve-derived properties (as opposed to the noisy measured value). The latent variables are marginalised over in the inference, and are constrained in virtue of the fact that they are all generated from the same underlying population distribution, which is modelled with a set of hyperparameters, themselves determined from the data. This approach has been shown to have better coverage statistics overall than the $\chi^{2}$ method traditionally employed, and leads to a reduction of mean squared errors for the recovered cosmological parameters by a factor of $\sim$ 2 - 3 when deployed on simulated data ~\citep{march2011,shariff2016}. Furthermore, the \bahamas{} framework can be used for principled Bayesian model comparison of the kind we perform in section~\ref{sec:model_comparion}, as it correctly marginalizes out all nuisance parameters. By contrast, the heuristic $\chi^2$ approach traditionally adopted is an approximation to the \bahamas{} likelihood, and cannot be used to compute Bayes factors or for formal Bayesian model comparison. 

The Bayesian methodology pioneered in~\cite{march2011} has been adopted and extended in several other papers, including e.g.,~UNITY~\citep{rubin2015}, Steve~\cite{hinton2018} and Simple-BayeSN~\citep{Mandel_2017}. (See also \cite{Nielsen16} for a profile likelihood interpretation.) Here we briefly summarize our Bayesian Hierarchical model, which builds on \bahamas{}, referring the reader to~\citep{march2011,shariff2016} for fuller details. 

We denote with a hat symbol observed quantities, in order to distinguish them from the latent (i.e., unobserved) variables in our model.
For each \SN\ $i$, the data $d_i$ can be summarised by a vector 
\begin{equation}
	d_{i} = \{\hat{z}_{{\rm hel}, i}, \hat{c}_{i},\hat{x}_{1_i},\hat{m}_{B_i}, \hat{\Sigma}_{C,i} \},
\end{equation}
where $\hat{z}_{{\rm hel}, i}$ is the observed heliocentric redshift, $\hat{m}_{B_i}$ is the observed peak B-band apparent magnitude, $\hat{x}_{1_i}$ and $\hat{c}_{i}$ are observed `stretch' and `colour' corrections, which are summary statistics of the lightcurve of the \SN\ obtained with the lightcurve fitter SALT2 \citep{guy2005,guy2007} during the standardization procedure. Furthermore, $\hat{\Sigma}_{C,i} = \text{Cov}(\hat{c}_{i}, \hat{x}_{1_i}, \hat{m}_{B_i})$ is a  $3\times3$ variance-covariance matrix that describes the measurement error on the observables. On the standard deviation scale, the measurement error for redshift for \SN\ with spectroscopic follow-up is $\sigma_z^{\rm sp_{SN}} \sim 5\times 10^{-3}$ when the redshift is determined from the \SN\ spectrum alone, and $\sigma_z^{\rm sp_{host}} \sim 5\times 10^{-4}$ when it is obtained from host-galaxy spectra~\citep{2008AJ....135.1766Z}. The redshift measurements are independent from each other and from all other observables. We discuss this uncertainty further in section~\ref{sec:pecvel_error}. (See also \cite{Calcino_2017} for the potentially important impact of systematic redshift errors as small as $\Delta z \sim 10^{-4}$.) 

In \bahamas{}, we introduce latent variables for each \SN\, in order to model each source of uncertainty according to its origin: measurement error, population scatter and intrinsic (residual) variability. A probabilistic hierarchical model is built as follows: each \SN\ has latent variables $M_{i}^\epsilon$, $x_{1_i}$ and $c_{i}$, representing the objects' ``true'' (i.e., noiseless) absolute magnitude, stretch correction and colour correction, respectively. These latent variables follow normal distributions, representing population variability of the SNe and parameterised by their means and variances: 
\begin{align}
    x_{1_i} &\sim \mathcal{N}\left(x_{1\star},R_{x_1}^2\right),\\
    c_i &\sim \mathcal{N}\left(c_{\star},R_c^2\right),\\
    M_i^\epsilon &\sim \mathcal{N}\left(M_{0}^\epsilon, \sigmares^{2}\right),
\end{align}
where $x_{1\star},c_{\star}$ and $M_{0}^\epsilon$ are the population means and $R_{x_1}^{2}, R_{c}^{2}$ and $\sigmares^{2}$ are the population variances, all of which are also estimated from the data. We collect the population-level parameters in a vector of variables $\paramspop \equiv \{x_{1\star},c_{\star}, M_{0}^\epsilon,R_{x_1}^{2}, R_{c}^{2}, \sigmares^{2}\}$. The intrinsic magnitude of each \SN\, $M_i$, is modified by applying the linear `Tripp relation' \citep{Tripp1998}, so that $M_i \rightarrow M_i^\epsilon \equiv M_i + \alpha {x}_{1_i} - \beta {c}_{i}$, where $\alpha$ and $\beta$ are nuisance parameters that control the slope of the stretch and colour correction, respectively. Therefore, $M_i^\epsilon$ is a linear function of $M_i$ that features a lower population variance, represented by $\sigmares^2$. In astrophysical parlance, $M_i^\epsilon$ is referred to as ``the corrected intrinsic magnitude'' of the \SN. Thanks to the standardization procedure, the residual standard deviation of the \SNe\ type Ia's corrected intrinsic magnitude can be sufficiently reduced so that they can be used as luminosity distance indicators. \cite{shariff2016} used the \bahamas\ model to determine the residual standard deviation of the corrected intrinsic magnitude, finding (from JLA data) a value $\sigmares \sim 0.104 \pm 0.005$, similar to (if somewhat smaller than) the value obtained with a $\chi^2$ method by \cite{betoule2014}. At the latent level, the apparent peak magnitude $m_{Bi}$ is related to the standardised intrinsic magnitude $M_i^\epsilon$ via the isotropic distance modulus of Eq.~\eqref{eq:dist_mod}:
\be
m_{Bi} = \mu_I(z_{{\rm hel},i}, {\zbar},_i, \pars_a) - \alpha {x}_{1_i} + \beta {c}_{i} + M_i^\epsilon.
\ee
Finally, the observed values of $\{ \hat{m}_{B_i}, \hat{x}_{1_i}, \hat{c}_{i} \}$ are modelled as normally distributed around their latent values, with variance-covariance matrix given by  $\hat{\Sigma}_{C,i}$. Additionally, a systematic errors covariance matrix that correlates different SNIe (for example, because of calibration uncertainties common between SNIe within the same survey) is included when available. Inference is based on the marginal distribution of the quantities of interest, $\pars_a$, which includes uncertainty at all levels of the hierarchy. 

It has become common practice to split the \SNe\ into two groups, based on their host-galaxy stellar mass~\citep{sullivan2010}, with the mass threshold between the two groups being around  value of $\log_{10}(M_{g}) = 10$, where $M_g$ is the host-galaxy mass measured in solar masses. \cite{smith2020} found a difference of up to 0.04 mag in the average intrinsic magnitude of the two groups, and application of the Tripp relation to the two groups separately may further reduce the residual dispersion in \SN\ absolute magnitudes. (See also~\cite{2021arXiv210205678T} for a similar result obtained using a Bayesian hierarchical approach.) However, \cite{2021ApJ...909...26B} have recently cast doubt on the robustness of this mass-step correction, which they ascribed instead to incorrect dust modeling. The ultimate origin of the mass step remains unclear, and it might  relate to stellar population age \citep{childress2014} and metallicity \citep{sullivan2010}. In any case, \cite{shariff2016} showed that adding a mass-step or a more general linear covariate as a function of host-galaxy mass has little impact on the ensuing cosmological parameters inference. Therefore, in this paper we do not adopt a mass-step correction. We refer the reader to \cite{march2011} and \cite{shariff2016} for the full mathematical detail of the \bahamas\ model, marginalization procedure and algorithms used for sampling the resulting posterior distribution.

\subsection{Data}
\label{sec:data}

The largest \SN{} compilation to date is the `Pantheon sample'~\citep{scolnic2018}, which contains 1048 spectroscopically confirmed \SN. This compilation includes 279 new \SN\ discovered by the Pan-STARRS1 (PS1) Medium Deep Survey \citep{panstarrs} in addition to the previous \SNe\ discovered by previous catalogues to create the total.

Unfortunately, the Pantheon sample only provides estimates (and associated uncertainties) of the distance modulus for each \SN\, but does not include the covariance matrices of both the measurement error for each \SN\ and the systematic covariance matrix across the whole data set. Because our Bayesian hierarchical model also requires these covariances over the light curve fit parameters as opposed to the covariance over distance modulus provided by the Pantheon data, we instead use
the smaller `Joint Light-Curve Analysis' (JLA) compilation \citep{betoule2014}. Very recently, a `Pantheon+' dataset has been presented in~\cite{PantheonPlus2022}, increasing the set of spectroscopically confirmed \SNe{} to 1550 in the redshift span $0.001$ to $2.26$. The data products of this larger set have not yet been released, so unfortunately we cannot use these data in our framework.
The JLA data contains 740 \SN\ including 374 \SN{} from the SDSS-II survey \citep{sdss_frieman,sdss_sako}, 239 from SNLS \citep{snls_conley,snls_sullivan}, a low-$z$ sample of 118 \SN\ at $z < 0.1$ which is comprised of numerous smaller surveys and nine Hubble Space Telescope \SN. These have been fit and standardised using the SALT2 Light-Curve Fitter \citep{guy2007}.

\begin{figure}
\includegraphics[width=\columnwidth]{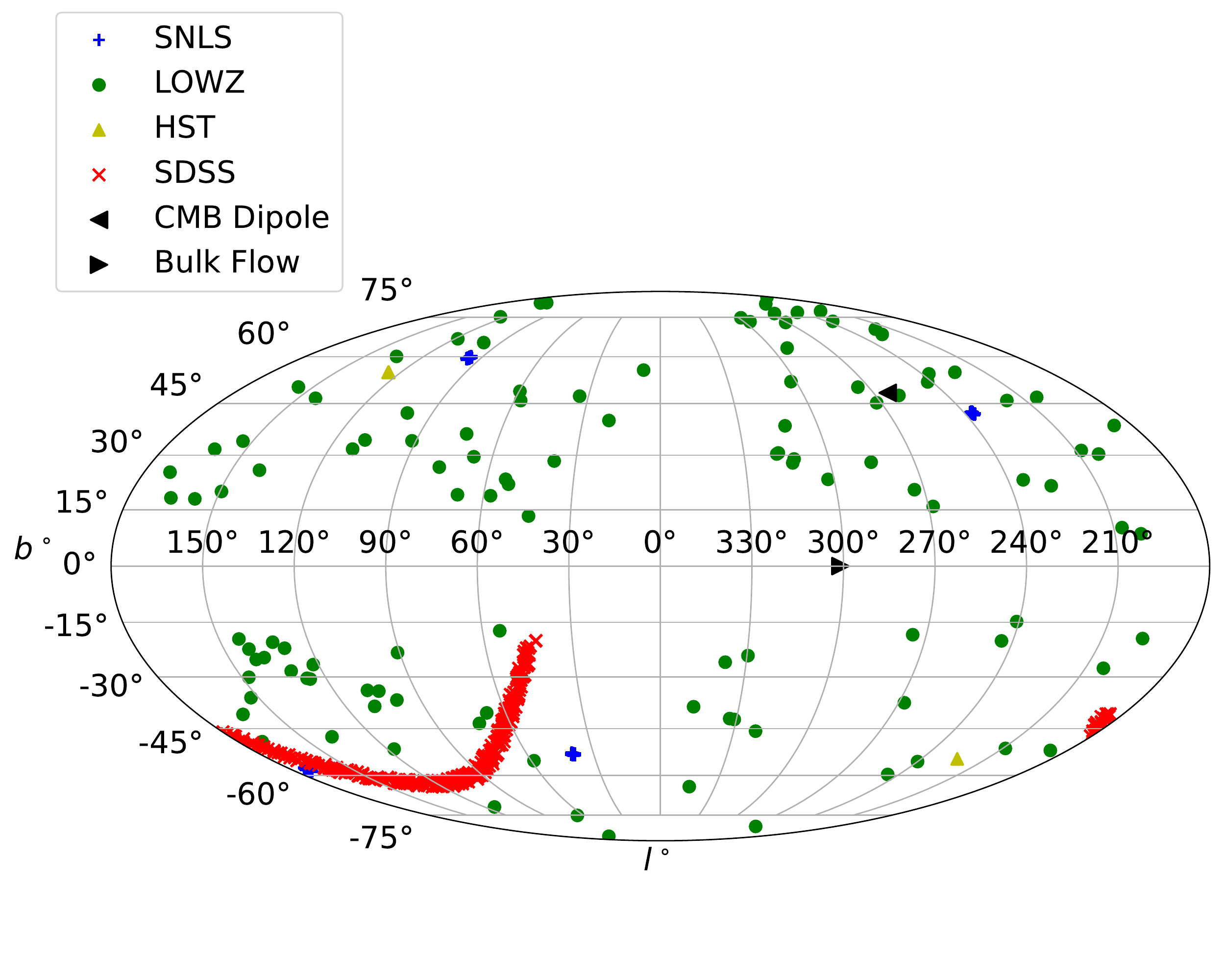}
\caption{A plot of the JLA \SNe\ showing their location in the sky in galactic coordinates, as well as the direction of the CMB dipole (left-pointing triangle) and bulk flow (right-pointing triangle), as determined by \protect\cite{RN339}. }
\label{fig:jla_mollweide}
\end{figure}

An overview of the distribution of JLA objects in the sky is provided in Fig. \ref{fig:jla_mollweide}. The long stripe in the lower left hemisphere is from the SDSS objects. It is clear to see that the distribution of the JLA objects in the sky is highly anisotropic.

\subsection{Accounting for Colour-dependent Selection Effects}
\label{sec:colour_selection}

Towards the high end of the redshift range of a given survey, \SNe\ that are intrinsically brighter (smaller $M_i$) or bluer (smaller $c_i$) are more likely to be observed and to be followed up spectroscopically to confirm their type. This selection bias must be accounted for to avoid bias in the estimates of the
cosmological parameters. Ignoring magnitude-based selection effects leads to estimates of the distance modulus that are biased low: at a given redshift, the average observed magnitude is smaller (i.e., the observed peak flux brighter) than the population mean, which leads to an estimated distance modulus that is biased low. This effect reduces or even obliterates the preference for a non-zero cosmological constant. Traditionally, this has been addressed by ``correcting back'' the estimates of the distance modulus by the average bias in each redshift bin, established with forward simulations of data subject to selection effects. This approach is adopted e.g., by ~\cite{betoule2014}. More recently, this method has been extended and refined with the so-called ``BEAMS with Bias Corrections'' (BBC) method~\citep{2017ApJ...836...56K}. The JLA analysis only corrected for magnitude-based selection effects, concluding that no additional correction were necessary for colour (\cite{betoule2014}, Fig.~11), despite observing a downward trend in observed colour with redshift for SNSL and SDSS. 

In the context of a Bayesian analysis, however, selection effects are treated differently: the posterior is conditional on the observed data \citep{2007ApJ...665.1489K}, which leads to a re-weighing factor increasing the statistical weight of \SNe\ that are less likely to be observed (see Eq.~\eqref{eq:like_selection_effects} below).  \cite{rubin2015} introduced a general formalism for the selection function that was further developed by \cite{hinton2018}. In practice, however, this formalism typically requires several simplifying assumptions that may be difficult to justify (e.g., a well-sampled \SN{} redshift distribution, a selection function that is described by a normal cumulative distribution function (CDF), independence of the selection probability from the underlying cosmology). We will present an improved selection effects treatment in an upcoming, dedicated work. 

\cite{RH16} argue that uncorrected-for colour-dependent selection effects remaining in the JLA data (after bias correction of the data) should be addressed by introducing a population colour mean that is both redshift- and survey-dependent. However, at the top level of the BHM we would like a \emph{physically-meaningful} population mean that describes the underlying, population-level latent mean colour, itself a reflection of the physical properties of the \SNe{}. Such a colour mean can be a function of redshift, to reflect evolution in the physical properties of \SNe\ with lookback time, but it cannot be survey-dependent, for clearly the latter dependency is caused by survey-specific selection effects and, in virtue of being survey-specific, {\em cannot} be the consequence of changing underlying physical properties of the \SNe\ being observed. 
Thus, by using population-level variables to address a survey-induced selection effect, the method advocated by \cite{RH16} goes against the physical interpretability of the BHM; we prefer the population-level variables to be tied to the physics of the \SN\ explosion mechanism rather than the survey-induced selection effects. Furthermore, as pointed out by  \cite{2017MNRAS.472..835D}, modeling residual colour drift with redshift as advocated by \cite{RH16} introduces undesirable degeneracies with cosmological parameters in the cosmographic expansion.

Here, we account for residual colour-based selection effects with an approximate method that captures the spirit of the correct Bayesian procedure. (We will present a more complete exposition in an upcoming work.) In general, we aim to base the likelihood function on the distribution of the data $d_i$ for \SN{} $i$ conditional on it having been observed, which is denoted by an indicator variable $I_i = 1$. ($I_i=0$ would indicate that \SN{} $i$ is not observed.) Denote by $\hat{D} = \{ \hat{d}_1, \dots, \hat{d}_n \}$ a random sample of \SNe\ where all \SNe\ have equal probability of being observed (i.e., there are no selection effects in $\hat{D}$). In the presence of selection effects, we observe a non-representative sample of $\nobs < n$ \SNe\ from $\hat{D}$, whereby a \SN{} $i$ is observed with probability given by the selection function $p(I_i=1|\hat{d}_i, \Psi)$, where $\hat{d}_i$ are the observed (noisy) data, and $\Psi$ are parameters describing the selection function (assumed known). The distribution of the observed $\hat d_i$, conditional on \SN{} $i$ being observed, is given by 
\be
p(\hat{d}_i | I_i=1, \Psi, \paramsBHM) = \frac{p(I_i=1|\hat{d}_i, \Psi)p(\hat{d}_i|\paramsBHM)}{p(I_i =1| \Psi, \paramsBHM)},
\label{eq:likelihood_with_selection}
\ee
where $\paramsBHM = \{\pars_a, \paramspop \}$ are the parameters of the hierarchical model (including the parameterization of the distance modulus, $\pars_a$, and the population-level distribution parameters, $\paramspop$). We assume that the selection probability conditional on the observed data, $\hat{d}_i$, i.e., $p(I_i=1|\hat{d}_i, \Psi)$ in the numerator of \eqref{eq:likelihood_with_selection}, does not depend on $\paramsBHM$ and thus is an ignorable constant if (as we assume here) the selection function and its parameter $\Psi$ are known. 
The quantity $p(\hat{d}_i|\paramsBHM)$ is the likelihood in the absence of selection effects, and the denominator gives the probability of observing a \SN{}, irrespective of the value of the data: 
\be \label{eq:sel_func_int}
p(I_i =1| \Psi, \paramsBHM) = \int {\rm d} \hat{d}_i p(I_i=1|\hat{d}_i, \Psi) p(\hat{d}_i | \paramsBHM).
\ee 

Omitting the ignorable multiplicative constants in the numerator of \eqref{eq:likelihood_with_selection} yields the likelihood function of $\Theta$ including selection effects for a sample of $\nobs$ observed \SNe{}, $\dobs \equiv \{\hat{d}_1, \dots, \hat{d}_\nobs \}$, 
\be \label{eq:like_selection_effects}
p(\dobs | \{I_i=1\}_{i=1}^{\nobs}, \Psi, \paramsBHM) \propto  \frac{p(\dobs | \paramsBHM)}{p(I=1| \Psi, \Theta)^\nobs},
\ee
where we have dropped the dummy index $i$ in the denominator. Therefore, selection effects are accounted for by dividing the likelihood function of the observed data, $p(\dobs | \paramsBHM)$, by a ``correction factor'' that gives the probability of making $\nobs$ observations.  

Thus far, we have been entirely general. Next we specify the form of the selection function, $p(I_i=1|\hat{d}_i, \Psi)$, entering in Eq.~\eqref{eq:sel_func_int}. The probability of a \SN{} being selected, and spectroscopically followed-up to determine its type, depends primarily on its magnitude and colour. (\SNe{} with larger stretch parameter $x_1$ are slower declining and thus remain visible and potentially detectable for longer, but this effect is subdominant.) The data correction procedure in \cite{betoule2014} should in principle account for both magnitude- and colour-based selection, but their discussion makes it clear that there are large uncertainties in the determination of the selection probability that enters their forward simulation of the data. For example, \cite{betoule2014} mention that the SDSS spectroscopic follow-up target selection favours intrinsically bluer \SNe{}, introducing complex colour-dependency in the selection function. In light of such difficult-to-simulate selection effects, we advocate a method that estimates any residual selection effect (after the data correction procedure of \cite{betoule2014}) directly from the observed data.   

We wish to account for residual colour-based selection effects that may remain in the data. Therefore, we assume that $p(I_i=1|\hat{d}_i, \Psi)$ depends only on $\hat{c}_i$ and $\hat{z}_{{\rm hel}, i}$, with the $N_s$ redshift bins for survey $s$ chosen as discussed in section~\ref{sec:data}, and factorize both the selection function and the likelihood in a product over redshift and survey bins, assumed independent of each other. Within each redshift, and survey bin, we allow a different selection function, which is derived below. With these assumptions, the probability of observing $\nobs$ \SNe\ is (with the shorthand notation $I_\nobs = 1$ denoting $\{I_1 = 1, \dots, I_\nobs = 1\}$):

\be \label{eq:sel_func_int_factorized}
p(I_\nobs=1| \Psi, \paramsBHM) = \prod_{s=1}^4\prod_{j=1}^{N_s} \left(\int {\rm d} \hat{d}_i p_{sj}(I_i=1|\hat{d}_i, \Psi) p_{sj}(\hat{d}_i | \paramsBHM)\right)^{N_{sj}} ,
\ee 
where $N_s$ is the number of bins for survey $s$ and $N_{sj}$ the number of observed \SNe{} in bin $sj$.  Within each redshift bin for survey $s$, we parameterize the selection function as a normal cumulative distribution function (CDF), and assume that we observe a \SN\ $i$ with colour $\ci$ with probability:
\be \label{eq:sel_funct}
p_{sj}(I_i=1|\hat{d}_i, \Psi)= \Phi\left(\frac{\cobs - \ci}{\sigmaobs}\right),
\ee
where  %
\be
\Phi(x) = \int_{-\infty}^x \norm_y(0,1) dy
\ee
is the CDF of a standard normal, and $\norm_y(\mu, \sigma^2)$ is a Gaussian distribution in $y$ with mean $\mu$ and variance $\sigma^2$.
In Eq.~\eqref{eq:sel_funct}, $\cobs$ is the colour value at which there is a 50\% probability  of observing a \SN{} {\em in redshift bin $j$ and for survey $s$}; $\sigmaobs$ denotes the width of the transition from the regime where all objects are observed, i.e. for $(\cobs - \ci)/\sigmaobs \ll 0$, to the regime where no objects are observed, where $(\cobs - \ci)/\sigmaobs \gg 0$, for the bin $sj$ being considered. In Eqs.~\eqref{eq:MC1}-\eqref{eq:MC2} below we show how to estimate the selection function parameters $\Psi = \{ (\cobs, \sigmaobs)\}$ ($s=1,\dots, 4, j=1,\dots. N_s$). As an approximation, we ignore uncertainty in the resulting estimates and assume the parameters are known exactly\footnote{We ignore uncertainty in our estimates of $\Psi$, and we defer the evaluation of the impact of this approximation to an upcoming, dedicated work.}.  

In principle, we would like to use the likelihood function of \cite{march2011} and \cite{shariff2016} as the second term of the integrand in Eq.~\eqref{eq:sel_func_int}. Since we only wish to account for residual colour-based selection effects, however, we ignore the part of the \bahamas\ likelihood that relates colour to magnitude via the Tripp linear relation (Eq. (C2) in \cite{march2011}), and instead only consider the distribution of colour values that one would obtain when integrating out the latent colour variables conditional on all other variables in the BHM, leading to the simple expression for the likelihood entering into Eq.~\eqref{eq:sel_func_int_factorized}:
\be \label{eq:likelihood_colour}
p_{sj}(\hat{d}_i | \paramsBHM) =  \norm_\ci(c_\star, R_c^2+\bar{\sigma}_{c, sj}^2),
\ee
where $\bar{\sigma}_{c,sj}$ is the average colour measurement error for the $n_{sj}$ \SNe{} in bin $sj$ (for simplicity, we assume all $n_{sj}$ \SNe{} in bin $sj$ have the same colour measurement error, given by $\bar{\sigma}_{c}$; we also ignore correlation between colour and stretch and magnitude) 

Eq.~\eqref{eq:likelihood_colour} features a redshift-independent conditional expectation of colour, described by $c_\star$.  
This formalizes the assumption of \cite{RH16} within \bahamas{} that the observed drift to bluer \SNe\ with redshift within a survey is a consequence of selection effects and not of a change in the underlying population colour distribution with redshift.

With the above elements, we can compute the probability of observing \SN{} $i$ in redshift bin $j$ for survey $s$ by integrating over its colour, $\ci$, obtaining \footnote{It is useful to recall that $\int_{-\infty}^{\infty}\Phi\left(\frac{\mu-x}{\sigma}\right) \norm_x(\nu, \tau^2) dx =  \Phi\left(\frac{\mu-\nu}{\sqrt{\sigma^2+\tau^2}}\right)$.}
\be \label{eq:correction_factor}
p_{sj}(I_i =1| \Psi, \paramsBHM) = \Phi\left( \frac{\cobs - c_\star}{\sqrt{(\sigmaobs)^2 + R_c^2 + \bar{\sigma}_{c, sj}^2}}\right).
\ee
For a given $sj$ bin, when $c_\star \ll \cobs$ selection effects are irrelevant, because the survey is seeing the entire colour population, and accordingly $p(I_i =1| \Psi, \paramsBHM) \rightarrow 1$ from Eq.~\eqref{eq:correction_factor}. However, $p(I_i =1| \Psi, \paramsBHM)$ becomes smaller for values of $c_\star > \cobs$, with the difference measured in units of the total standard deviation, i.e., when the survey is preferentially seeing the bluer part of the population because of colour-based selection bias. In this case, the observed distribution of the $N_{sj}$ objects in the $sj$ bin deviates from the latent distribution, and the likelihood that ignores selection effects would incorrectly penalize this value of $c_\star$. According to Eq.~\eqref{eq:like_selection_effects}, the correction factor in the denominator of Eq.~\eqref{eq:like_selection_effects} increases the weight given to observed \SNe{} with $c_\star > \cobs$.

Finally, there remains the issue of determining the value of $\Psi$, the selection function parameters. Ideally, one would do so from forward simulation of surveys, but this is unpractical for our purposes, and unfeasible for the low-$z$ sample, which is obtained from a collection of telescopes with poorly understood selection functions. Furthermore, as noted earlier, colour-based selection effects might escape {\em ab initio} modeling of this kind, as indicated in e.g.~\cite{betoule2014}. As an alternative, we estimate the value of $\Psi$ in each redshift- and survey-bin $sj$, by matching the first and second moment of the empirical colour distribution within the bin to the marginal distribution of colour based on the right-hand-side of \eqref{eq:like_selection_effects}, understood as a distribution over observed colour values $\ci$ within each bin. Let $m_C(t)$ denote the moment generating function of the random variable $C$ (the observed colour), defined as 
\be \label{eq:moments_function}
m_C(t) \equiv \int_{-\infty}^{\infty}e^{ty}f_C(y)dy,
\ee
where $f_C(c)$ in our case is given by (in the bin under consideration)
\be \label{eq:fc}
f_C(c) = \frac{\norm_c(c_\star, R_c^2+\bar{\sigma}_{c,sj}^2) \Phi\left(\frac{\cobs - c}{\sigmaobs}\right)}{p_{sj}(I_i =1| \Psi, \Theta)},
\ee
with denominator given by Eq.~\eqref{eq:correction_factor}.
 Our strategy, known as the \emph{method of moments}, is to analytically compute the first two moments of the distribution, set them equal to the empirical moments, and solve the resulting system of equations to obtain estimates of  
 the selection function parameters in each survey and redshift bin, $\{\cobs, \sigmaobs\}$. Details of the calculation are presented in Appendix~\ref{app:moments}, where we show that the moment generating function is given by Eq.~\eqref{eq:moments_function_result}, its first moment, $dM_C(t)/{dt}|_{t=0}$, by Eq.~\eqref{eq:moments_mean}, and its second moment, $d^2M_C(t)/{dt^2}|_{t=0}$, by Eq.~\eqref{eq:moments_var}. Specifically, in each bin, we set the first moment equal to the empirical mean, and set the second central moment equal to the empirical variance: 
\begin{align} \label{eq:MC1}
& \frac{dM_C(t)}{dt}{\Big \vert}_{t=0}  = \langle \ci \rangle, \\ 
& \frac{d^2M_C(t)}{dt^2}{\Big \vert}_{t=0} - \frac{dM_C(t)}{dt}{\Big \vert}_{t=0}^2 = \frac{1}{1-N_{sj}} \sum_{i=1}^{N_{sj}}(\ci - \langle \ci \rangle)^2,  \label{eq:MC2}
\end{align}
where $\langle \ci \rangle = \frac{1}{N_{sj}}\sum_{i=1}^{N_{sj}} \ci$ is the empirical mean. We set the population mean and standard deviation to $c_\star = -0.0022$ and $R_c = 0.0758$, the empirical mean and standard deviation from the lowest two bins of the SDSS, SNLS and Low-$z$ surveys, where colour-based selection effects are expected to be negligible. We then solve the resulting coupled equations to determine $\{\cobs, \sigmaobs\}$.

In cases where ${N_{sj}}$ is small $( \sim10$) and the sample variance is small, the method of moments yields an estimate of $\sigmaobs$ near or equal to zero.
This is similar to a well-known pathology in the estimation of the shape parameter in the skew-normal distribution (see e.g.~\cite{2012arXiv1203.2376A}). A simple solution is to impose a lower cutoff to the value of $\sigmaobs$. We chose a cutoff value $\sigmaobs>0.01$, since any smaller value leads to a selection function indistinguishable from a step function. Tests of our method of moments on simulated data are provided in Appendix~\ref{app:moments}. 

Because our selection effect model assumes in Eq.~\eqref{eq:likelihood_colour} that the colour observations are independent, we set the corresponding covariances in the variance-covariance matrix for the systematic effects to zero. This has a minor effect on our estimates of the cosmological parameters, as those terms are subdominant with respect to other off-diagonal terms. 

\begin{figure*}
    \centering
\includegraphics[width = 2\columnwidth]{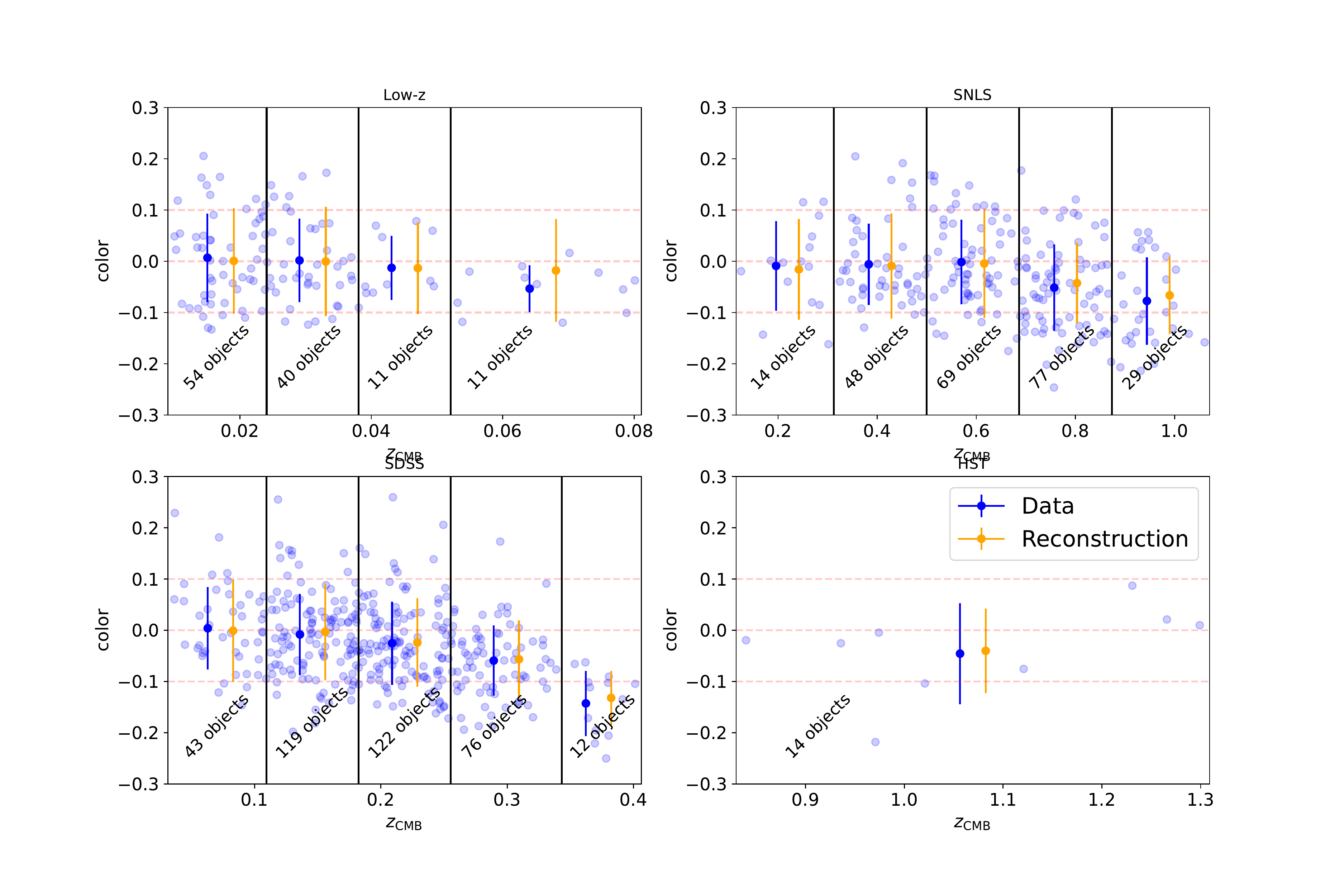}
     \caption{Colour-based selection function in binned JLA data, with redshift bins boundaries indicated by the vertical black lines.The blue circles are the individual \SNe, the blue errorbars represent the data mean and standard deviation within each top-hat bin, while the orange errorbars give the mean and standard deviation of simulated data from the model using the reconstructed selection function in that bin (shifted horizontally for ease of comparison).}
     \label{fig:survey_reconstructions}
\end{figure*}

We apply the method of moments to estimate $\Psi$ to each the four sub-surveys separately, and we verify its accuracy by simulating colour observations from the model of Eq.~\eqref{eq:likelihood_colour}, assuming a Normal constant-in-redshift latent colour distribution with mean $c_\star = 0$ and standard deviation $R_c = 0.1$. We then apply the reconstructed selection function in each redshift bin, and compare the resulting distribution of simulated \SNe\ with the observed sample within that bin. This comparison is shown in Fig.~\ref{fig:survey_reconstructions}. For the simulation study and the real JLA data in Section~\ref{sec:param_reconstruction}, we divided each sub-survey in 5 approximately equal-spaced redshift bins, with the following exceptions: the HST data is treated in one single bin owing to the the small number of \SN{} in HST. For the same reason, in the Low-z survey data, the two highest-redshift bins are combined into a single bin. For the SDSS survey, the right edge of the $4^{\rm th}$ bin has its boundary shifted 0.015 in redshift space to the right, in order to account for a discontinuity in the data, where the population of objects have a lower colour and are separated by a gap in redshift. The agreement between real data and simulation is excellent, thus validating our approach.

\subsection{A New Derivation of Peculiar Velocity Corrections}
\label{sec:peculiar_velocity}

 Our motion w.r.t. the CMB frame is measured precisely by the temperature dipole observed in CMB anisotropies. The most accurate result is from~\cite{collaboration2018planckI}, giving a velocity $\vsol = 369.82 \pm 0.11$ km/s in the direction $l = 264.021^\circ\pm0.011^\circ, b=48.253^\circ\pm 0.005^\circ$. This induces in the non-relativistic limit a redshift correction~\citep{Davis2011} 
\be
\zsol \approx - \frac{\vsol}{c} (\ncmb \cdot \nsn) \lesssim 10^{-3},
\ee
where $\ncmb$ is a unit vector in the direction of the CMB dipole and $\nsn$ is a unit vector in the direction of the \SN. Given the small uncertainties in the measurement for $\vsol$ and the CMB dipole direction, we can consider $\zsol$ as known exactly, and thus neglect measurement error on this quantity (as it is $\sim 3\times10^{-7}$). Estimating $\zSN = \vpecSN/c$ requires knowledge of the peculiar velocity of the \SN\ in the CMB frame, $\vpecSN$. This can be measured either from a peculiar velocity field survey (for example, by using the Fundamental Plane (FP) relation or the Tully-Fisher (TF) relation to measure the distance to a galaxy,
and then subtracting from the observed velocity the expansion component obtained from the Hubble-Lema\^itre law) or derived from linear perturbation theory applied to a smoothed density field. The latter approach has a long history, originally having been used to predict the peculiar velocities of FP and TF samples \citep{Hudson1993, Strauss1995, DavisNusserWillick1996} but more recently applied to peculiar velocity data that include \SNe{} type Ia.
 \citep{RiessDavisBaker1997, Radburn-SmithLuceyHudson2004, PikeHudson2005,  RN224, TurnbullHudsonFeldman2012, 10.1093/mnras/stv547, RN345, Lilow:2021omg, StahldeJaegerBoruah2021}.

While the peculiar velocity of the \SN\ becomes rapidly negligible for $z\gtrsim0.1$, it is important for local objects ($z \ll 0.1$), where it can be significant w.r.t. the expansion velocity (up to $\sim 30\%$) and where it leads to much larger changes in the apparent magnitude, due to the steeper gradient of the distance modulus at low redshift. For example, at $z=0.01$ an uncorrected \SN\ peculiar velocity $\vpecSN$ induces a redshift systematic error $\delta z = \vpecSN/c$, which corresponds to a significant change in the theoretical distance modulus $\delta \mu \approx \frac{d\mu}{dz}\delta z \approx 5/\ln(10)(\delta v/(c z))  = 0.14$~mag for $\vpecSN = 200$ km/s. To avoid difficulties with peculiar velocities, earlier \SN\ cosmological analyses routinely adopted a lower redshift cutoff $z_{\rm cut}$, removing \SNe\ below  $z_{\rm cut}$; for example, \cite{Kessler2009} used $z_{\rm cut} = 0.02$; \cite{2007ApJ...659...98R} used $z_{\rm cut} = 0.023$. Recently, \cite{2020arXiv201005765H} estimated the impact of uncorrected peculiar velocities on the Pantheon sample from numerical N-body simulations, and recommended a cutoff $z_{\rm cut} = 0.02$ to protect against significant bias to cosmological parameters. However, a better way that does not discard useful data at low redshift is to assign uncertainties that scale with distance, as we do here.  

The JLA sample contains 37 \SNe\ with $\zhel < 0.02$, and 110 with $\zhel < 0.05$, for which an appropriate treatment of peculiar velocities is required if they are to be used in the cosmological analysis -- particularly in our case, where we wish to use them to constrain a local dipole in the expansion. To first order in redshift
 Eq.~\eqref{eq:redshift_relation} gives 
\be \label{eq:zbar_linear}
\zbar = \zhel - \zsol - \zSN,
\ee
meaning that the redshift of a comoving observer, $\zbar$, is obtained from the measured heliocentric redshift by subtracting our local dipole ($\zsol$) and the redshift due to the \SN\ peculiar velocity, $\zSN$. 

The model used in~\cite{betoule2014} to estimate $\zSN$ has been criticised by~C19, who highlighted potential bulk flow velocity discontinuities at $z=0.04$, pointed out that peculiar velocity corrections arbitrarily disappear beyond $200/h$ Mpc ($z \sim 0.067$, the limit of the galaxy density field measurements from which the peculiar velocities were derived) and that the residual uncorrelated velocity dispersion of $\sigma_v = 150$ km/s might be underestimated. While~RH20 pointed out technical flaws with the analysis of~C19, it is important in the light of this valid criticism to revisit the issue of low-redshift peculiar velocity corrections here. 

To this end, in this work we replace the peculiar velocity corrections used by~\cite{betoule2014} -- which rested on the IRAS PSCz catalogue from~\cite{1999MNRAS.308....1B} -- with the more recent ones obtained by~\cite{10.1093/mnras/stv547}. We follow \cite{RN339,RN345}, who carried out a thorough comparison between density reconstruction from galaxy redshift surveys and kernel smoothing of peculiar velocity data methods. We adopt here their peculiar velocity field inferred from 69,160 galaxies from the 2M++ galaxy redshift catalogue~\citep{10.1111/j.1365-2966.2011.19233.x}. The catalogue covers almost the entire sky (with the notable exception of the plane of the galaxy),  is highly complete out to 200$/h$ Mpc ($z \sim 0.067$) in the region covered by 6dF and SDSS, and out to 125$/h$ Mpc ($z \sim 0.041$) in the region covered by 2MRS.   
We thus remove\footnote{Differently from C19, we do not remove the magnitude bias corrections made to the JLA \SNe, as they are important to account for selection effects, nor do we neglect the contribution of peculiar velocities uncertainty to the covariance matrix, which we re-derive for our case.} the \SN\ peculiar velocity corrections for the low-$z$ JLA sample that are in common with the A2 sample of \cite{RN339} (107 objects), and replace them with new values obtained as follows. 

The radial peculiar velocity for a \SN\ at comoving distance $r$ and direction $\nsn$ is obtained from the luminosity-weighted density field $\vgal$ as 
\be \label{eq:flow_model}
\vpec(r, \nsn, \Flowpars) = \nsn \cdot \left( \beta_v \vgal(r, \nsn) + \Vext \right) 
\ee
where $\Flowpars = \{\beta_v, \Vext \}$, with a rescaling parameter $\beta_{v} = 0.411 \pm 0.020$ and external residual bulk flow velocity (in Galactic Cartesian coordinates) $ \Vext = [52 \pm 20, -163 \pm 21,   49 \pm 16] $ km/s (how we treat and propagate the uncertainties in these values is addressed in section \ref{sec:pecvel_error}). We follow the methodology of \cite{RN339}, with the difference that we only use the SFI$++$ peculiar velocity sample (therefore not including A2 \SNe{} data) in order to avoid circularity (i.e.\ using \SNe\ data to predict the peculiar velocity correction for the same \SNe\ data).

\begin{figure*}
\includegraphics[width=2\columnwidth, trim={0 12cm 0 2cm},clip]{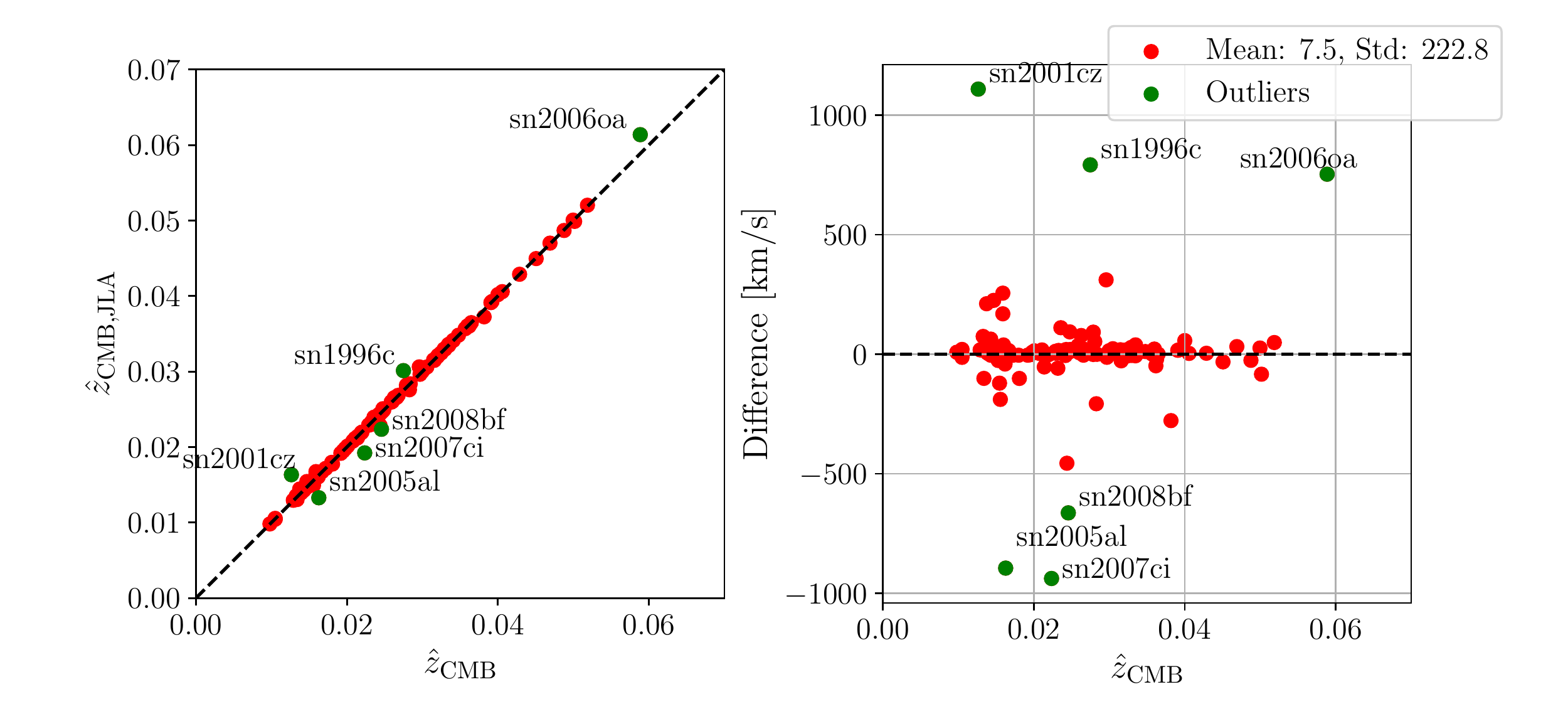}
\caption{Left panel: comparison of the CMB restframe redshift used in this work for the low-redshift sample (horizontal axis, 107 \SNe) with that of the JLA analysis (vertical axis). There are 6 \SNe\ (highlighted in green) with a difference exceeding 3 standard deviations around the identity line (dashed). Right panel: the same comparison but showing the fractional differences between redshifts on the vertical axis.}
\label{fig:zCMB}
\end{figure*}

\begin{figure}
\includegraphics[width=\columnwidth, trim={2cm 20cm 18cm 3cm},clip]{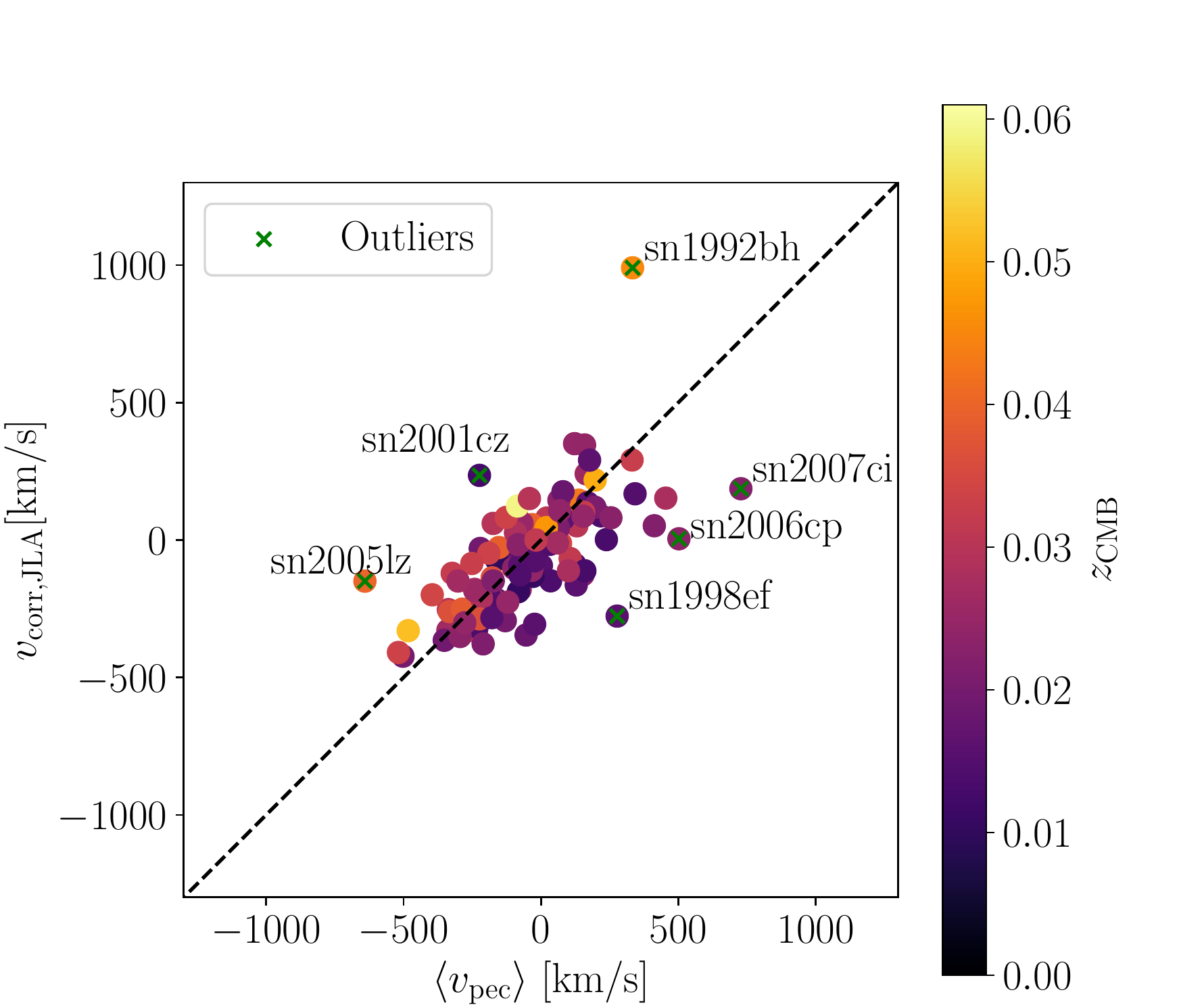}
\caption{Comparison of the peculiar velocity corrections used in this work (horizontal axis) with those adopted in the JLA analysis (vertical axis) for the 107 \SNe\ in the A2 low-redshift sample common to the JLA data. The colour coding gives the CMB restframe redshift used in this work. Outliers (defined as more than 3 standard deviations away from the dashed identity line) are labeled.}
\label{fig:pec_vel}
\end{figure}

We do not wish to use the distance modulus information from a \SN{} at this stage of the analysis, only its observed redshift in the CMB restframe, $\zhatCMB \equiv \hat{z}_{\text{hel}} - \zsol$. Firstly, the observed redshift in the CMB restframe is corrected to the average redshift of the group to which the host galaxy belongs. Differently from JLA, we correct the CMB restframe redshift for {\em all} host galaxies, including those in clusters and poorer groups.
This is necessary to suppress the highly non-linear velocity contribution to the observed redshift, and it leads to deviations of a few percent in $\zhatCMB$ in most low-$z$ \SNe, compared with the value used by JLA, see Fig.~\ref{fig:zCMB}. However, there are 6 \SNe\ that show much larger changes in their CMB frame redshift, up to $\sim 30\%$ (highlighted in green in Fig.~\ref{fig:zCMB}); two of them (\textbf{sn2007ci} and \textbf{sn2001cz}) are in common with the outliers in peculiar velocity, identified in Fig.~\ref{fig:pec_vel}. It is worth noting that \cite{Carr2021} reviewed Type Ia \SNe{} literature and discovered that some \SN\ had been reported with incorrect redshifts and/or positions due to misprints. Two of the outliers in Fig.~\ref{fig:pec_vel}, \textbf{sn2008bf} and \textbf{sn1996c}, are among the \SN{} identified by~\cite{Carr2021} as requiring updates: \textbf{sn2008bf} suffers from an uncertain host galaxy identification, while \textbf{sn1996c} had an incorrect sky location. Many other \SN\ had their redshifts and positions updated by~\cite{Carr2021}, but their dataset is not publicly available yet and therefore we cannot make use of their findings. 

We compute the expected peculiar velocity by marginalizing over the unknown comoving distance of the \SN{}, $r$:  
\begin{equation} \label{eq:average_pecvel}
    \langle \vpec \rangle = \int d r p(r| c\zhatCMB) \vpec(r, \nsn),
\end{equation}
where $\vpec$ is computed self-consistently from the flow model, Eq.~\eqref{eq:flow_model}, and $p(r|\zhatCMB)$ is the probability density function (pdf) for $r$ given the observed redshift. This can be linked via a variable transformation to the pdf for the true (latent) CMB redshift of the \SN, $\zCMB(r)$, via:
 \begin{equation}\label{eqn:p_r}
    p(r|c \zhatCMB) = p(c \zCMB(r)|c \zhatCMB )\bigg|\frac{\partial c \zCMB(r)}{\partial r}\bigg|,
\end{equation}
where from Eq.~\eqref{eq:CMB_redshift_relation}
\begin{equation}\label{eqn:comoving2redshift}
\zCMB(r) = \zbar(r) + (1+\zbar(r))\vpec(r, \nsn)/c,
    %z_{\text{pred}}(r) = z_{\text{cos}}(r) + [1+z_{\text{cos}}(r)]\frac{V_r(r)}{c}.
\end{equation}
and the cosmological redshift at comoving distance $r$, $\zbar(r)$, is computed for the $\Lambda$CDM concordance model, with parameters as in Table~\ref{table:priors}. Note that the transformation between $r$ and $\zCMB$ may not be unique due to the existence of triple-valued regions. However, in practice we checked that the reconstructed velocity field indeed leads to unique transformation between $r$ and $\zCMB$ for the parameters under consideration. Nevertheless, it can lead to `flat' regions in redshift space (i.e, where $\frac{\partial \zCMB}{\partial r} \approx 0$), which result in large uncertainties in the expected peculiar velocity. 

The 2M++ reconstruction employs linear perturbation theory to predict the velocities. As shown in \citet{10.1093/mnras/stv547}, this leads to an uncertainty due to non-linearities of $\sigvNL = 150$ km/s. Therefore, assuming a Gaussian uncertainty, we can write the probability of the predicted redshift in the CMB frame for a \SN\  at comoving distance $r$ given its observed redshift transformed in the CMB frame in Eq.~\eqref{eqn:p_r} as:
\begin{equation}\label{eqn:p_cz}
    p(c \zCMB(r)| c \zhatCMB) =\norm(c\zhatCMB, (\sigvNL)^2)
\end{equation}

 As a check for the robustness of our method, we also estimated the predicted peculiar velocity using an iterative prescription. In this alternative method, we start from the observed CMB restframe redshift as an initial, rough approximation for the comoving distance (under the fiducial \lcdm{} assumption), taking $\vpec = 0$. In subsequent iterations, an updated estimate for the comoving distance is obtained using Eq.~\eqref{eqn:comoving2redshift} and the velocity estimate we get from the reconstruction at the given comoving distance. This step is repeated until convergence. The iterative method gives broadly consistent results as the marginalization-based method described above. However, the iterative prescription may underestimate the uncertainty in the vicinity of triple-valued regions, and therefore we elected to use the marginalization method instead.

Our new peculiar velocity corrections obtained from Eq.~\eqref{eq:average_pecvel} are compared against those used in \cite{betoule2014} in Fig.~\ref{fig:pec_vel}, which are obtained from the JLA data products via\footnote{In Eq.~\eqref{eq:vcorrJLA} we use the notation introduced in this paper but we notice that our $\zbar$ is (confusingly) denoted ``$\zCMB$'' in the \cite{betoule2014} and associated data products.} 
\be \label{eq:vcorrJLA}
v_{\rm corr, JLA} = c \left( \frac{1+\zhel}{(1+\zbar)(1+\zsol)} -1\right).
\ee 
The correlation coefficient between our peculiar velocity corrections and that used in \cite{betoule2014} is only $\sim 0.60$. There are also 6 \SNe\ (labeled in Fig.~\ref{fig:pec_vel}) that are more than 3 standard deviations of the sample away from the identity line. Several reasons can explain the differences between our peculiar velocities and those used in JLA: firstly, the density field used to predict peculiar velocities in JLA was based on the IRAS PSCz survey, which is likely to be noisier than 2M++ everywhere. Secondly, in the JLA analysis only galaxies in rich clusters are corrected to the mean redshift of the cluster, whereas here we correct the CMB restframe redshift for all host galaxies, including those in clusters and poorer groups. 

 For \SNe\ beyond $z=0.067$ (in the SNLS, SDSS and high-$z$ samples), the relative importance of the peculiar velocity corrections diminishes as the recession velocity of the Hubble flow increases, so the detailed treatment becomes less important. In the JLA analysis, peculiar corrections have been set to 0 beyond redshift $z \sim 0.067$, the limit of the galaxy survey from which said corrections were derived. We do the same here, noting that beyond redshift $z \sim 0.067$ the exact treatment of peculiar velocity correction becomes less crucial since their relative importance diminished quickly.  
 
 \subsection{Peculiar Velocity Correction Uncertainties} 
 \label{sec:pecvel_error}
 
As we have replaced the peculiar velocities of JLA with our own, new estimates, we also update their error analysis associated with the peculiar velocities treatment. First, we remove the diagonal term from the JLA statistical covariance matrix representing the uncertainty from peculiar velocity corrections. We also remove the ``pecvel'' contribution to the systematic covariance matrix. We replace them with the following terms, flowing from our updated treatment of peculiar velocity corrections. 

From Eq.~\eqref{eq:zbar_linear}, the covariance of the cosmological redshift value for \SNe\ $i,j$ is given by (without including the negligible error in $\zsol$) is given by:
\begin{align} \label{eq:z_covariance}
\Xi^2_{z,ij} & =  \delta_{ij}^{\rm K} \sigma^2_{z,i}  + \Sigma^2_{z,ij},  \\
\Sigma^2_{z,ij}  & = \delta_{ij}^{\rm K}  \left(\sigma^2_{\rm NL} + \sigma^2_{\rm 2M++}(\zCMB) + \sigma^2_{v,i} \right)/c^2  + C^{\rm flow}_{ij}/c^2,
\end{align}
where $\sigma_{z,i}$ is the spectroscopic redshift measurement uncertainty, $\sigma_{{\rm NL}} = 150$ km/s is the uncertainty in the peculiar velocity due to non-linearities, $\sigma_{\rm 2M++}$ is the redshift-dependent uncertainty due to survey incompleteness, $\sigma_{v,i}$ is the standard deviation of the average peculiar velocity prediction, Eq.~\eqref{eq:average_pecvel}, and $C^{\rm flow}_{ij}$ is the correlated covariance coming from uncertainty in the flow model ($\delta_{ij}^{\rm K} $ is the Kronecker delta). We address each term in turn. 

For the 107 low-$z$ \SNe{} in our analysis, the largest reported statistical uncertainty in their redshift measurement in the JLA data release is $\max_i \sigma_{z,i}^{\rm JLA} = 0.0014$, corresponding to a velocity uncertainty of 420 km/s.  The JLA data release also has $\sigma_{z,i}^{\rm JLA}=0$ for 10 of the 107 low-$z$ \SNe{} and $\sigma_{z,i}^{\rm JLA}<0$ for 42 \SN{}. 
In order to resolve the issue of 0 or negative redshift uncertainties and to be conservative, we adopt the following prescription for the standard deviation of the spectroscopic uncertainty:
\be \label{eq:z_floor}
\sigma_{z,i} = \max(\sigma_{z,i}^{\rm JLA}, 5\cdot 10^{-4}).
\ee
 where the floor value of $5\cdot 10^{-4}$ represents the typical uncertainty of spectroscopic redshift determination from host galaxies spectra.

The term $\sigma^2_{{\rm NL}} = 150$ km/s represents uncertainty in the linear velocity prediction due to unaccounted-for non-linearities, which we fix at the value recommended in ~\cite{10.1093/mnras/stv547}. However, the uncertainty in our reconstructed peculiar velocity increases with redshift, an effect that was ignored in previous work: firstly, the predicted peculiar velocities for tracers near the outer edge of the 2M$++$ catalogue (200 Mpc/$h$) have larger uncertainty because of unaccounted-for structures outside of the survey limits, as well as because of lack of coverage beyond 125 Mpc/$h$ for part of the sky~\citep{Hollinger:2021hwx}; secondly, the noise increases at larger distances due to the smaller number of galaxies with larger weights that are used to represent the density field~\citep{Lilow:2021omg}. We capture these effects via the redshift-dependent term:
\be 
\sigma_{\rm 2M++}(\zCMB) = 
\begin{cases}
\sigma_1 \zCMB & \mbox{ for } \zCMB< z_{400}, \\
\sigma_1 z_{400} & \mbox{ for } \zCMB \geq z_{400},
\end{cases}
\ee
where $z_{400} = 0.138$ is the redshift corresponding to a radial comoving distance of 400$/h$ Mpc, and $\sigma_1$ is chosen so that the total peculiar velocity rms beyond $z_{400}$, i.e. $(\sigma^2_{{\rm NL}} + \sigma_{\rm 2M++}^2(z_{400}))^{1/2}$, equals $380$ km/s. This prescription also approximately matches the \lcdm{} prediction at the 2M++ boundary, $z=0.067$, where $(\sigma^2_{\rm NL} + \sigma^2_{\rm 2M++}(0.067))^{1/2} = 227$ (km/s).
This is in contrast with the original JLA analysis which uses a redshift-independent 150 km/s uncertainty throughout the redshift range.

The term $ \sigma^2_{v,i}$ is the variance of $\vpec$ under the distribution given by Eq.~\eqref{eqn:p_r}, i.e. 
\be 
\sigma^2_{v,i} =  \langle (\vpec)^2 \rangle -  \langle \vpec \rangle^2.
\ee 
Finally, we translate the redshift covariance of Eq.~\eqref{eq:z_covariance} into a magnitude covariance via linear propagation of errors using the isotropic distance modulus of Eq.~\eqref{eq:dist_mod}, i.e. 
\be \label{eq:mag_covariance}
\sigma^2_{m,ij} =  \Sigma^2_{z,ij} \frac{\partial\mu_I}{\partial\zbar}{\bigg\rvert}_{\begin{subarray}{l} \zhel = \hat{z}_{{\rm hel}, i} \\ \zbar=\zbar_i \end{subarray}} 
\frac{\partial\mu_I}{\partial\zbar}{\bigg\rvert}_{\begin{subarray}{l} \zhel = \hat{z}_{{\rm hel}, j} \\ \zbar=\zbar_j \end{subarray}} 
 + \delta_{ij}^{\rm K} \sigma^2_{z,i}\left(\frac{\partial\mu_I}{\partial\zhel}{\bigg\rvert}_{\begin{subarray}{l} \zhel = \hat{z}_{{\rm hel}, i} \\ \zbar=\zbar_i \end{subarray}} \right)^2
\ee
where $\zbar_i$ is computed from Eq.~\eqref{eq:zbar_linear}, and the distance modulus derivatives are evaluated at the fiducial cosmological parameter values given in Table~\ref{table:priors}.

So far, we have considered a fixed value for the flow parameters, $\Flowpars = \{\beta_v, \Vext \}$ entering in Eq.~\eqref{eq:flow_model}. The uncertainties in the inferred flow parameters lead to correlated uncertainties in the peculiar velocities which needs to be accounted for, and that in previous work are usually considered a source of systematic error. Our parameterised flow model allows us to translate them into a statistical error, as follows. In order to estimate the covariance coming from uncertainty in the flow model parameters, we draw $10^4$ posterior samples of the flow parameters $\Flowpars_k$ ($k=1, \dots, 10^4$) from the fitted flow model, using the method of \citet{RN345}, and we calculate the average peculiar velocity, $\langle {\vpec},_{i} \rangle$ for all the 107 \SNe\ in our low-$z$ sample from those samples. We then estimate the covariance of the average peculiar velocity between \SN\ $i$ and $j$ as $C_{ij}^{\rm{flow}} = \text{Cov}(\langle {\vpec,}_{i} \rangle, \langle {\vpec,}_{j} \rangle)$, where the covariance matrix is computed from the $k$ samples. 
Since the value of $ \sigma_{v,i}$ above varies among the $k$ samples (although the variation is small, $\sol 10\%$), we use the average of $ \sigma_{v,i}$  from the $10^4$ posterior samples. In accord with terminology used in the literature, we call this term the ``systematic uncertainty'', although as noted above we have actually translated it into a statistical uncertainty. 
There are no changes to the $x_1$ and $c$ terms of the covariance and these are left unchanged from the original JLA analysis.

\begin{figure*}
    \centering
    \includegraphics[width=\columnwidth]{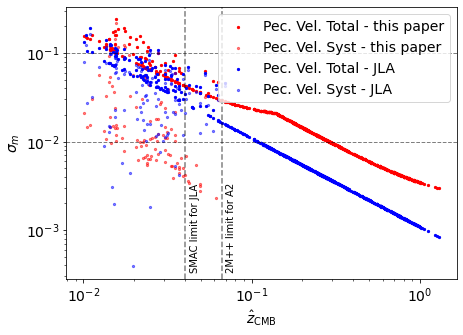}
    \includegraphics[width=\columnwidth]{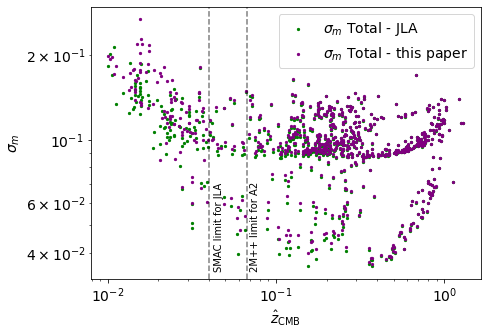}
    \caption{Left panel: statistical and systematic (diagonal entries of the covariance matrix only) uncertainties in the apparent magnitude induced by peculiar velocity corrections for our re-analysis compared with JLA. 
    Right panel: 
    total diagonal $\sigma_m$ for our reanalysed and original JLA data. These components include the statistical error from the SALT2 fits and peculiar velocities as well systematics from several other components that are outlined in \protect\cite{betoule2014}. }
    \label{fig:total_sigma_m_diagonal}
\end{figure*}

We show in the left panel of Fig.\,\ref{fig:total_sigma_m_diagonal} the square root of the diagonal entries of the peculiar velocities covariance matrix, translated into magnitude covariance,i.e.\ $\sigma_{m,ii}$ in Eq.~\eqref{eq:mag_covariance}). The right panel shows the square root of the diagonal entries of the total magnitude covariance matrix (including all other magnitude uncertainties). Our values are compared with the original JLA values on the same figure. The largest difference in the total value of $\sigma_m$ appears in the low redshift range, where our re-analysis modifies the associated peculiar velocities, which are dominant in this redshift range. 
In general, the net effect is to increase the statistical uncertainty while decreasing the systematic uncertainty with respect to the JLA analysis: at the median redsfhit of the low-$z$ sample, $z=0.0243$, the average diagonal $\sigma_m$ due to statistical uncertainty in the peculiar velocities is $0.076$ in our analysis (vs $0.045$ JLA), while the average systematic diagonal error is $0.020$ in our work (vs $0.039$ JLA). Overall, the total magnitude uncertainty due to peculiar velocities is increased by $\sim 30\%$ in our analysis (at the median redshift) compared to JLA. 
In the right panel of Fig.\,\ref{fig:total_sigma_m_diagonal}, we compare the total uncertainties on the apparent magnitude (including all statistical and systematic uncertainties) between this work and the JLA analysis, showing that our magnitude uncertainties are generally larger, especially at low redshifts where the new peculiar velocity uncertainties dominate the error budget. 

\subsection{Choice of Priors}
\label{sec:prior_selection}

As always in a Bayesian analysis, particular attention must be paid to priors, especially in the present case where we are interested in performing not only parameter inference but also model comparison (for an overview of the issue, see e.g.~\cite{2008ConPh..49...71T}).

The priors for the cosmological parameters $\pars_1$ and for the other parameters in the hierarchical model are chosen as in~\cite{shariff2016}, to which we refer for fuller details. A summary is provided in Table \ref{table:priors}. For the cosmographic expansion parameters $\pars_2$, the $\mathrm{Uniform}(0,2)$ priors in $\Omega_m, \Omega_\Lambda$ translate into the following non-uniform prior for the deceleration parameter $q_0 = \Omega_m/2 - \Omega_\Lambda$: 
\be
p(q_0) =
\begin{cases}
\frac{2}{3}(q+2) & \mbox{ for } -2 \leq q_0 < -1, \\
2/3 & \mbox{ for } -1 \leq q_0  <  0, \\
\frac{2}{3}(1-q) & \mbox{ for } 0\leq q_0 \leq 1, \\
0 & \mbox { otherwise}.
\end{cases}
\ee
Given that the likelihood's support is almost entirely within the region $-1.0 \lesssim q_0 \lesssim 0$, we choose the slightly simplified, uniform prior $q_0 \sim \mathrm{Uniform}(-2,1)$. Since the jerk $j_0$ and spatial curvature $\Omega_\kappa$ appear in the degenerate combination $j_0 - \Omega_\kappa$ in the second-order term in the cosmographic expansion, Eq.~\eqref{eq:cosmographic_dl}, we adopt a uniform prior on this combination, with ranges motivated by the range of physically plausible curvature values.  

The dipole amplitude parameter, $D_\mu$, is a positive quantity for which the obvious choices of priors are a uniform or log-uniform prior. The lower boundary of the former is naturally 0, while for the latter a lower cutoff must be imposed in order for the prior to be proper (i.e., normalizable). This however is problematic for the model selection outcome: as the likelihood becomes flat (i.e., insensitive to the value of $D_\mu$) once $D_\mu$ drops below a certain threshold, the marginal posterior becomes equal to the marginal prior for arbitrarily lower values of $D_\mu$ with a log-uniform prior. This has an influence on the Bayes factor, as a larger prior range (i.e., a lower $D_\mu$ cut-off in a log-uniform prior) leads to a less favourable model selection outcome for the anisotropic model. In order to avoid this difficulty, we choose a uniform prior on $D_\mu$ itself. The upper prior cutoff is set by the characteristic scale expected by a dipole signal. This could be gleaned from a theoretical model, or, in a phenomenological approach such as ours, guided by the order of magnitude of previous upper limits on the parameter, which is of order $\sim 10^{-3}$ (see Table \ref{table:compiled_author_results}). Such upper limits can however be considerably relaxed in the case of a dipole that is decaying with redshift, leading to upper limits of order $\sim 10^{-1}$ even when no dipole is present (see our simulated case of Fig.~\ref{fig:simulated_lcdm_vanilla_inference_anisotropy}). In order to accommodate such a scenario, we choose a uniform prior $D_\mu \sim \mathrm{Uniform(0,0.2)}$. 

We choose to sample the area of the sky in a uninformative manner since we do not have any prior belief of the directions a dipole might be pointing to. Requiring rotational invariance on the surface of the 2-sphere leads to a uniform distribution on the Galactic longitude of the dipole vector, $l_d\sim \mathrm{Uniform}(0, 2\pi)$ (in radians), and a uniform distribution on the cosine of the latitude of the dipole vector, $\cos(b_d) \sim \mathrm{Uniform}[0,1]$, with $b_d \in [0, \pi/2]$. Flipping the sign of $b_d$ is equivalent to the transformation $l_d \rightarrow l_d + \pi\, \text{mod}\, 2\pi $ and $D_\mu \rightarrow -D_\mu$. Hence in order to cover the possibility of a dipole pointing in a direction in the southern Galactic hemisphere, we extend the dipole amplitude to negative values, and therefore our prior is  modified to $D_\mu \sim \mathrm{Uniform(-0.2,0.2)}$.  Similar considerations lead to a prior for the dipole amplitude on the deceleration parameter  $D_{q_0} \sim \mathrm{Uniform(-30,30)}$.

For the prior on the exponential scale parameter, $S$, we need to select a lower boundary (lest $D_\mu$ becomes unidentifiable and to stop pathologies associated with $S=0$), which we take to be the scale of the lowest redshift \SNe\ in our data, namely $S=0.01$; for the upper boundary, we take $S=0.1$ as it is known that the bulk flow does not disappear at least out to $z \sim 0.067$. In summary, our prior is thus  $S \sim \textrm{Uniform}[0.01,0.1]$.

\section{Simulations and Tests of Methodology}
\label{sec:sim_param_reconstruction}

\subsection{Simulated Data}
\label{sec:simulating_data}
\begin{table*}
\caption{Supernova population parameters, cosmological parameters and dipole model parameters adopted in this work, together with the prior choices and fiducial values for simulation studies. `SD' stands for standard deviation}
\begin{tabular}{l l l l}
  \toprule
  \SN\ population distributions and covariates & Symbol & Fiducial value & Prior distribution\\
  \midrule
  Mean absolute magnitude of \SN\ & $M_0 $& -19.3 & $M_0 \sim \mathrm{Normal}(-19.3,2^2)$ \\
  Residual scatter of \SN\ magnitude after corrections & $\sigmares$ & 0.1 & $\sigmares \sim \mathrm{InvGamma(0.003,0.003)}$  \\
  Coefficient of stretch covariate& $\alpha$ & 0.14 & $\alpha \sim \mathrm{Uniform(0,1)}$ \\
  Coefficient of colour covariate & $\beta$ & 3.2 & $\beta \sim \mathrm{Uniform(0,4)}$ \\
  Mean of stretch  & $x_*$ & 0.0 & $x^{*} \sim \mathrm{Normal}(0,10^2)$ \\
  Mean of colour  & $c_*$ & 0.0  & $c^{*} \sim \mathrm{Normal}(0,1^2)$ \\
  SD of stretch distribution & $R_x$ & 1.0 & $R_x \sim \mathrm{LogUniform}(-5,2)$\\
  SD of colour distribution & $R_c$ & 0.1 & $R_c \sim \mathrm{LogUniform}(-5,2)$  \\
  \toprule
  Parameters controlling the expansion history & \\ 
  \midrule
  Matter energy density & $\omegam$ & 0.3&  $\omegam \sim \mathrm{Uniform}(0,2)$  \\
  Dark energy density & $\Omega_{\Lambda}$ & 0.7 & $\Omega_{\Lambda} \sim \mathrm{Uniform}(0,2)$  \\
  Deceleration Parameter & $q_0$ & -0.55  &  $q_0 \sim \mathrm{Uniform}(-2,1) $\\
  Jerk and spatial curvature & $j_0 - \Omega_K$ & 1 &  $j_0 - \Omega_K \sim \mathrm{Uniform}(-2,2) $\\
  Hubble-Lema\^itre constant [km/s/Mpc]& $H_0$ & 72 & Fixed \\
   \toprule
  Anisotropic expansion parameters &   \\
    \midrule
  Galactic longitude of dipole [rad]& $l_d$ & $\{ -, 4.60\}$ & $l_d \sim \mathrm{Uniform}(0, 2\pi)$  \\
  Galactic latitude of dipole [rad] & $b_d$ & $\{-, 0.84 \} $  & $\cos(b_d) \sim \mathrm{Uniform}(0, 1)$, $b_d \in [0, \pi/2]$\\
  Amplitude of dipole on $\mu$ & $D_\mu$ & \{ 0, 0.02\} & $D_\mu \sim
  \mathrm{Uniform}(-0.2,0.2)$ \\
  Amplitude of dipole on $q_0$ & $D_{q_0}$ & \{ 0, 10\} & $D_{q_0} \sim
  \mathrm{Uniform}(-30,30)$ \\
    Exponential scale of dipole & $S$ & \{ --, 0.026\} & $S\sim \mathrm{Uniform}(0.01,0.10)$ \\
\bottomrule
\label{table:priors}
\end{tabular}
\end{table*}

To test the ability of our setup to recover the anisotropy parameters within the \bahamas{} framework, we forward simulate data from our model in a similar manner to \cite{march2011}, adding however JLA-like Galactic coordinates to our simulated \SNe\ in order to carry out inference on a potential dipole.  The fiducial values for the parameters are listed in Table \ref{table:priors}. We consider a case with no dipole ($D_\mu = 0$) and a case with a large dipole ($D_\mu = 0.02$), pointing in the approximate direction of the CMB dipole and with an exponential scale, $S=0.026$, matching the value preferred by the results of C19. The steps to generate the simulated data are as follows:
\begin{enumerate}
\item Draw a value for the latent CMB restframe redshift, $z_{\text{CMB}, i}$, for each \SN. $\zhel$ is assumed to equal this CMB redshift. We draw 740 objects at the same redshifts as the real JLA data to ensure the binning of the data is the same when applying selection effects correction on simulated data as it is on real data. Notice that our simulations do not include the issue of peculiar velocity corrections, which are assumed to have already been performed. 

\item Compute $\mu_{i}(\hat{z},\FullPars)$ using the fiducial values for our chosen cosmology, whether \lcdm{} or the Cosmographic expansion.

\item Apply the dipolar modulation to the distance modulus using \eqref{eq:dipolar_modulation}.

\item Draw the latent parameters $x_{1i}$, $c_{i}$ and $M_{i}$ from the normal distributions, $x_{1i} \sim N(x_{*},R_{x}^2)$, $c_{i} \sim N(c_{*},R_{c}^2)$ and $M_{i} \sim N(M_{0}, \sigmares{}^2)$ respectively.

\item Compute $m_{Bi}$ using $x_{1_i}$, $c_i$, $M_i$ and the Phillip's relation equation in \eqref{eq:dist_mod}.

\item Draw the value of the standard deviations $\sigma_{x_{1i}}$, $\sigma_{c_i}$, and $\sigma_{m_i}$, from the appropriate normal distributions fitted to the errors in the JLA data and use them to construct the $3 \times 3$ covariance matrix for each SNe as $C = \mathrm{diag}(\sigma_{c_i}^2,\sigma_{x_{1i}}^{2},\sigma_{m_i}^2)$.

\item Draw the observed SALT2 parameters from $\hat{x}_{1i}\sim N(x_{1i}, \sigma_{x_{1i}}^2)$, $\hat{c}_{i} \sim N(c_{i},\sigma_{ci}^2)$ and $\hat{m}_{Bi} \sim N(m_{Bi},\sigma_{mi}^2)$.

\item Apply the selection function on the colour values drawn in the previous step. We use the values of $\sigmaobs$ and $\cobs$ inferred from the real JLA data in section \ref{sec:colour_selection} for the redshift bins the \SN\ falls in. If a given \SN\ is not selected, we cycle back to step (iv) and redraw that \SN. The process ends when all 740 objects are selected.

\item Generate positions for \SN\ in the sky which match the positions of the JLA data. The non-isotropic distribution of the data has an effect on our ability to constrain a dipole, so it is important to match the real JLA \SN\ positions for a realistic simulation. 
\end{enumerate}

\begin{figure*}
\includegraphics[width=2\columnwidth]{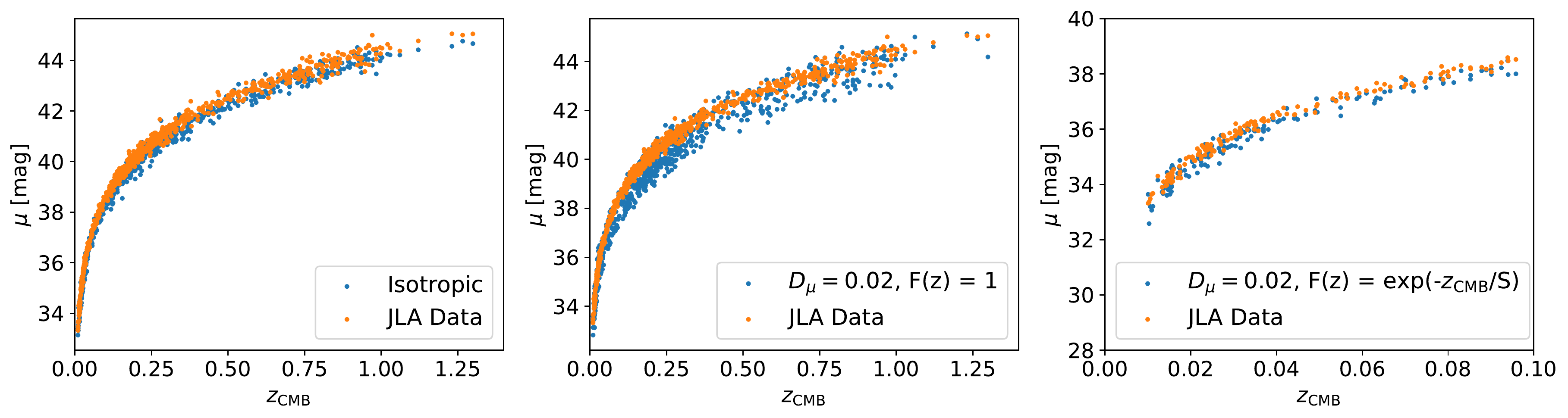}
\caption{Simulated \SN\ data generated using the JLA dataset as a reference point. The simulations assume a \lcdm{} model and no anisotropy in the distance modulus for the left most model. A dipole of value $D_\mu = 0.02$ is present for the second plot. The third plot also has this value of the dipole, but restricted to a local scale ($z \sim 0.1$) by multiplying the dipole term by the function $F(z) = \exp{(-z/0.026)}$. For this third plot the redshift has been truncated to only show the redshift range where the dipole is noticeable.}
\label{fig:simulated_magnitude_redshift}
\end{figure*}

\begin{figure*}
\includegraphics[width=2\columnwidth]{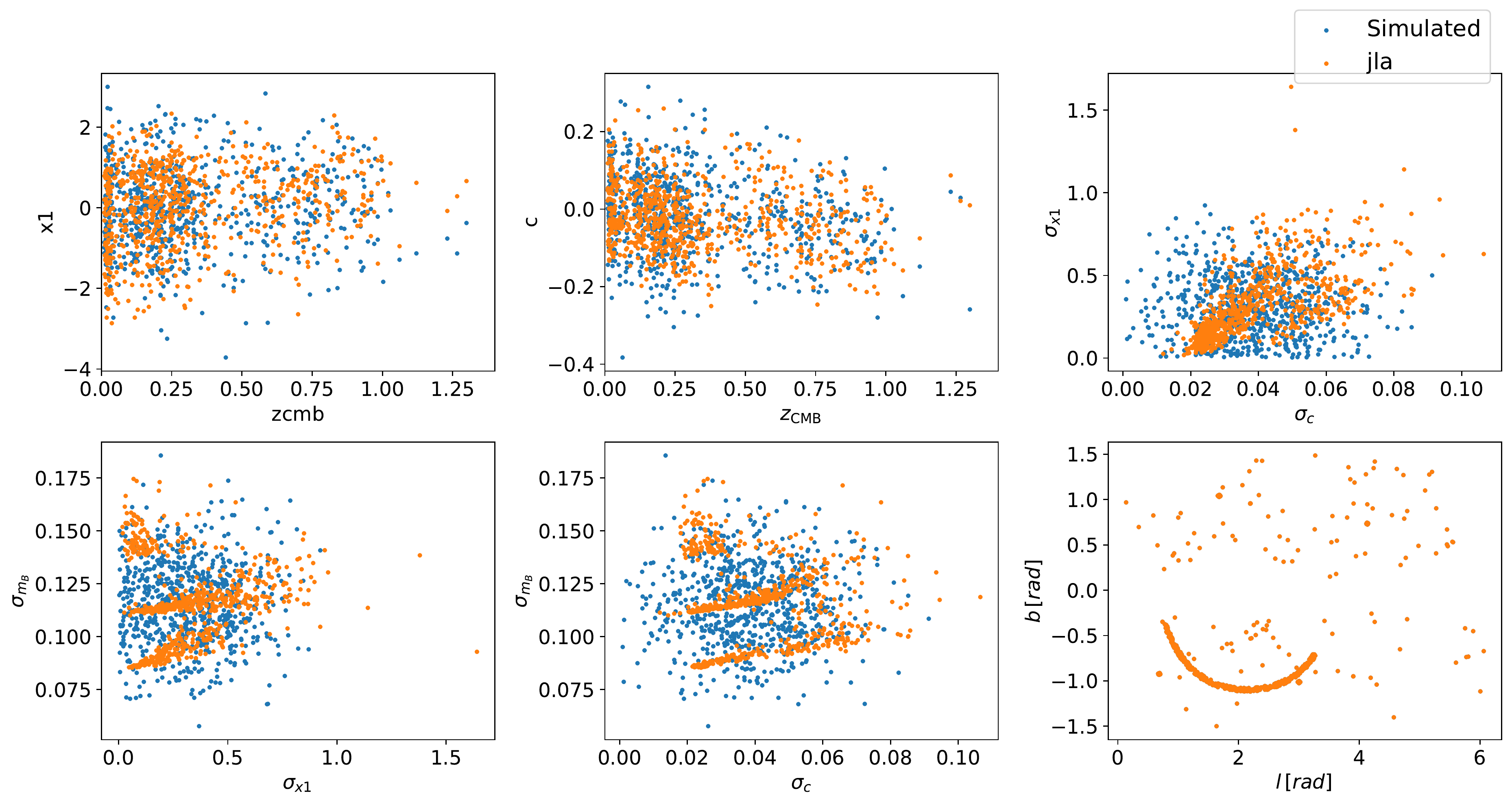}
\caption{Simulated \SN\ data generated using the JLA dataset as a reference point, assuming \lcdm{} as a fiducial model: distribution of colour ($c$), stretch ($x_1$) and the respective standard deviation entering the observational error covariance matrix, as well as the \SN's sky coordinates.}
\label{fig:simulation_distributions}
\end{figure*}

An example of the simulated data, compared with the JLA dataset, is shown in Fig. \ref{fig:simulated_magnitude_redshift} and \ref{fig:simulation_distributions}, for a \lcdm{} realisation.
Some banding of the data can be seen of the plots for $\sigma_{x_{1}}$, $\sigma_{c}$, and $\sigma_{m_B}$. This results from the different survey components comprising the JLA data. We can see that our simulated data does not capture this banding in the errors because we draw our errors from Gaussian distributions, but this approximation has no quantitative effect on our simulations. 
The well defined `stripe' in the plot of the galactic coordinates of the data is from the SDSS component of the JLA data.

We test the ability of our method to retrieve the true fiducial values listed in Table \ref{table:priors} from the simulated data under the 7 different scenarios below: 
\begin{enumerate}
\item \textbf{\lcdm{}-Isotropic} simulates an isotropic universe ($D_\mu = 0$) from \lcdm{} and the inference also assumes a \lcdm{} cosmology. 
\item \textbf{\lcdm{}-D} simulates a dipole ($D_\mu = 0.02$) with $F(z) = 1$ from \lcdm{} and the inference also assumes a \lcdm{} cosmology. 
\item \textbf{\lcdm{}-D-exp} simulates a dipole ($D_\mu = 0.02$) with $F(z) = \exp{(-z/S)}$ from \lcdm{}  and inference also assumes a \lcdm{} cosmology, with the additional free parameter $S$. 
\item \textbf{Cosmographic-Isotropic} simulates an isotropic universe ($D_\mu = 0$) from a cosmographic expansion and inference also assumes a cosmographic expansion. 
\item \textbf{Cosmographic-D} simulates  a dipole with $F(z) = 1$ from a cosmographic expansions and inference also assumes a cosmographic expansion. 
\item \textbf{Cosmographic-D-exp} simulates a dipole ($D_\mu = 0.02$) with $F(z) = \exp{(-z/S)}$ and inference also assumes a cosmographic expansion, with the additional free parameter $S$.
\item  \textbf{Cosmographic-D-exp*} simulates data as in {\lcdm{}-D-exp}  but the reconstruction adopts the cosmographic expansion instead. This serves to assess the bias in parameter reconstruction from the cosmographic expansion when the reconstruction model is miss-specified. 
\end{enumerate}

For each parameterisation we generate 10 realisations of the data; each realisation contains 740 \SN\ objects to be similar in size to the JLA data. The posterior results are averaged over the 10 realisations for each parameterisation when we reconstruct the parameters in section \ref{sec:param_reconstruction}. 
We do not use the Gibbs sampler of \cite{shariff2016} for posterior sampling, but rather adopt \texttt{PyMultiNest} \citep{buchner2014}, an implementation of the Nested Sampling algorithm \texttt{Multinest} \citep{feroz2008,feroz2009,feroz2013}.  The benefit is that we can also compute the Bayesian evidence which we will use for Bayesian model comparison in section~\ref{sec:model_comparion}.

\subsection{Parameter Reconstruction from Simulations}
\label{sec:param_reconstruction}
\begin{table*}

\caption{Difference between the 1D marginal posterior mean and the true value used to simulated JLA-data, averaged over 10 data realizations each,  $\Delta x \equiv x_{\rm true} - \bar{x}$ for each parameter $x$. In parenthesis, the difference is expressed in units of the 1D posterior standard deviation.The top (middle) section shows the case where \lcdm{} (Cosmographic expansion) is assumed both for the simulation and the inference; the bottom row assumes \lcdm{} for the simulation and the the Cosmographic expansion for the inference -- a case of model miss-specification.}
\resizebox{\textwidth}{!}{\begin{tabular}{lllllllll}
  \toprule
Model  & $\Delta q_{0}$ & $\Delta (j_{0}-\Omega_\kappa)$ & $\Delta \Omega_{M}$ & $\Delta \Omega_{\Lambda}$&$\Delta l(\mathrm{rad)}$ & $\Delta b(\mathrm{rad)}$ & $|\Delta D_\mu|  (\times 10^{-2})$ &$\Delta S (\times 10^{-2})$ \\
  \midrule
  \multicolumn{9}{c}{Simulation: \lcdm{}; reconstruction: \lcdm{}} \\ 
  \midrule 
  (ia) \lcdm{}-Isotropic, $F(z) = 0$ &  - & - & 0.03(0.51$\sigma$) & 0.05(0.36$\sigma$) & -  & - & < $1.00 \times 10^{-1}$(2$\sigma$) & - \\
  (ib) \lcdm{}-Isotropic, $F(z) = \exp(-z/S)$ &  - & - & 0.03(0.51$\sigma$) & 0.05(0.36$\sigma$) & -  & - & < $4.96 \times 10^{-1}$ (2$\sigma$) & < 9.29 (2$\sigma$) \\
  (ii) \lcdm{}-D  & - & - & -0.02(0.19$\sigma$) & 0.02(0.10$\sigma$) & -0.003(0.20$\sigma$) & -0.003(0.25$\sigma$) & 0.036(0.13$\sigma$) & - \\
  (iii) \lcdm{}-D-exp  & - & - & 0.03(0.91$\sigma$) & 0.07(0.72$\sigma$)& 0.02(0.20$\sigma$) & -0.02(0.17$\sigma$) & -0.16(0.28$\sigma$) & -0.10(0.21$\sigma$) \\ 
    \midrule
  \multicolumn{9}{c}{Simulation: Cosmographic; reconstruction: Cosmographic} \\ 
  \midrule 
  (iv) Cosmographic-Isotropic, $F(z) = 0$  &  -0.02(0.20$\sigma$) & 0.03(0.01$\sigma$) & - & - & - & - & - & - \\
  (v) Cosmographic-D & -0.06(0.56$\sigma$) & 0.27(0.58$\sigma$) & - & - & -0.0116(0.37$\sigma$) & -0.003(0.26$\sigma$) & -0.01(0.35$\sigma$) & - \\
  (vi) Cosmographic-D-exp & -0.01(0.10$\sigma$) & -0.05(0.14$\sigma$) & - & - & -0.04(0.20$\sigma$) & 0.01(0.08$\sigma$) & -0.13(0.30$\sigma$) & 0.16(0.37$\sigma$)  \\ 
  \midrule
  \multicolumn{9}{c}{Simulation: \lcdm{}; reconstruction: Cosmographic} \\
  \midrule 
  (vii) {Cosmographic-D-exp} & -0.11(1.05$\sigma$) & -0.63(1.37$\sigma$) & - & - & 0.11(0.60$\sigma$) & 0.038(0.37$\sigma$) & -0.128(0.44$\sigma$) & -0.165(0.24$\sigma$) \\
    \bottomrule
\label{table:simulated_results}
\end{tabular}}
\end{table*}

We use \bahamas{} to construct posterior distributions (averaged over 10 data realizations) for the set of cosmological parameters $\pars_1$ (i.e., \lcdm{}) or $\pars_2$ (i.e., cosmographic expansion) and anisotropy parameters $\{l_d,b_d,D_\mu,S\} $. Although they are sampled over during reconstruction, \SNe\  population parameters and SALT2 coefficients are numerically marginalised over in corner plots and not visualised as they are not the focus of this paper. The difference between the 1D marginal posterior mean (averaged over realizations) and the true value of each parameter is displayed in Table \ref{table:simulated_results}. We observe that in all cases except for  the scenario {Cosmographic-D-exp*} the difference is a fraction of a standard deviation, hence entirely within realization and sampling noise. The model misspecification of {Cosmographic-D-exp*}, however, does lead to shifts of up to $\sim 1.4 \sigma$ in the reconstructed cosmographic parameters, a reflection of the fact that the data have been generated under a different model, namely \lcdm{}, than has been assumed in the reconstruction. However, the difference for the anisotropy parameters remains below $0.5\sigma$.

\begin{figure*}
   % \begin{subfigure}%
        \includegraphics[width = 0.8\columnwidth]{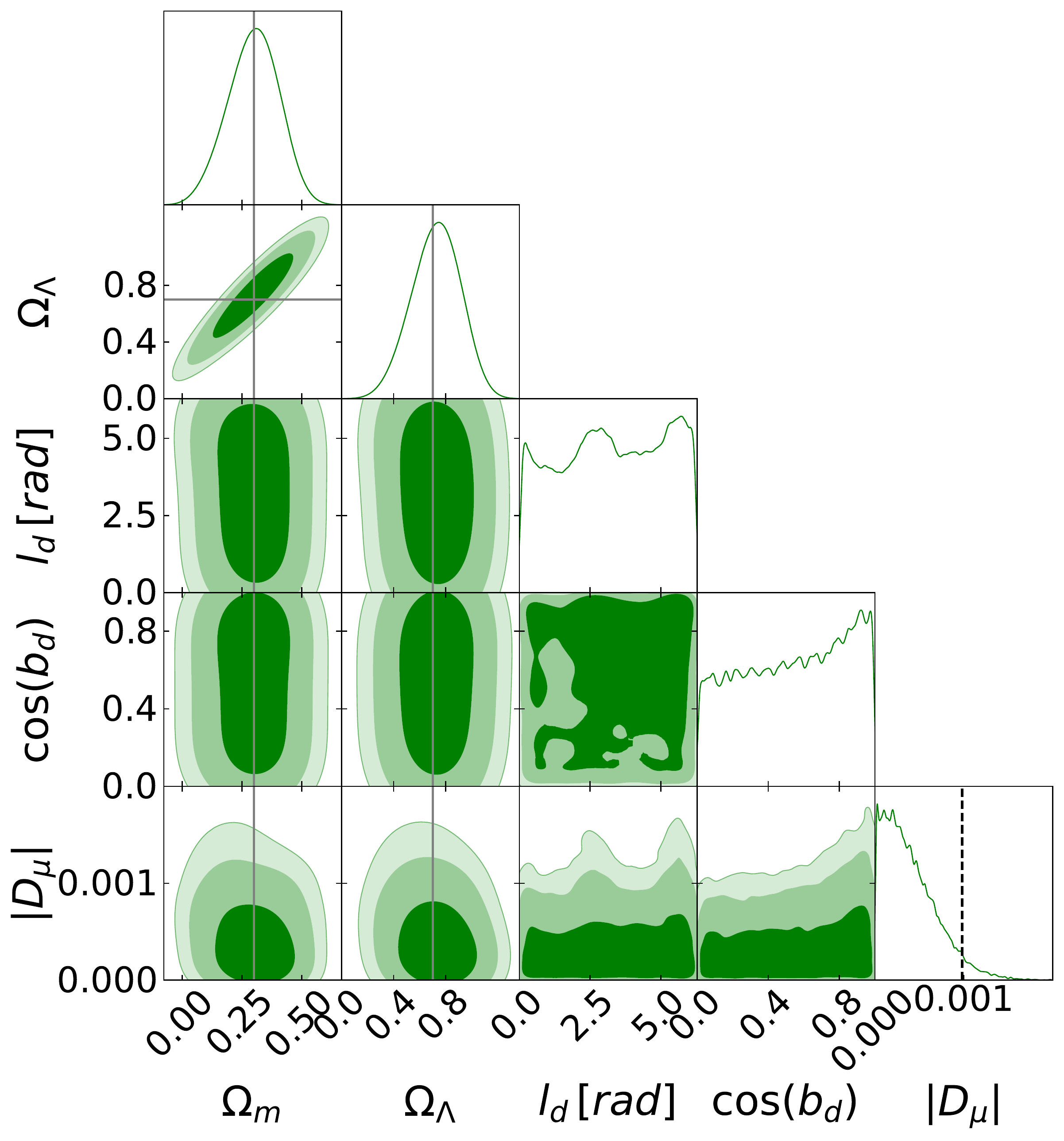}
    %\end{subfigure}
    %\begin{subfigure}
        \includegraphics[width = \columnwidth]{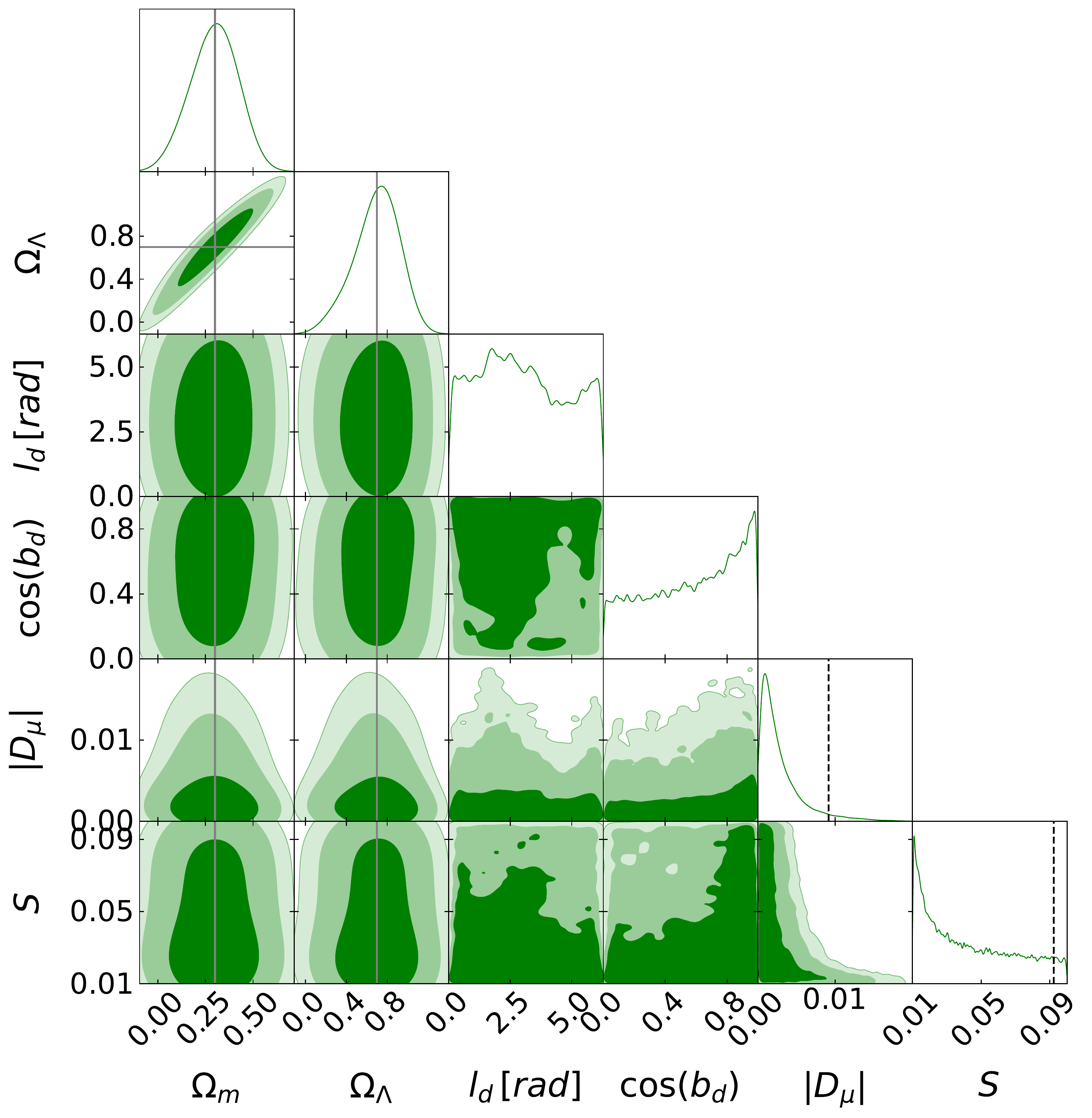}
    %\end{subfigure}
    \caption{Posterior distributions from simulated JLA-like data, averaged over 10 data realizations. The simulations assume a \lcdm{}  isotropic universe ($D_\mu = 0$); the reconstructions allow for a dipole in the distance modulus with $F(z) = 1$ (left) and $F(z)= \exp{(-z/S)}$ (right). Contours in 2D plots enclose joint 1, 2 and 3 $\sigma$ highest posterior density (HPD) credible regions; the vertical dashed line in the 1D marginals for $D_\mu$ and $S$ delimits the 2$\sigma$ upper limit. Notice the different scales in the axis of $D_\mu$ between the two cases.}%
        \label{fig:simulated_lcdm_vanilla_inference_anisotropy}%
\end{figure*}

It is instructive to investigate the expected constraints on the dipole amplitude when the simulated data are from a isotropic universe (scenario \lcdm{}-Isotropic), shown in Fig.~\ref{fig:simulated_lcdm_vanilla_inference_anisotropy}. The 1-sided $95\%$ upper limit we can place on $D_\mu$ for this simulated data is $D_\mu \leq 8.08 \times 10^{-4}$ (left panel). When however we introduce the additional freedom of an exponential scale parameter $S$, the constraints in $D_\mu$ degrade by two orders of magnitude, as a small value of $S$ confines any anisotropy to very small redshifts where the statistical power of our data is small and therefore degenerate with many values of $D_{\mu}$ which leads to the entire prior space on $D_\mu$ being well explored right up to the prior edge (right panels of Fig.~\ref{fig:simulated_lcdm_vanilla_inference_anisotropy}.).  The 95\% upper bound becomes $D_\mu < 4.36 \times 10^{-2}$ and $S < 4.56 \times 10^{-2}$. Qualitatively similar results apply for the Cosmographic expansion. 

The above simulations do not include colour-based selection effects for simplicity. A simulation study including colour-based selection appears in Appendix~A (see Fig.~\ref{fig:reconstruction_w_selectioneffects}) and shows that averaging the posterior distributions across replicates yields 2-dimensional and 1-dimensional marginal distributions that are centered on the true values of the parameters.

\section{Results}
\label{sec:results}
\subsection{Parameter Inference}
\label{sec:parameter_inference}

\begin{figure*}
    \includegraphics[width = 0.9\columnwidth]{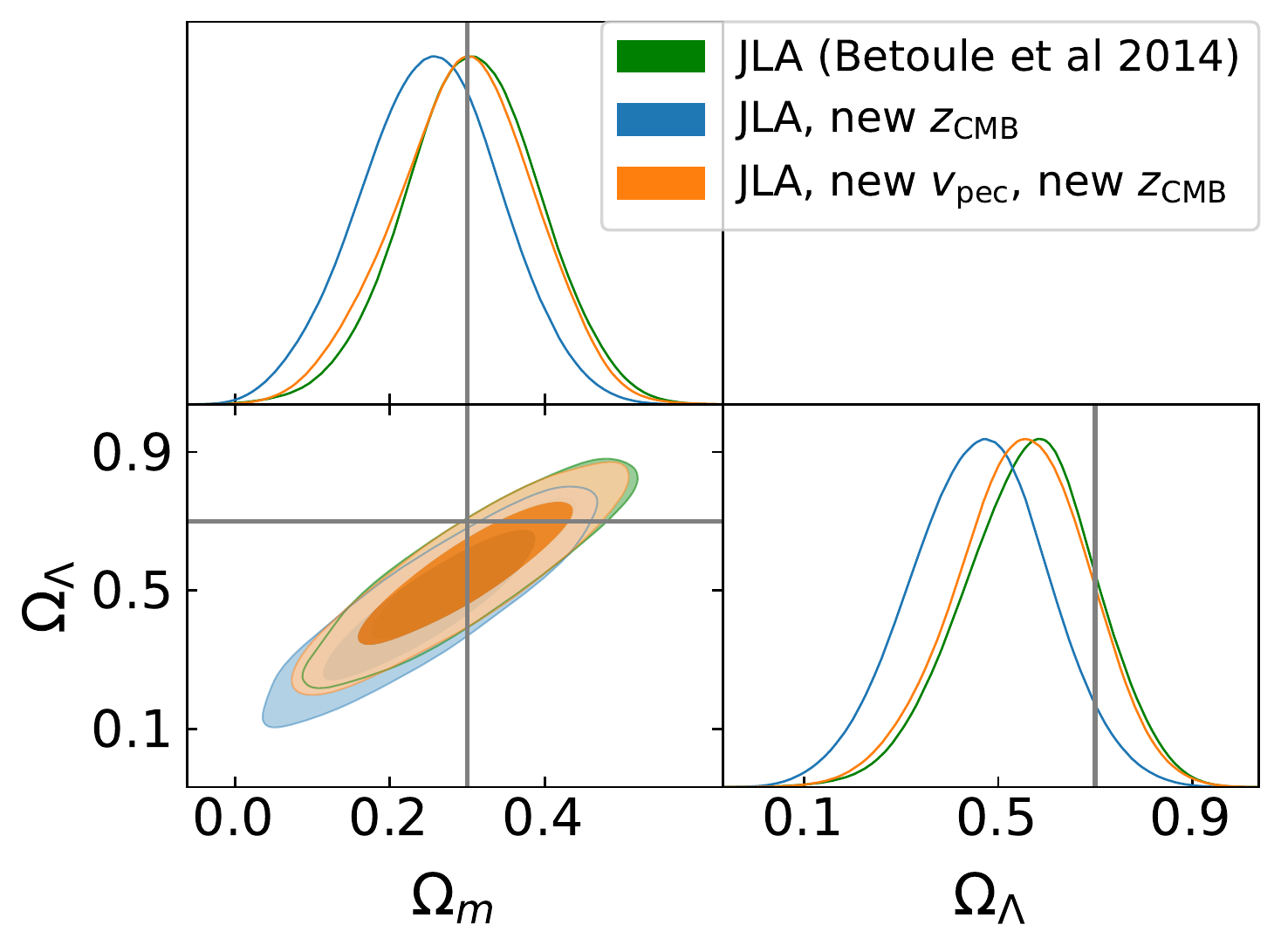}
    \includegraphics[width = 0.9\columnwidth]{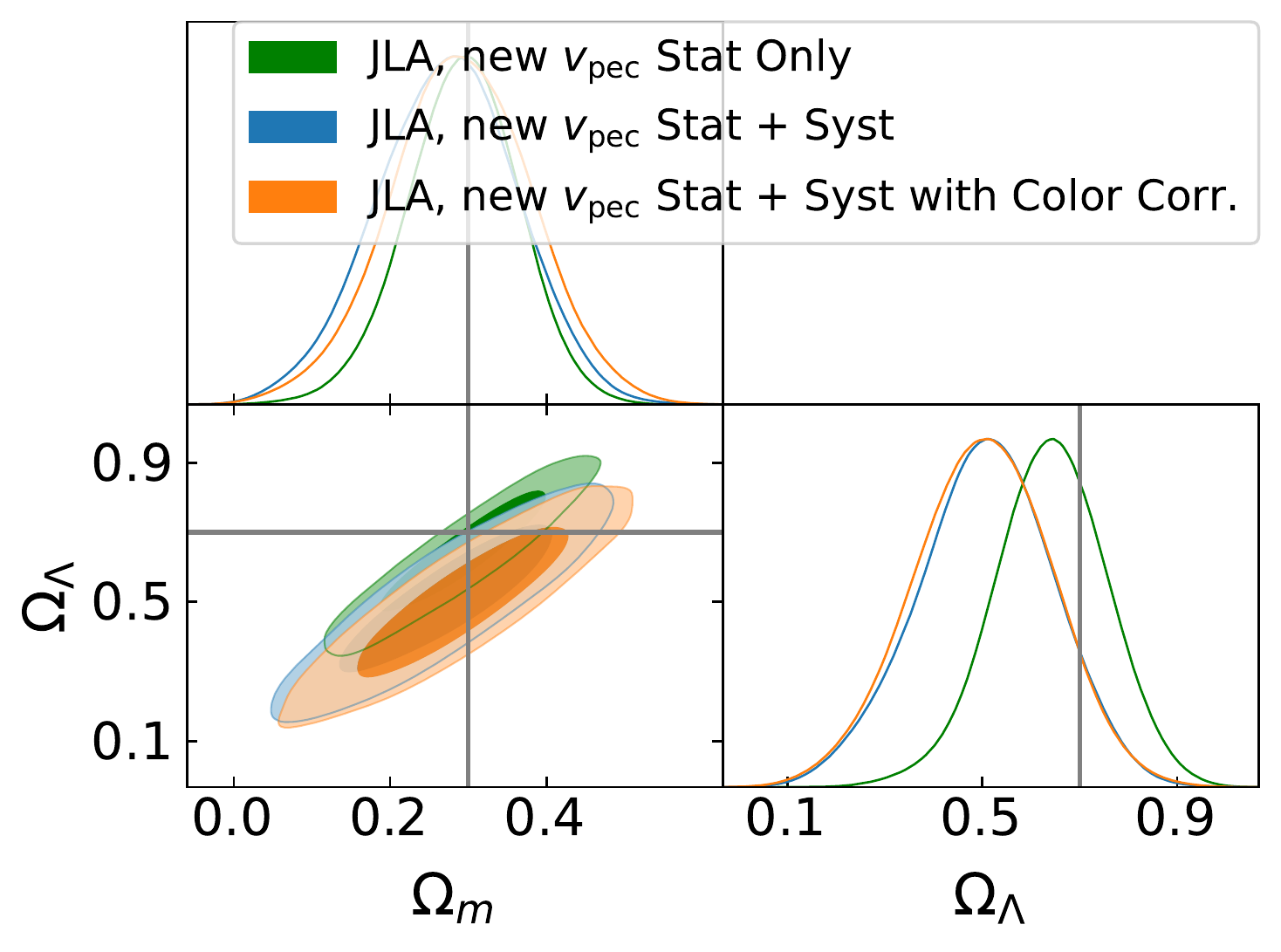}

    \caption{Impact of our new treatment of peculiar velocities and colour selection effects correction on the constraints on $\Omega_m, \Omega_\Lambda$ for the case of an isotropic universe. Left panel: comparison of marginal posteriors  $\Omega_m, \Omega_\Lambda$ using the old JLA peculiar velocity uncertainties and $\zCMB$ (green) to the ones obtained using our new, group-corrected values of the CMB restframe redshifts (blue), and additionally replacing the JLA peculiar velocity uncertainties with our newly derived ones (orange). Right panel: posterior using only the statistical covariance matrix (green), adding the systematic covariance matrix (including our new peculiar velocity covariances, blue) and further adding colour selection effects corrections (orange). In this panel, we adopt our newly derived, group-corrects CMB restframe redshifts. }%
    \label{fig:covmats_comparison}
\end{figure*}

We begin by presenting the impact of our new treatment of peculiar velocities and colour selection effects correction on the constraints on $\Omega_m, \Omega_\Lambda$ for the case of an isotropic universe, as shown in Fig.~\ref{fig:covmats_comparison}. In the left panel, we  compare the constraints using the old JLA peculiar velocities and $\zCMB$ as in \cite{betoule2014} (green) to the ones obtained using our new, group-corrected values of the CMB restframe redshifts (blue), and additionally replacing the JLA peculiar velocities with our newly derived ones (orange). The constraints on the parameters of our models from JLA data are summarised in Table \ref{table:final_jla_results}, for both the \lcdm{} model and the Cosmographic expansion. In the top section we also investigate the impact of our newly derived values for $\zCMB$, peculiar velocity corrections and colour-based selection effects on $\Omega_m$, $\Omega_\Lambda$ constraints in an isotropic universe. Starting from the same treatment as \cite{betoule2014}, we find $\Omega_m = 0.306 \pm 0.087$, in good agreement with the value in \cite{betoule2014}, $\Omega_m = 0.295 \pm 0.034$ but with significantly larger uncertainty, perhaps on account of the different statistical approach. When replacing the CMB restframe redshifts used in \cite{betoule2014} with the ones presented here, we find  $\Omega_m = 0.253 \pm 0.089$, a shift of about half a standard deviation according to our uncertainty, but of $1.5\sigma$ in units of the standard deviation quoted by \cite{betoule2014}. The effect of using the new peculiar velocity corrections (with their newly derived associated covariance matrix) while maintaining the value of CMB restframe redshift from \cite{betoule2014} results in a more modest shift, $\Omega_m = 0.297 \pm 0.089$ (case 2M++, old $\zCMB$). When using both the new redshift values and our newly derived peculiar velocity corrections in combination, we obtain $\Omega_m = 0.273 \pm 0.090$. All these results do not use our new colour-based selection effects corrections; once those are included, the constraint on the matter density shifts back to a value close to the original JLA analysis, namely $\Omega_m = 0.290 \pm 0.091$ (but notice the larger uncertainty on our result).

In the right panel of Fig.~\ref{fig:covmats_comparison}, we observe a shift in the posterior towards lower $\Omega_\Lambda$ and larger $\Omega_m$ when adding the systematic covariance matrix (including our new peculiar velocity covariances) to the statistical covariance matrix, as already noticed by \cite{shariff2016}. Further adding the correction for colour selection effects shifts the posterior only slightly.

\begin{table*}

\caption{Parameter constraints from 1D marginal posterior distributions from JLA data, and outcome of Bayesian model comparison (last column) for the models and data considered.  The second column indicates the peculiar velocity corrections adopted: `PSCz' is the same treatment as in Betoule et al 2014; `2M++' indicates the new treatment in this paper; `PSCz, new $\zCMB$' uses the peculiar velocity corrections of Betoule et al 2014 but our newly derived CMB restframe redshift, while `2M++, old $\zCMB$' uses our newly derived peculiar velocity corrections but the same $\zCMB$ as Betoule et al 2014. 
The last column is the difference between the model's log-evidence and log-evidence of the \lcdm{}-Isotropic or Cosmographic-Isotropic case (with or without colour correction, as appropriate), i.e., the log of the Bayes factor. A negative value indicates that the model under consideration is disfavoured w.r.t. the isotropic universe. The $\star$ on some of the data for $\Delta \ln{(Z)}$ are to indicate that the calculation involved a comparison for a Cosmographic expansion with a restricted prior on the deceleration parameter of $q_{0} \sim [-2,0)$ which limits it to an accelerating universe only.}

\resizebox{\textwidth}{!}{
\begin{tabular}{llllllllllll}
 \toprule
  \multirow{2}{*}{Model} & $\vpec$ &Selection&\multirow{2}{*}{$\Omega_{m}$} & \multirow{2}{*}{$\Omega_{\Lambda}$} &\multirow{2}{*}{$l_d~\mathrm{[rad]}$} & \multirow{2}{*}{$b_d~\mathrm{[rad]}$} & \multirow{2}{*}{$|D_\mu|$} &Dipole&Quantity & \multirow{2}{*}{$\Delta \ln{(Z)}$} \\
  &correction& correction&&&&&&scale, $S$&modulated&& \\
  \midrule
  \lcdm{}-Isotropic & PSCz & No & $ 0.306 \pm 0.087 $&$ 0.562 \pm 0.135$ & - &  - &  - & - & - & \\ %$ 62.11 \pm 0.01 $  \\
 \lcdm{}-Isotropic & PSCz, new $\zCMB$ & No & $ 0.253 \pm 0.089 $&$ 0.460 \pm 0.139$& - &  - &  - & - & - & \\%  $ 50.65 \pm 0.01 $  \\
  \lcdm{}-Isotropic &  2M++, old $\zCMB$  & No &$ 0.297 \pm 0.089 $&$ 0.547 \pm 0.137$ & - &  - &  - & - & - & \\ % $64.29 \pm 0.01 $  \\
  
  \lcdm{}-Isotropic & 2M++ & No &$ 0.273 \pm 0.090 $&$ 0.508 \pm 0.140$ & - &  - &  - & - & - &$ 0.0$  \\% 57.19 \pm 0.01 $ & 99.97 \\
  \lcdm{}-Isotropic & 2M++ & Yes &$ 0.290 \pm 0.091 $&$ 0.500 \pm 0.141$ & - &  - &  - & - & - &  $0.0$ \\  %$ 254.18 \pm 0.01 $ & 889.78\\ 
  CDM only-Isotropic        & 2M++ & No & $ 0.037 \pm 0.028 $& $0$ & - &  - &  - & - & - &  $-5.65$ \\  %$ 249.33 \pm 0.01 $ 
  CDM only-Isotropic        & 2M++ & Yes & $ 0.037 \pm 0.028 $& $0$ & - &  - &  - & - & - &  $-4.85$ \\  %$ 249.33 \pm 0.01 $ 

  \midrule 
  
  \lcdm{}-D & 2M++ & No &$ 0.263 \pm 0.090 $&$ 0.493 \pm 0.140$ & Unconstrained &  Unconstrained &  $<5.75 \times 10^{-4} \, (95\%)$ & - & $\mu$ &$ -8.68 \pm 0.01 $  \\
  \lcdm{}-D & 2M++ & Yes &$ 0.285 \pm 0.094 $&$ 0.490 \pm 0.143$ & Unconstrained &  Unconstrained &  $<5.93 \times 10^{-4} \, (95\%)$ & - & $\mu$ &$ -6.81 \pm 0.01 $  \\

  \lcdm{}-D-Exp & 2M++ & No &$ 0.269 \pm 0.087 $&$ 0.499 \pm 0.137$ & Unconstrained &  Unconstrained &  $<9.81 \times 10^{-3} \, (95\%)$ &  $<8.75 \times 10^{-2}\, (95\%)$ & $\mu$ &$ -5.27 \pm 0.08 $  \\
  \lcdm{}-D-Exp & 2M++ & Yes &$ 0.286\pm 0.091 $&$ 0.487 \pm 0.142$ & Unconstrained &  Unconstrained &  $<1.05 \times 10^{-2} \, (95\%)$ &  $<8.71 \times 10^{-2}\, (95\%)$ & $\mu$ &$ -3.49 \pm 0.01 $  \\
  
 \toprule
     	\multirow{2}{*}{$\downarrow$}&  \multirow{2}{*}{$\downarrow$} & \multirow{2}{*}{$\downarrow$}&\multirow{2}{*}{$q_0$} & \multirow{2}{*}{$j_0 - \Omega_\kappa$} & \multirow{2}{*}{$\downarrow$} & \multirow{2}{*}{$\downarrow$} &  \multirow{2}{*}{$\downarrow$} & \multirow{2}{*}{$\downarrow$}& \multirow{2}{*}{$\downarrow$}&\multirow{2}{*}{$\downarrow$}\\
\\
\midrule 
 Cosmographic-Isotropic & PSCz &No & $-0.352 \pm 0.092 $&$ 0.114 \pm 0.391$ & - &  - &  - & - & - &  \\ %61.76 \pm 0.01  \\
 
 Cosmographic-Isotropic & PSCz, new $\zCMB$ & No & $ -0.288 \pm 0.095 $ & $ -0.132 \pm 0.370 $ & - &  - &  - & - & - &  \\ %50.06 \pm 0.01 $ \\
 Cosmographic-Isotropic & 2M++, old $\zCMB$ & No & $-0.342 \pm 0.093 $& $ 0.071 \pm 0.390 $ & - &  - &  - & - & - & \\ %63.87 \pm 0.01 $ \\
 
 Cosmographic-Isotropic & 2M++ & No &$-0.320 \pm 0.094 $&$ -0.254 \pm 0.383 $ & - &  - &  - & - & - & $ 0.0$  \\ %54.09 \pm 0.02 $ \\
 Cosmographic-Isotropic & 2M++ & Yes &$-0.302 \pm 0.090 $&$ -0.023 \pm 0.357 $ & - &  - &  - & - & - & $ 0.0 $  \\ % 253.10\pm 0.01 $ & 889.39\\
 Cosmographic-($q_0 = 0$) & 2M++ & No &$ 0  $& $-0.953 \pm 0.140$ & - &  - &  - & - & - &  $-3.90 \pm 0.06^{\star}$ \\  %$ 254.18 \pm 0.01 $  \\
 Cosmographic-($q_0 = 0$) & 2M++ & Yes &$ 0  $& $-0.881 \pm 0.134$ & - &  - &  - & - & - &  $-3.28 \pm 0.09^{\star}$  \\  %$ 254.18 \pm 0.01 $  \\
 Cosmographic-($q_0 \ge 0$) & 2M++ & No &$ < 0.067 \times 10^{-2} \, (95\%) $& $-0.922 \pm 0.141$ & - &  - &  - & - & - &  $-7.39 \pm 0.05^{\star}$\\  %$ 254.18 \pm 0.01 $  \\
 Cosmographic-($q_0 \ge 0$) & 2M++ & Yes &$ < 0.074 \times 10^{-2} \, (95\%) $& $-0.922 \pm 0.144$ & - &  - &  - & - & - &  $-7.00 \pm 0.09^{\star}$ \\  %$ 254.18 \pm 0.01 $  \\

\midrule  
 Cosmographic-D & 2M++ & No &$-0.307 \pm 0.093 $& $ -0.085 \pm 0.375 $ & Unconstrained & Unconstrained & $<6.43 \times 10^{-4} \, (95\%)$ & - &  $\mu$ & $ -6.36 \pm 0.02 $ \\
 Cosmographic-D & 2M++ & Yes &$-0.291 \pm 0.095 $& $ -0.070 \pm 0.371 $ & Unconstrained &  Unconstrained &  $<6.32 \times 10^{-4} \, (95\%)$  & - &  $\mu$ & $ -6.47 \pm 0.02 $ \\
 
 Cosmographic-D-Exp & 2M++ & No &$-0.307 \pm 0.089 $&$ -0.069 \pm 0.353 $ & Unconstrained &  Unconstrained &   $<1.00 \times 10^{-2} \, (95\%)$ & $<8.68 \times 10^{-2} \, (95\%)$ &  $\mu$ & $ -2.96 \pm 0.05 $ \\
 Cosmographic-D-Exp & 2M++ & Yes &$-0.293 \pm 0.094 $&$ -0.048 \pm 0.362 $ & Unconstrained &  Unconstrained &  $<9.63\times 10^{-3} \, (95\%)$ & $<8.94 \times 10^{-2} \, (95\%)$ &  $\mu$ & $ -3.26 \pm 0.01$ \\

   \toprule
	\multirow{2}{*}{$\downarrow$}&  \multirow{2}{*}{$\downarrow$} & \multirow{2}{*}{$\downarrow$}&\multirow{2}{*}{$q_0$} & \multirow{2}{*}{$j_0 - \Omega_\kappa$} & \multirow{2}{*}{$\downarrow$} & \multirow{2}{*}{$\downarrow$} &  \multirow{2}{*}{$|D_{q}|$} & \multirow{2}{*}{$\downarrow$}& \multirow{2}{*}{$\downarrow$}&\multirow{2}{*}{$\downarrow$}\\
\\
\midrule 

Cosmographic-D & 2M++ & No &$-0.305 \pm 0.094 $&$ -0.083 \pm 0.376 $ & Unconstrained &  Unconstrained &  $<6.06 \times 10^{-2} \, (95\%)$& - &  $q_0$ & $-7.12  \pm 0.02$ \\
Cosmographic-D & 2M++ & Yes &$-0.286 \pm 0.094 $&$ -0.084 \pm 0.367 $ & Unconstrained &  Unconstrained &  $<6.29 \times 10^{-2} \, (95\%)$ & - &  $q_0$ & $ -7.45 \pm 0.08$ \\
 
 Cosmographic-D-Exp & 2M++ & No &$-0.316 \pm 0.097 $&$ -0.040 \pm 0.370 $ & Unconstrained &  Unconstrained &  $<22.46\, (95\%)$ & $<5.59 \times 10^{-2} \, (95\%)$ &  $q_0$ & $ -1.22 \pm 0.02 $ \\
 Cosmographic-D-Exp & 2M++ & Yes &$-0.296 \pm 0.093 $&$ -0.043 \pm 0.359 $ & Unconstrained &  Unconstrained & $<23.45 \, 95\%$ & $<6.00 \times 10^{-2} \, (95\%)$ &  $q_0$ & $ -1.42 \pm 0.02 $ \\
 
\label{table:final_jla_results}
\end{tabular}}

\end{table*}

\begin{figure*}
        \includegraphics[width = 0.833\columnwidth]{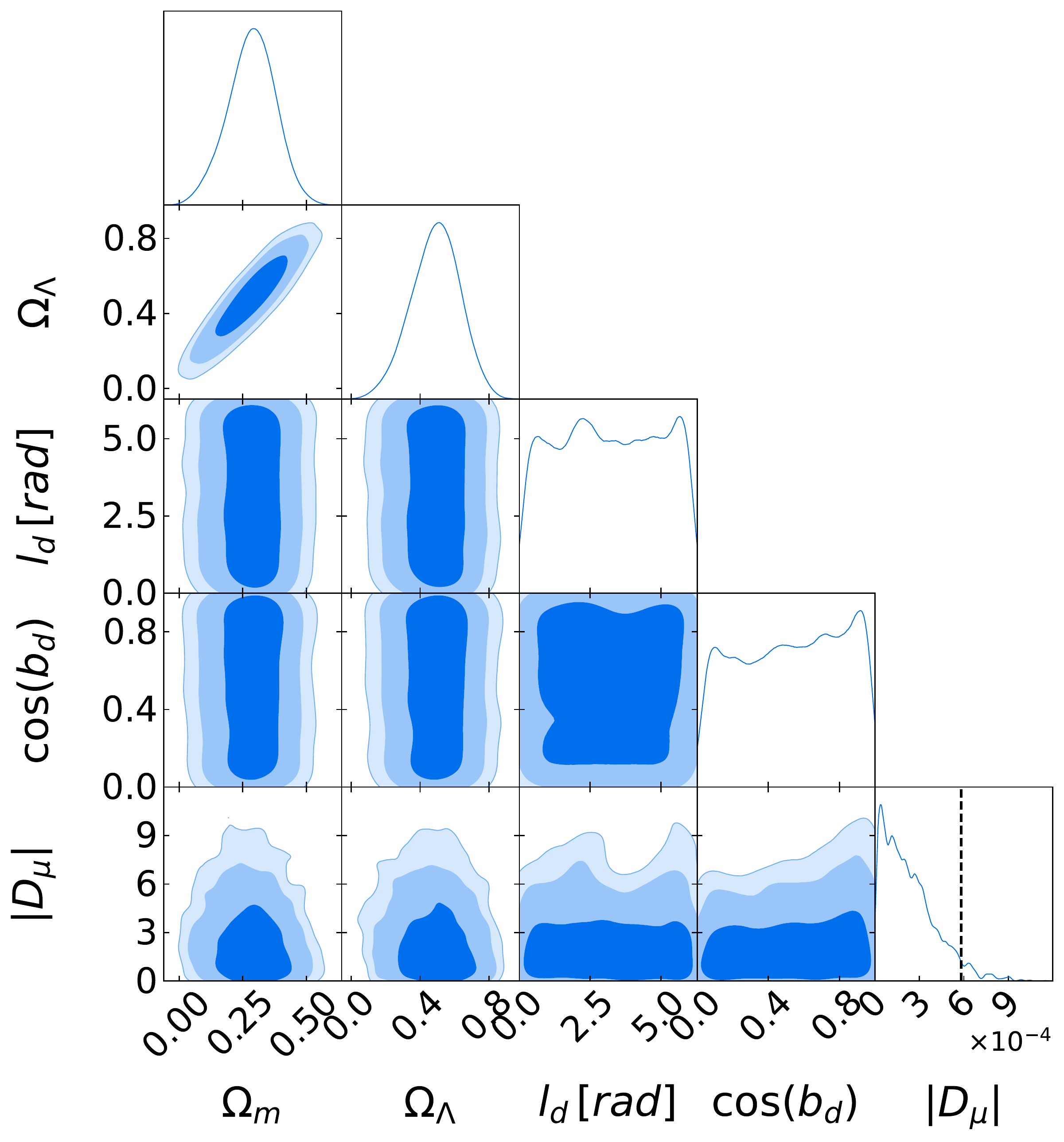}
        \includegraphics[width = \columnwidth]{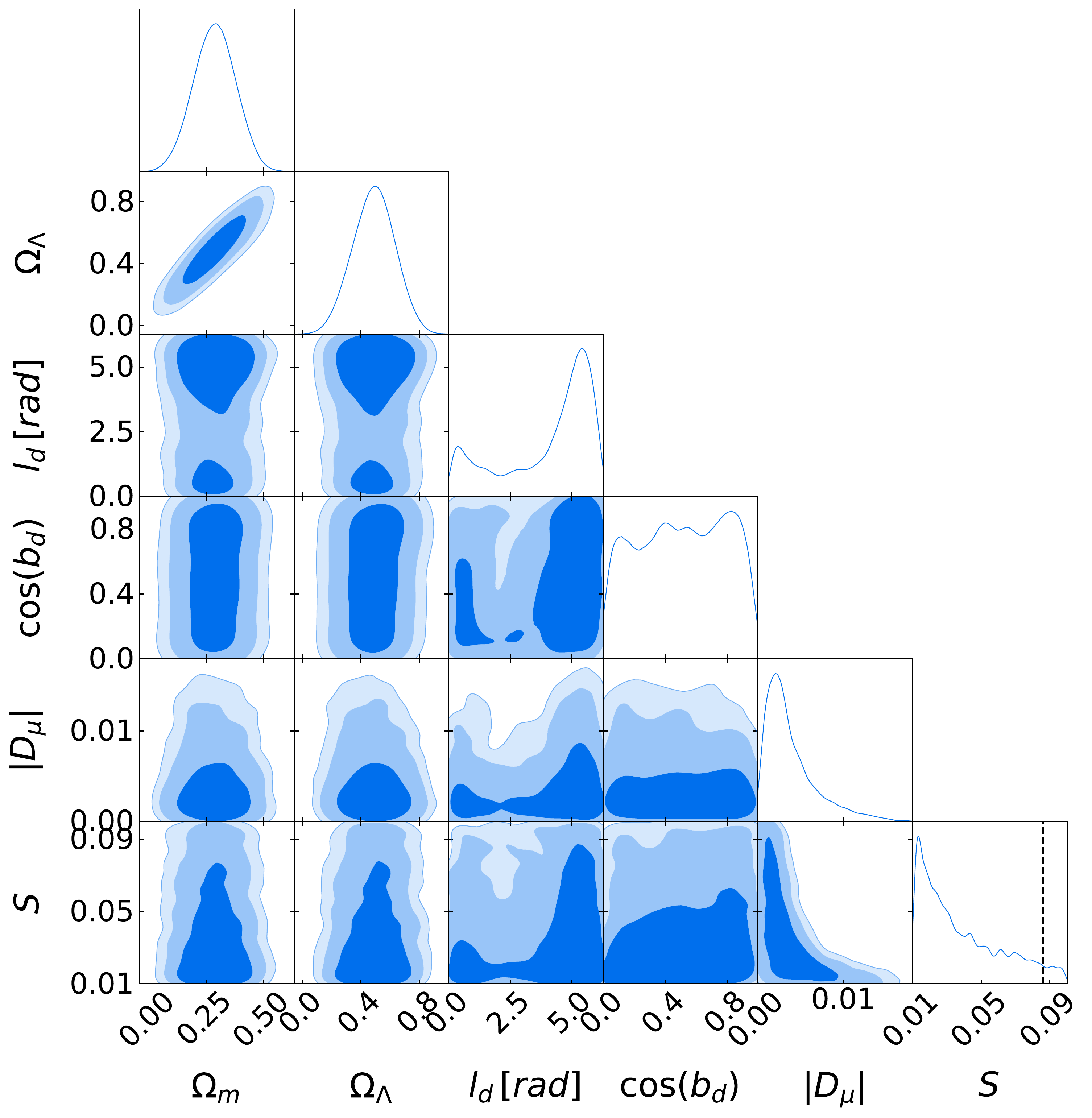}
    \caption{Posterior inference allowing for a distance modulus dipole from JLA data, assuming the \lcdm{} model with $F(z)= 1$ (left) and $F(z)= \exp{(-z/S)} $  (right), including colour-based selection effects correction. Contours in 2D plots enclose joint 1, 2 and 3 $\sigma$ HPD credible regions; the vertical dashed line in the 1D marginals for $D_\mu$ and $S$ delimits the 2$\sigma$ upper limit.}%
    \label{fig:jla_LCDM}
\end{figure*}

\begin{figure*}
    \includegraphics[width = 0.833\columnwidth]{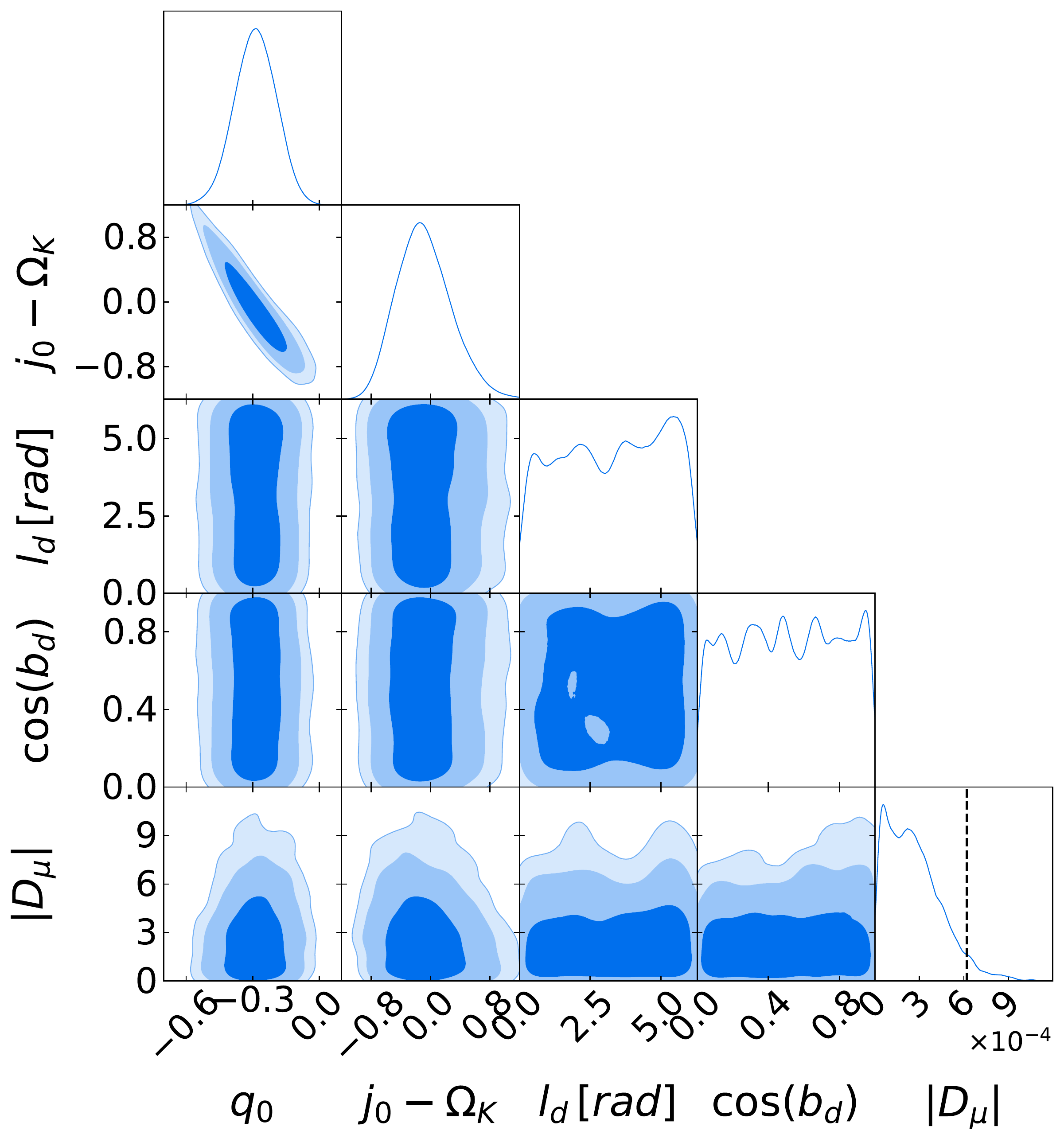}
    \includegraphics[width = \columnwidth]{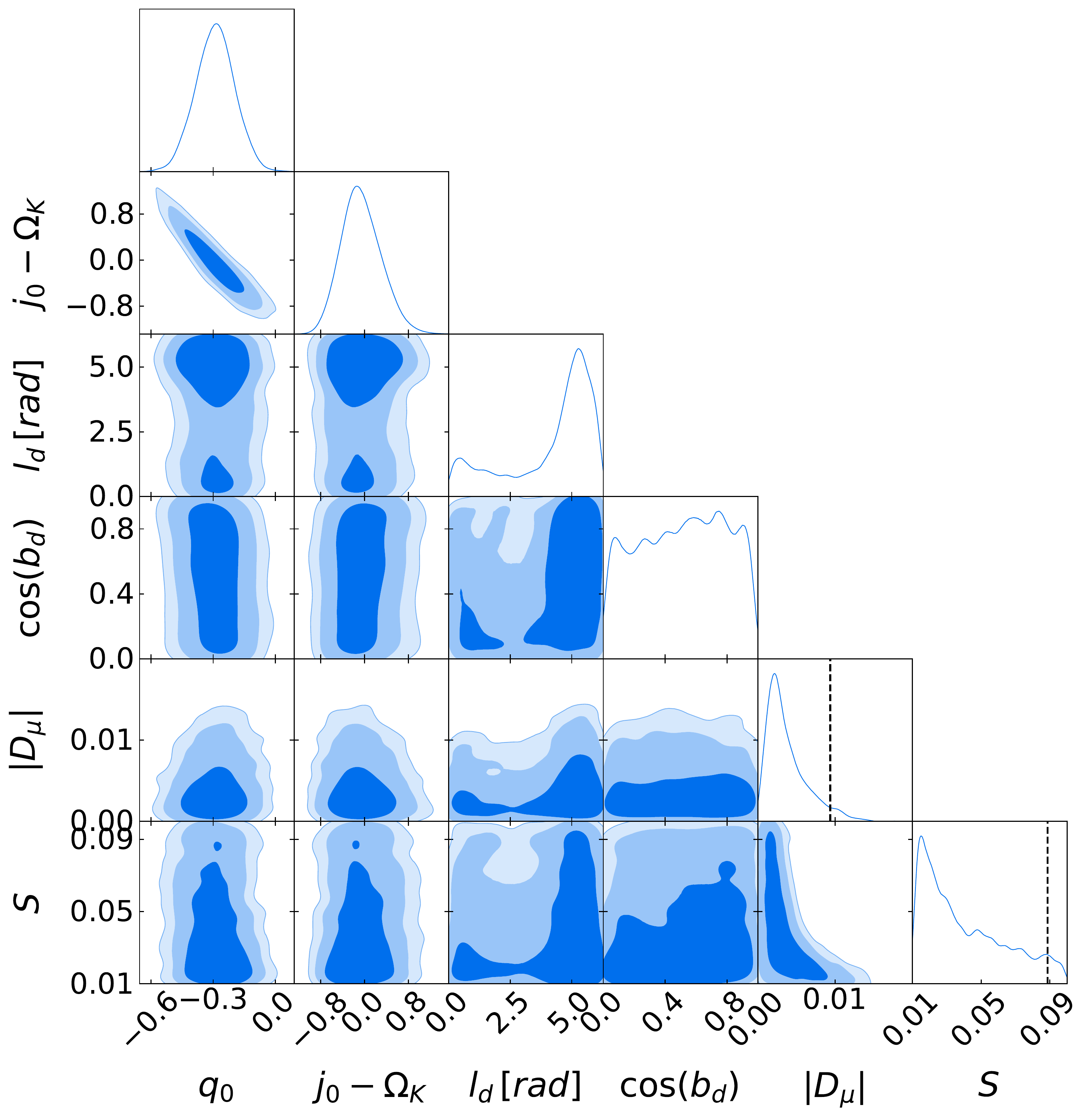}
\caption{Posterior inference allowing for a distance modulus dipole from JLA data, assuming the Cosmographic expansion model with $F(z)= 1$ (left) and $F(z)= \exp{(-z/S)} $  (right), including colour-based selection effects correction. Contours in 2D plots enclose joint 1, 2 and 3 $\sigma$ HPD credible regions; the vertical dashed line in the 1D marginals for $D_\mu$ and $S$ delimits the 2$\sigma$ upper limit.}%
\label{fig:jla_cosmographic}%
\end{figure*}

%LCDM dipole parameters 
The second section of Table \ref{table:final_jla_results} presents our constraints on the distance modulus dipole parameters, also comparing the impact of using the colour selection effect correction (as indicated in the third column), which is found to be quite minor on all constraints. Posterior 1D and 2D distributions are shown in Fig. \ref{fig:jla_LCDM} for the \lcdm{} model and in Fig.~\ref{fig:jla_cosmographic} for the Cosmographic expansion. The posterior distribution for the dipole $|D_\mu|$ in the left panel of Fig. \ref{fig:jla_LCDM} for the \lcdm{} model with $F(z) = 1$ peaks at 0,  and we set a 1-tailed $95\%$ upper bound of $|D_\mu| < 5.93 \times 10^{-4}$  (95.45\% probability) -- a factor of $\sim 2$ tighter than the limits derived by \cite{lin2016} from the same data, namely  $|D_\mu| < 1.98 \times 10^{-3}$. The dipole direction is correspondingly unconstrained.  

In the second case, shown in the right panel of Fig. \ref{fig:jla_LCDM}, a scale function of the form $F(z) = \exp{(-z/S)}$ is used to constrain the dipole to local region. As expected from our simulations, we find a degeneracy between the dipole parameter and its scale, whose effect is to degrade the upper limits on the dipole amplitude to $ |D_\mu| < 1.05 \times 10^{-2}$ for the \lcdm{} case. We can see $|D_{\mu}|$ is well explored again, right up to its prior edge because of this degeneracy, which contributes to these weaker constraints. The limits on the dipole scale $S$ are also weak, with the 1D marginal distribution stretching all the way to the prior boundary ($S=0.10$), but peaking near the lower prior boundary. The very weak preference for a non-zero dipole (seen in the peak away from 0 in the 1D marginal distribution) could be an indication of a residual effects of the bulk flow, which points broadly in the same direction as the more prominent peak in the $l_d$ posterior distribution. Such departures from perfect isotropy are weak, and not dissimilar from what we observed in our isotropic universe simulations (Fig.~\ref{fig:simulated_lcdm_vanilla_inference_anisotropy}, right panel) -- hence they can be easily ascribed to the result of random noise. 
With regards to constraints on cosmological parameters, with respect to the \lcdm{}-Isotropic case, we observe only a very mild shift in their value as a consequence of the introduction of a potential dipole in the model, $|\Delta\omegam|=0.005$ and $|\Delta \omegade|=0.006$. Both are shifts of less than $0.1$ standard deviations of the posterior, and are similar in scale for the case with $F(z) = 1$.  A similar result is seen in the Cosmographic case with  $|\Delta q_{0}| = 0.009$ which is also quite minor relative to the posterior standard deviation.

%Cosmographic expansion results 
For the isotropic Cosmographic expansion, the posterior mean for the deceleration parameter increases from $q_0 = -0.352 \pm 0.092$ when using the old PSCz velocity corrections and $\zCMB$ to $q_0 = -0.302 \pm 0.090$ when updating both velocity corrections and CMB restframe redshifts to the new values we derived here (the error indicates the standard deviation of the posterior, not the uncertainty on the mean). When introducing the possibility of a dipole, the posterior mean for $q_0$ hovers around $-0.30$, depending on the dipole model and whether we adopt the colour selection effects corrections. These values are quite a bit larger than the expectation under \lcdm{}, namely $q_0 = -0.55$, but  not as large as the results reported in Table 2 of \cite{colin2019} ($q_0 = -0.157$ in our notation), who ascribed the shift of the deceleration parameter towards 0 to evidence for an anisotropic universe. Our analysis (with its improved treatments of peculiar velocities and colour selection effects) disagrees with those conclusions: the marginal posterior probability (obtained from our posterior samples, not a Gaussian approximation to the posterior) for  $q_0 \geq 0$ (i.e., no acceleration) for the $F(z)=1$ case with selection effects corrections and a distance modulus dipole is $9.3\times 10^{-4}$. We return on this question from the point of view of Bayesian model comparison in section~\ref{sec:model_comparion}.

The constraints on the dipole parameters for the Cosmographic expansion model are qualitatively similar to those presented for the \lcdm{} model, as shown in Fig.~\ref{fig:jla_cosmographic} and detailed in the central two sections of Table~\ref{table:final_jla_results}. For the $F(z)=1$ case, the posterior dipole amplitude peaks at 0 and we set a $95\%$ upper limit $|D_{q_0}| < 6.32 \times 10^{-4}$. There is no significant evidence for a dipole moment in the Cosmographic expansion case under our data and models. A qualitatively similar picture holds for the $F(z)= \exp{(-z/S)}$ case, albeit with weaker limits on the dipole amplitude owing to the degeneracy with the scale parameter $S$.

\begin{figure*}%
    \includegraphics[width = 0.833\columnwidth]{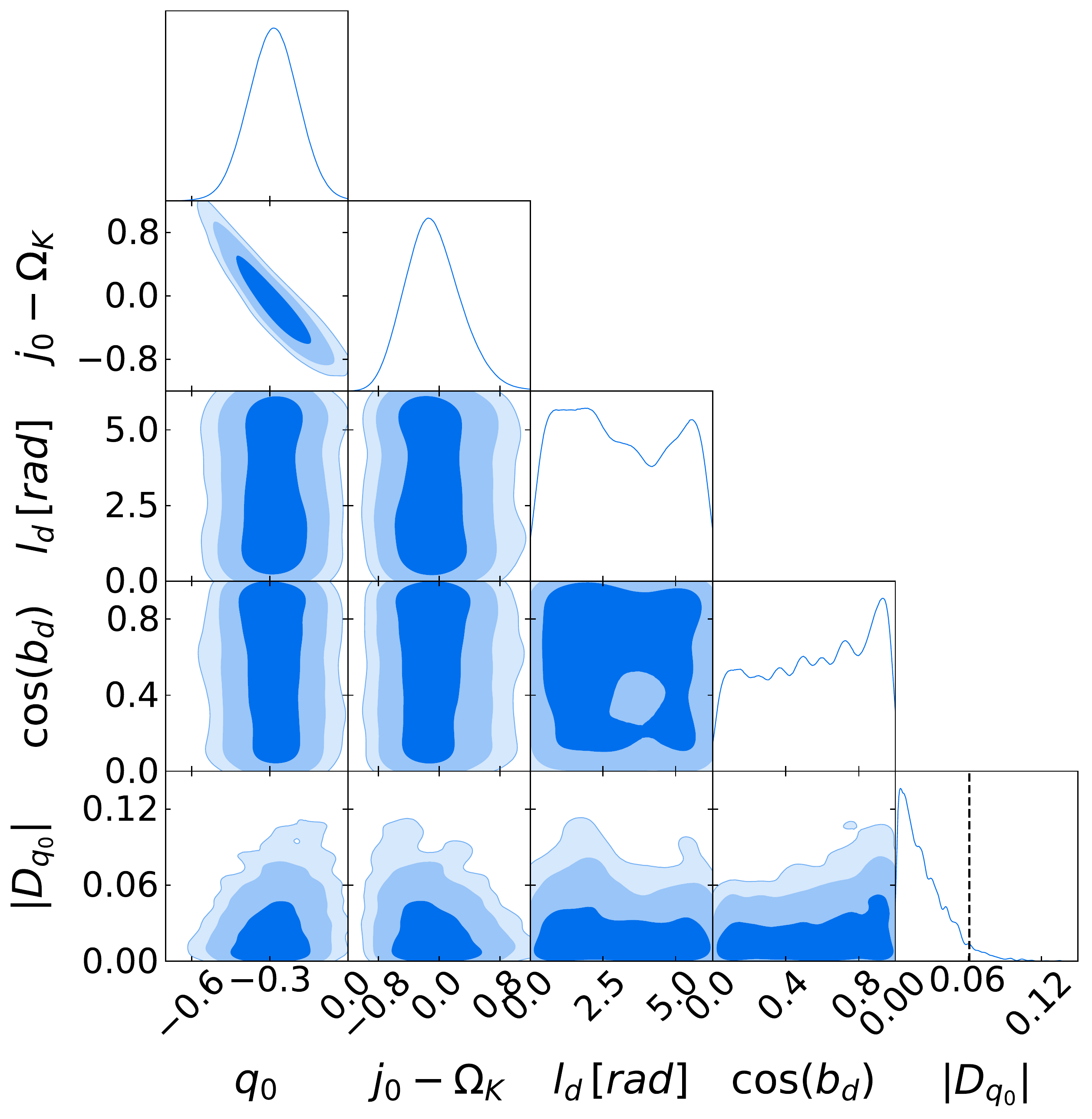}
    \includegraphics[width = \columnwidth]{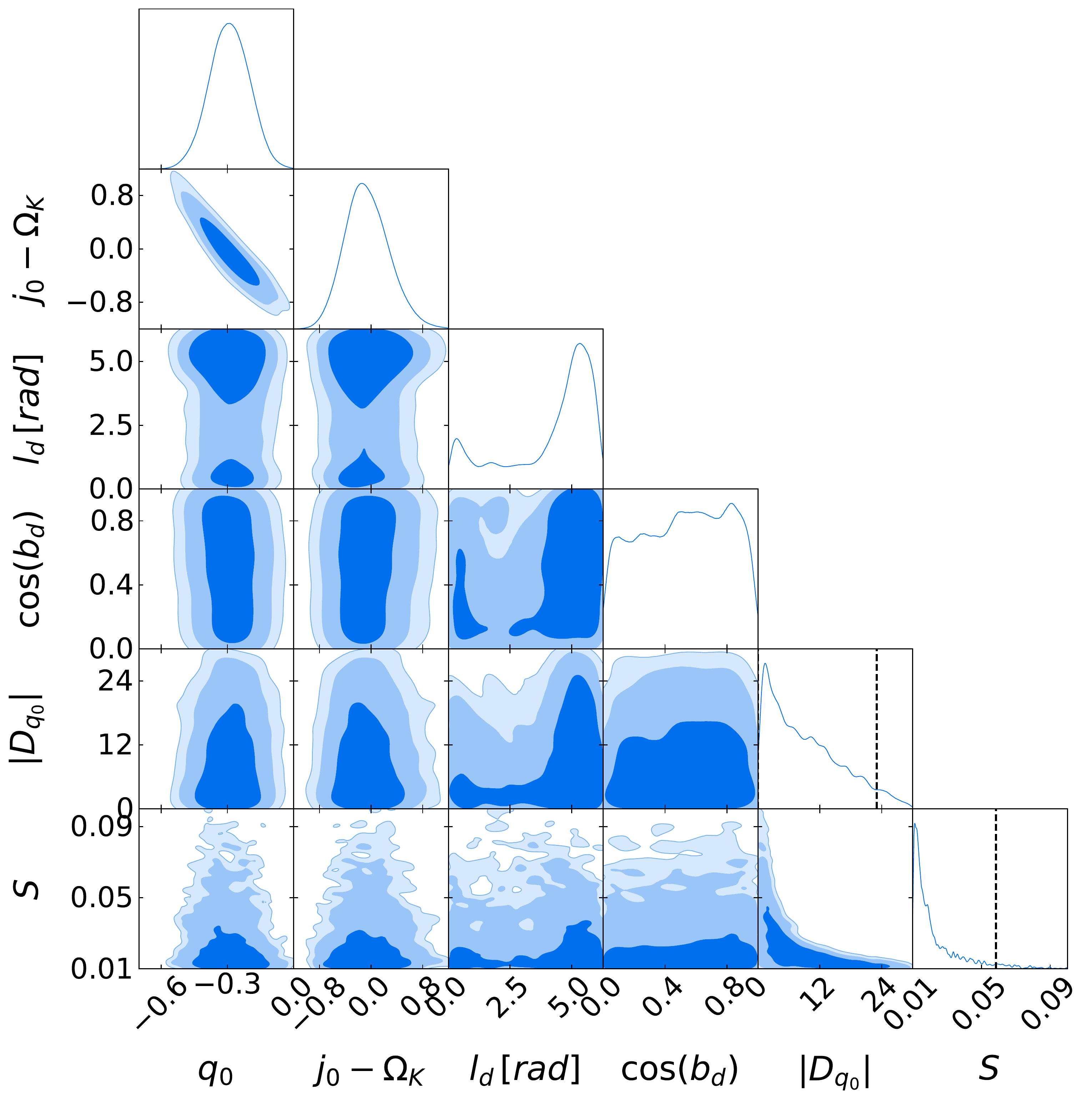}
    \caption{Posterior inference allowing for a distance modulus dipole from JLA data, assuming the Cosmographic expansion model with $F(z)= 1$ (left) and $F(z)= \exp{(-z/S)} $  (right), including colour-based selection effects correction. Contours in 2D plots enclose joint 1, 2 and 3 $\sigma$ HPD credible regions; the vertical dashed line in the 1D marginals for $D_{q_0}$ and $S$ delimits the 2$\sigma$ upper limit. The dipole here is modelled directly on the deceleration parameter rather than the distance modulus, similar to C19.}%
    \label{fig:jla_dipole_exp_scale_on_deceleration_cosmographic_no_pecvel}%
\end{figure*}

For a more direct comparison with the results of C19, we have also investigated the same model as C19, where the dipole modulation is applied to the deceleration parameter $ q_{0} $, as in Eq.~\eqref{eq:dipolar_modulation_q}, rather than to the distance modulus $ \mu $. Differently from C19,  we did {\em not} remove the bias corrections to the magnitude, we kept the direction of the dipole free (as opposed to being fixed in the CMB dipole direction), used our new peculiar velocity corrections and CMB restframe redshifts (as opposed to heliocentric redshifts) and applied our new colour selection effects. The resulting posteriors are shown in Fig. \ref{fig:jla_dipole_exp_scale_on_deceleration_cosmographic_no_pecvel} and constraints presented in the bottom section of Table \ref{table:final_jla_results}. 

We see qualitatively similar results to the case where we model the dipole on $q_{0}$ in the Cosmographic expansion. For the $F(z)=1$ case, the posterior dipole amplitude again peaks at $0$ and we set a $ 95\% $ upper limit $|D_{q_0}| < 6.29 \times 10^{-2}$. A qualitatively similar picture holds for the $ F(z)= \exp{(-z/S)} $ case with a preference for a small $S$ value. This is in contrast with \cite{colin2019}, who saw a preference of a scale value $S=0.0262$, a likely consequence of their removal of peculiar velocity correction and the use of heliocentric redshifts. 

\begin{table*}
\caption{Comparison of our results with previous searches for a dipolar modulation. We only include comparable results that utilize the ``dipole fitting'' approach (as opposed to the ``hemisphere comparison''  method). Reported detections (at higher than $95\%$ significance) are highlighted in boldface. The third column `$v_{\rm pec}$ corr'states whether peculiar velocity corrections were adopted, and if so from which data set they were derived. It should be noted here, previous significant detections involved data that had no peculiar velocity corrections.}
\resizebox{\textwidth}{!}{\begin{tabular}{l@{\qquad}lllllll@{\qquad}}
  \toprule
  Reference & Data & $v_{\rm pec}$ corr & $l_d$ (deg) & $b_d$ (deg) & Dipole amplitude & Quantity modulated \\
  \midrule
  \cite{cooke2010}  & Union & None & $309$ & $43$ & $ 0.14 \pm 0.12$& $d_L$ \\ 
  \midrule
  \cite{2016MNRASLin2} & Union2.1 &  None & $171.8_{-42.0}^{+42.0}$ & $9.9_{-20.3}^{+20.3}$ & $0.160 \pm 0.115$ & $\Omega_{m}$ \\
  \midrule 
  \cite{mariano2012} & Union2 & \bf None & $309_{-18}^{+18}$ & $-15.1_{-11.5}^{+11.5}$ & \boldmath{$(1.3 \pm 0.6) \times 10^{-3}$} & $\mu$ \\
  %\cite{mariano2012} & Keck + VLT & \bf None & $316_{-110}^{+107}$ & $-5_{-60}^{+45}$ & \boldmath{$(1.02 \pm 0.25) \times 10^{-3}$}& $\mu$ \\
    \cite{yang2014} & Union2.1 & \bf None & $307.1_{-16.2}^{+16.2}$ & $-14.3_{-10.1}^{+10.1}$ & \boldmath{$(1.2 \pm 0.5) \times 10^{-3}$}& $\mu$ \\ 
  \cite{wang2014} & Union2.1+GRB& \bf None & $309.2_{-15.8}^{+15.8}$ & $-8.6_{-10.5}^{+10.5}$ & \boldmath{$(1.37 \pm 0.57) \times 10^{-3}$}& $\mu$ \\ 
  \cite{lin2016}  & JLA & PSCz & $316_{-110}^{+107}$ & $-5_{-60}^{+41}$ & $ < 1.98 \times 10^{-3}\,(95\%) $& $\mu$ \\
  \cite{sun2018} & Union2.1 & \bf None & $309_{-15.7}^{+15.5}$ & $-8.9_{-9.8}^{+11.2}$ & \boldmath{$(1.46 \pm 0.56) \times 10^{-3}$} & $\mu$ \\
  \cite{sun2018} & Constitution& None & $67.0_{-66.2}^{+66.5}$ & $-0.6_{-26.3}^{+25.2}$ & $ (4.4 \pm 5.0) \times 10^{-4}$ & $\mu$ \\
  \cite{sun2018} & JLA & PSCz & Unconstrained & Unconstrained & Unconstrained & $\mu$ \\
  \cite{Sun_2018} & Pantheon & 2M++ & $329_{-28}^{+101}$ & $37_{-21}^{+52}$ & $(3.7_{-3.7}^{+2,5}) \times 10^{-4}\,(95\%)$ & $\mu$ \\
  \cite{2019MNRAS.486.5679Z} & Pantheon & 2M++ & $306.00_{-125.01}^{+82.95}$ & $-34.20_{-54.93}^{+16.82}$ & $ < 1.16 \times 10^{-4}\,(95\%) $ & $\mu$ \\
  This work  & JLA & 2M++ & Unconstrained & Unconstrained & $ < 5.93 \times 10^{-4} (95\%) $& $\mu$, $F(z) = 1$ \\
  This work  & JLA & 2M++ & Unconstrained & Unconstrained & $ < 1.05 \times 10^{-2} (95\%) $& $\mu$, $F(z) = \exp{(-z/S)}$ \\

  \midrule 
 \cite{colin2019} & JLA & {\bf None} & $264.021$ (fixed)& $48.253$ (fixed) & \boldmath{$-8.03_{-2.05}^{+2.05}$} & $q_0$, $F(z) = \exp(-z/S)$ \\
  \cite{Rubin_2020} & JLA & {\bf None} & $264.021$ (fixed)& $48.253$ (fixed) & $-8.65^{+2.2}_{-2.6}$ & $q_0$, $F(z) = \exp(-z/S)$\\
  \cite{Rubin_2020} & JLA & PCSz & $264.021$ (fixed)& $48.253$ (fixed)& $-1.1^{+3.2}_{-3.4}$ & $q_0$, $F(z) = \exp(-z/S)$\\
   This work   & JLA & 2M++ & Unconstrained & Unconstrained & $ < 6.32 \times 10^{-2}$ ($95\%$) & $q_{0}$, $F(z) = 1$ \\
   This work  & JLA & 2M++ &  Unconstrained &  Unconstrained & $ < 13.68$ ($95\%$)& $q_{0}$, $F(z) = \exp(-z/S)$ \\
  \bottomrule
\label{table:compiled_author_results}
\end{tabular}}
\end{table*}

\subsection{Bayesian Model Comparison}
\label{sec:model_comparion}

We compare the isotropic expansion model to the alternatives featuring a dipole via Bayesian model comparison, and report the difference in the log of the Bayesian evidence (i.e., the log of the Bayes factor) in Table \ref{table:final_jla_results}: 
\be 
\Delta \ln (Z) = \ln B_D - \ln B_I,
\ee
where $B_I$ is the Bayesian evidence for the isotropic model (either \lcdm{} or Cosmographic expansion) and $B_D$ is the evidence for a model featuring a dipole, with priors as in Table~\ref{table:priors}. A value of $\Delta \ln (Z) < 0$ indicates a preference for the isotropic model. According to the Jaynes' scale for the strength of evidence, values of $|\Delta \ln (Z) | = 2.5 (5.0)$ correspond to moderate (strong) evidence for one of the models being compared. For equal prior probability for the models, the quantity $\exp(\Delta \ln (Z)) = B_D/B_I$ gives the posterior odds between the isotropic and the dipole expansion models, which are approximately 12:1 (150:1) for moderate (strong) evidence (see \cite{2008ConPh..49...71T} for details on Bayesian model comparison). When considering the numerical odds derived from the Bayesian evidence, it is important to bear in mind that these can be very sensitive to the choice of prior distribution, particularly the prior on the parameters that only appears in the more complicated model (i.e., the model featuring the dipole). While we believe that our choices of prior distributions are well justified in Section~\ref{sec:prior_selection}, researchers that make other choices for their prior distributions may compute odds ratios that differ from those that we report.

In Table~\ref{table:priors} we only carry out pair-wise model comparisons between models that use the same data and same treatment of colour selection effects, for comparing models with different data and/or assumptions about the data generating process would be meaningless. The evidences (and associated uncertainties) are estimated with \texttt{PyMultinest}, which was run with 400 live points and an evidence tolerance of 0.5. We observe that the isotropic model is favoured over all others: in the \lcdm{} scenario, the constant dipole model is disfavoured with odds ranging from 900:1 to almost 6000:1, depending on the adoption of colour selection effects. The model with a dipole falling off with redshift is also disfavoured, albeit with smaller odds ranging between 32:1 and 194:1. 

A similar pictures holds for comparison between models in the Cosmographic expansion case. Here, the odds against the anistropic models are generally smaller than in \lcdm, owing to the smaller parameter space volume ratio between posterior and prior, which control the strength of the Occam's razor effect in favour of the isotropic model. We also note that when introducing a dipole scale parameter $S$, despite the larger number of free parameters in this model w.r.t. the case where the dipole is constant in redshift, the Bayes factor against it is actually smaller than the constant-in-redshift dipole case. This can be explained by noting that the introduction of $S$ as a free parameter leads to much less stringent limits on the dipole amplitude because of the degeneracy explained above. Since $S$ itself is only weakly constrained, the Occam's razor effect for these two parameters is weakened, leading to a weaker preference for the isotropic model. As a consequence, the Cosmographic-D-Exp model where the dipole is on the deceleration parameter is only very mildly disfavoured (odds smaller than 5:1) w.r.t. the isotropic model -- a consequence of the fact that the additional parameters for this model cannot be strongly constrained, and hence the posterior odds remain approximately equal to the prior odds. 

As another case, we reproduce the setup used in C19, namely removing peculiar velocity corrections entirely, using heliocentric redshifts (i.e., $\zhel$ instead of $\zbar$ in Eq.~\eqref{eq:zbar_linear}, a choice that imprints the dipole due to the solar system's motion onto the data, as pointed out by~\cite{Rubin_2020}) and removing from the covariance matrix all uncertainties associated with peculiar velocity corrections.  In this setup, we compare the evidence for an isotropic Cosmographic expansion with that of a dipolar modulation of the form $F(z) = \exp{(-z/S)}$ either on the deceleration parameter, Eq.~\eqref{eq:dipolar_modulation_q} (as in C19), or on the distance modulus, Eq.~\eqref{eq:dipolar_modulation}. We adopt a Gaussian prior with a standard deviation of $10^{\circ}$ on the dipole direction, centred on the bulk flow direction from~\cite{RN345}, namely $l_{bf} =  301^\circ\pm4^\circ, b_{bf} = 0^\circ\pm3^\circ$ (in excellent agreement with the results of \cite{Said:2020epb}, obtained using the Fundamental Plane relation, namely $l_{bf} =  304^\circ \pm 4^\circ, b_{bf} = 1^\circ \pm 4^\circ$). When the dipole is modelled on the deceleration parameter, as in C19, the Cosmographic anisotropic model is still disfavoured with respect to the isotropic one, with odds of approximately 17:1 ($\Delta \ln(Z) = -2.84 \pm 0.08$). Although the anisotropic model achieves a better quality of fit by absorbing the dipole in the data, from an Occam's razor persepective it remains disfavoured due to its additional, unwarranted model complexity. We observe a similar effect (if stronger) when the dipole is modelled instead on distance modulus $\mu$, with odds of approximately 150:1  ($\Delta \ln(Z) = 5.03 \pm 0.03$) in favour of the isotropic model. We can repeat this comparison in the isotropic \lcdm\ case and compare that with an anisotropic \lcdm\ model with $F(z) = \exp{(-z/S)}$, finding $\Delta \ln(Z) = -4.92 \pm 0.15$, which again favours the isotropic model with odds of 136:1.

Another interesting question is the strength of evidence in favour of an accelerating universe in the isotropic expansion case. The Bayes factor between the isotropic \lcdm{} model and an isotropic model with no dark energy ($\Omega_\Lambda = 0$) disfavours the latter with odds in excess of 120:1 (including selection effects corrections). For completeness (and to compare the above Bayesian model comparison results with a hypothesis testing approach), we have also computed the log-likelihood difference for the best-fit parameter values:
\be
\Delta \ln (L) =  \ln \hat{L}_{\Omega_\Lambda=0} - \ln \hat{L}_{\Lambda\text{CDM}}
\ee
where $\hat{L}_{\Lambda\text{CDM}}$ is the maximum likelihood value for the \lcdm{} model and  $\hat{L}_{\Omega_\Lambda=0}$ is the maximum likelihood value for a universe with no dark energy. We find $\Delta {\ln (L)} = -6.35$ (with selection effects correction), which we translate into a $p$-value using Chernoff theorem (as the hypothesis being tested, $\Omega_\Lambda = 0$, lies on the boundary of the allowed parameter space), obtaining a $p$-value of $5.9 \times 10^{-3}$ for the hypothesis that $\Omega_\Lambda=0$.

In the phenomenological Cosmographic expansion setting, the accelerating isotropic model, with a uniform prior on $q_0 \in [-2,0)$ is favoured with odds of approximately 26:1 when compared to a coasting universe, i.e., $q_0=0$. The accelerating universe is preferred with odds of almost $\sim$ 1100:1 when  compared with a decelerating model, i.e. one with a uniform prior $q_0 \in (0,1]$. 

\section{Summary and Conclusions}
\label{sec:conclusions}

We have revisited the question of a dipolar anisotropy in the expansion of the universe, and derived new constraints on a possible dipolar modulation from \SN\ data. Our approach builds on the Bayesian hierarchical model \bahamas{}, which has been extended to include a new approximate correction for residual colour-based selection effects. We have also upgraded the treatment of peculiar velocities and host galaxy redshifts from the original JLA paper, by adopting state-of-the-art flow models constrained using the 2M++ galaxy catalogue. Finally, we have improved the treatment of both statistical and systematic uncertainties pertaining to peculiar velocity corrections -- the dominant source of error for $z \lesssim 0.1$ \SN, which are all-important for a robust, accurate and precise measurement of anisotropy in the local expansion. 

We did not find any evidence for a deviation from isotropy, either in the framework of \lcdm{} or in phenomenological Cosmographic expansion.  We placed tight constraints on the amplitude of a possible dipole both in the distance modulus and on the deceleration parameter. Our upper bounds are more stringent by a factor of $\sim 2$ than the results previously obtained from the same data sets with a comparable approach. We note that all previous searches that have claimed a significant detection of anisotropy have neglected peculiar velocity corrections.

We have used the framework of Bayesian model comparison to evaluate the Bayes factor between models featuring a dipole and an isotropically expanding universe (both in \lcdm{} and in the Cosmographic expansion). We found moderate to strong Bayesian evidence against an anisotropic expansion. We have also evaluated the evidence in favour of acceleration, finding that a non-zero cosmological constant is preferred, using JLA \SNe ~alone, by odds of 120:1, a result corroborated by a more traditional $p$-value approach based on a frequentist hypothesis test, which rejects $\Omega_\Lambda = 0$ with a $p$-value of $5.9\times 10^{-3}$. In the Cosmographic expansion, a decelerating universe is disfavoured with odds of almost $1100:1$ w.r.t. an accelerating one.  We conclude that the preferred model remains the \lcdm{} isotropically expanding model.

Our work does not address the important question of the so-called ``$H_0$ tension'' -- the fact that the value of the Hubble-Lemaître constant obtained from local distance ladder measurements shows a highly statistical significant discrepancy with respect to the the smaller value obtained via CMB anisotropies at high redshift (see, e.g., \cite{DiValentino:2021izs} for a review). \SNe\ type Ia, on their own, only provide a relative distance measurement, given the exact degeneracy between $H_0$ and $M_0$. However, in combination with calibrated distance indicators, such as Cepheid variables, they deliver an absolute distance scale that is used to obtain tight constraints on $H_0$, see~\cite{RN1573} for the latest results of the SH0ES collaboration, and \cite{Freedman2021}, who finds that calibrating \SN\ via the tip of the Red Giant Branch (TRGB) instead, the tension with the CMB anisotropies-determined value is no longer significant. The novel peculiar velocity corrections presented here have potentially a bearing on this important problem, in that a change in the estimated peculiar velocity implies a horizontal shift of the \SN\ in the Hubble diagram -- potentially even correlated across several \SNe. This in turn could affect the linear fit that gives, via the slope, the value of $H_0$. While we leave the investigation of this effect to future work, we do provide our new corrections to the JLA data publicly (see Data Availability section at the end), so that others wishing to use it for studies of $H_0$ can do so.

In this work we adopted the JLA compilation since we were unable to use the most recent Pantheon sample, owing to the full covariance matrix (including systematics) not being publicly available. We plan to apply our new framework, including up-to-date peculiar velocity corrections, to the Foundation sample~\citep{FoundationSNe1, FoundationSNe2}, together with the recent data release of the Dark Energy Survey \citep{des2018a}. In the near future, we expect to be able to obtain even tighter constraints on possible anisotropy in the expansion from upcoming, larger \SN\ surveys like the one that will be delivered by the Vera Rubin Observatory Legacy Survey of Space and Time (LSST) \citep{lsstsciencecollaboration2009lsst}. 

\section*{Acknowledgements}
RT was partially supported by STFC grants ST/P000762/1 and ST/T000791/1. WR was supported by PhD scholarship by the Science and Technology Facilities Council (STFC), UK and partially supported by a Quantitative Research grant provided by G Research, which is gratefully acknowledged.  MJH acknowledges support from an NSERC Discovery Grant.

\section*{Data Availability}
A Python implementation of BAHAMAS for public used in this article is being worked on and will be made available soon. The data products generated from our analysis from the article is available from \url{https://zenodo.org/record/5854639}

%%%%%%%%%%%%%%%%%%%%%%%%%%%%%%%%%%%%%%%%%%%%%%%%%%



\bibliographystyle{mnras}
\bibliography{main} % if your bibtex file is called example.bib

\begin{thebibliography}{}
\makeatletter
\relax
\def\mn@urlcharsother{\let\do\@makeother \do\$\do\&\do\#\do\^\do\_\do\%\do\~}
\def\mn@doi{\begingroup\mn@urlcharsother \@ifnextchar [ {\mn@doi@}
  {\mn@doi@[]}}
\def\mn@doi@[#1]#2{\def\@tempa{#1}\ifx\@tempa\@empty \href
  {http://dx.doi.org/#2} {doi:#2}\else \href {http://dx.doi.org/#2} {#1}\fi
  \endgroup}
\def\mn@eprint#1#2{\mn@eprint@#1:#2::\@nil}
\def\mn@eprint@arXiv#1{\href {http://arxiv.org/abs/#1} {{\tt arXiv:#1}}}
\def\mn@eprint@dblp#1{\href {http://dblp.uni-trier.de/rec/bibtex/#1.xml}
  {dblp:#1}}
\def\mn@eprint@#1:#2:#3:#4\@nil{\def\@tempa {#1}\def\@tempb {#2}\def\@tempc
  {#3}\ifx \@tempc \@empty \let \@tempc \@tempb \let \@tempb \@tempa \fi \ifx
  \@tempb \@empty \def\@tempb {arXiv}\fi \@ifundefined
  {mn@eprint@\@tempb}{\@tempb:\@tempc}{\expandafter \expandafter \csname
  mn@eprint@\@tempb\endcsname \expandafter{\@tempc}}}

\bibitem[\protect\citeauthoryear{{Akrami}, {Fantaye}, {Shafieloo}, {Eriksen},
  {Hansen}, {Banday}  \& {G{\'o}rski}}{{Akrami} et~al.}{2014}]{akrami2014}
{Akrami} Y.,  {Fantaye} Y.,  {Shafieloo} A.,  {Eriksen} H.~K.,  {Hansen} F.~K.,
   {Banday} A.~J.,   {G{\'o}rski} K.~M.,  2014, \mn@doi [\apjl]
  {10.1088/2041-8205/784/2/L42}, \href
  {https://ui.adsabs.harvard.edu/abs/2014ApJ...784L..42A} {784, L42}

\bibitem[\protect\citeauthoryear{{Amanullah} et~al.,}{{Amanullah}
  et~al.}{2010}]{amanullah2010}
{Amanullah} R.,  et~al., 2010, \mn@doi [\apj] {10.1088/0004-637X/716/1/712},
  \href {http://adsabs.harvard.edu/abs/2010ApJ...716..712A} {716, 712}

\bibitem[\protect\citeauthoryear{{Andrade}, {Bengaly}, {Alcaniz}  \&
  {Santos}}{{Andrade} et~al.}{2018a}]{andrade2018}
{Andrade} U.,  {Bengaly} C.~A.~P.,  {Alcaniz} J.~S.,   {Santos} B.,  2018a,
  \mn@doi [\prd] {10.1103/PhysRevD.97.083518}, \href
  {http://adsabs.harvard.edu/abs/2018PhRvD..97h3518A} {97, 083518}

\bibitem[\protect\citeauthoryear{Andrade, Bengaly, Santos  \& Alcaniz}{Andrade
  et~al.}{2018b}]{Andrade2018eta}
Andrade U.,  Bengaly C. A.~P.,  Santos B.,   Alcaniz J.~S.,  2018b, \mn@doi
  [Astrophys. J.] {10.3847/1538-4357/aadb90}, 865, 119

\bibitem[\protect\citeauthoryear{Antoniou \& Perivolaropoulos}{Antoniou \&
  Perivolaropoulos}{2010}]{Antoniou_2010}
Antoniou I.,  Perivolaropoulos L.,  2010, \mn@doi [Journal of Cosmology and
  Astroparticle Physics] {10.1088/1475-7516/2010/12/012}, 2010, 012

\bibitem[\protect\citeauthoryear{{Azzalini} \& {Arellano-Valle}}{{Azzalini} \&
  {Arellano-Valle}}{2012}]{2012arXiv1203.2376A}
{Azzalini} A.,  {Arellano-Valle} R.~B.,  2012, arXiv e-prints, \href
  {https://ui.adsabs.harvard.edu/abs/2012arXiv1203.2376A} {p. arXiv:1203.2376}

\bibitem[\protect\citeauthoryear{Bennett et~al.,}{Bennett
  et~al.}{2013}]{bennett2013}
Bennett C.~L.,  et~al., 2013, The Astrophysical Journal Supplement Series, 208,
  20

\bibitem[\protect\citeauthoryear{{Bernal}, {C{\'a}rdenas}  \& {Motta}}{{Bernal}
  et~al.}{2017}]{2017PhLB..765..163B}
{Bernal} C.,  {C{\'a}rdenas} V.~H.,   {Motta} V.,  2017, \mn@doi [Physics
  Letters B] {10.1016/j.physletb.2016.12.008}, \href
  {https://ui.adsabs.harvard.edu/abs/2017PhLB..765..163B} {765, 163}

\bibitem[\protect\citeauthoryear{{Betoule} et~al.,}{{Betoule}
  et~al.}{2014}]{betoule2014}
{Betoule} M.,  et~al., 2014, \mn@doi [\aap] {10.1051/0004-6361/201423413},
  \href {https://ui.adsabs.harvard.edu/abs/2014A&A...568A..22B} {568, A22}

\bibitem[\protect\citeauthoryear{{Boruah}, {Hudson}  \& {Lavaux}}{{Boruah}
  et~al.}{2020a}]{RN339}
{Boruah} S.~S.,  {Hudson} M.~J.,   {Lavaux} G.,  2020a, arXiv e-prints, \href
  {https://ui.adsabs.harvard.edu/abs/2020arXiv201001119B} {p. arXiv:2010.01119}

\bibitem[\protect\citeauthoryear{Boruah, Hudson  \& Lavaux}{Boruah
  et~al.}{2020b}]{RN345}
Boruah S.~S.,  Hudson M.~J.,   Lavaux G.,  2020b, \mn@doi [Monthly Notices of
  the Royal Astronomical Society] {10.1093/mnras/staa2485}, 498, 2703

\bibitem[\protect\citeauthoryear{{Branchini} et~al.,}{{Branchini}
  et~al.}{1999}]{1999MNRAS.308....1B}
{Branchini} E.,  et~al., 1999, \mn@doi [\mnras]
  {10.1046/j.1365-8711.1999.02514.x}, \href
  {https://ui.adsabs.harvard.edu/abs/1999MNRAS.308....1B} {308, 1}

\bibitem[\protect\citeauthoryear{{Brout} \& {Scolnic}}{{Brout} \&
  {Scolnic}}{2021}]{2021ApJ...909...26B}
{Brout} D.,  {Scolnic} D.,  2021, \mn@doi [\apj] {10.3847/1538-4357/abd69b},
  \href {https://ui.adsabs.harvard.edu/abs/2021ApJ...909...26B} {909, 26}

\bibitem[\protect\citeauthoryear{{Brout} et~al.,}{{Brout}
  et~al.}{2022}]{PantheonPlus2022}
{Brout} D.,  et~al., 2022, arXiv e-prints, \href
  {https://ui.adsabs.harvard.edu/abs/2022arXiv220204077B} {p. arXiv:2202.04077}

\bibitem[\protect\citeauthoryear{{Buchner} et~al.,}{{Buchner}
  et~al.}{2014}]{buchner2014}
{Buchner} J.,  et~al., 2014, \mn@doi [\aap] {10.1051/0004-6361/201322971},
  \href {http://adsabs.harvard.edu/abs/2014A%26A...564A.125B} {564, A125}

\bibitem[\protect\citeauthoryear{Cai \& Tuo}{Cai \& Tuo}{2012}]{Cai_2012}
Cai R.-G.,  Tuo Z.-L.,  2012, \mn@doi [Journal of Cosmology and Astroparticle
  Physics] {10.1088/1475-7516/2012/02/004}, 2012, 004

\bibitem[\protect\citeauthoryear{Calcino \& Davis}{Calcino \&
  Davis}{2017}]{Calcino_2017}
Calcino J.,  Davis T.,  2017, \mn@doi [Journal of Cosmology and Astroparticle
  Physics] {10.1088/1475-7516/2017/01/038}, 2017, 038

\bibitem[\protect\citeauthoryear{Campanelli, Cea, Fogli  \& Marrone}{Campanelli
  et~al.}{2011}]{campanelli2011}
Campanelli L.,  Cea P.,  Fogli G.~L.,   Marrone A.,  2011, \mn@doi [Phys. Rev.
  D] {10.1103/PhysRevD.83.103503}, 83, 103503

\bibitem[\protect\citeauthoryear{Carr, Davis, Scolnic, Said, Brout, Peterson
  \& Kessler}{Carr et~al.}{2021}]{Carr2021}
Carr A.,  Davis T.~M.,  Scolnic D.,  Said K.,  Brout D.,  Peterson E.~R.,
  Kessler R.,  2021, The Pantheon+ Analysis: Improving the Redshifts and
  Peculiar Velocities of Type Ia Supernovae Used in Cosmological Analyses,
  \mn@doi{10.48550/ARXIV.2112.01471}, \url {https://arxiv.org/abs/2112.01471}

\bibitem[\protect\citeauthoryear{Carrick, Turnbull, Lavaux  \& Hudson}{Carrick
  et~al.}{2015}]{10.1093/mnras/stv547}
Carrick J.,  Turnbull S.~J.,  Lavaux G.,   Hudson M.~J.,  2015, \mn@doi
  [Monthly Notices of the Royal Astronomical Society] {10.1093/mnras/stv547},
  450, 317

\bibitem[\protect\citeauthoryear{{Chambers} et~al.,}{{Chambers}
  et~al.}{2016}]{panstarrs}
{Chambers} K.~C.,  et~al., 2016, arXiv e-prints, \href
  {https://ui.adsabs.harvard.edu/abs/2016arXiv161205560C} {p. arXiv:1612.05560}

\bibitem[\protect\citeauthoryear{Childress, Wolf  \& Zahid}{Childress
  et~al.}{2014}]{childress2014}
Childress M.~J.,  Wolf C.,   Zahid H.~J.,  2014, \mn@doi [Monthly Notices of
  the Royal Astronomical Society] {10.1093/mnras/stu1892}, 445, 1898

\bibitem[\protect\citeauthoryear{{Colin}, {Mohayaee}, {Rameez}  \&
  {Sarkar}}{{Colin} et~al.}{2019a}]{Nielsen19}
{Colin} J.,  {Mohayaee} R.,  {Rameez} M.,   {Sarkar} S.,  2019a, arXiv
  e-prints, \href {https://ui.adsabs.harvard.edu/abs/2019arXiv191204257C} {p.
  arXiv:1912.04257}

\bibitem[\protect\citeauthoryear{Colin, Mohayaee, Rameez  \& Sarkar}{Colin
  et~al.}{2019b}]{colin2019}
Colin J.,  Mohayaee R.,  Rameez M.,   Sarkar S.,  2019b, \mn@doi [Astronomy &
  Astrophysics] {10.1051/0004-6361/201936373}, 631, L13

\bibitem[\protect\citeauthoryear{{Conley} et~al.,}{{Conley}
  et~al.}{2011}]{snls_conley}
{Conley} A.,  et~al., 2011, \mn@doi [\apjs] {10.1088/0067-0049/192/1/1}, \href
  {https://ui.adsabs.harvard.edu/abs/2011ApJS..192....1C} {192, 1}

\bibitem[\protect\citeauthoryear{{Cooke} \& {Lynden-Bell}}{{Cooke} \&
  {Lynden-Bell}}{2010}]{cooke2010}
{Cooke} R.,  {Lynden-Bell} D.,  2010, \mn@doi [\mnras]
  {10.1111/j.1365-2966.2009.15755.x}, \href
  {http://adsabs.harvard.edu/abs/2010MNRAS.401.1409C} {401, 1409}

\bibitem[\protect\citeauthoryear{{DES Collaboration} et~al.,}{{DES
  Collaboration} et~al.}{2018}]{des2018a}
{DES Collaboration} et~al., 2018, arXiv e-prints, \href
  {http://adsabs.harvard.edu/abs/2018arXiv181102375D} {}

\bibitem[\protect\citeauthoryear{{Dam}, {Heinesen}  \& {Wiltshire}}{{Dam}
  et~al.}{2017}]{2017MNRAS.472..835D}
{Dam} L.~H.,  {Heinesen} A.,   {Wiltshire} D.~L.,  2017, \mn@doi [\mnras]
  {10.1093/mnras/stx1858}, \href
  {https://ui.adsabs.harvard.edu/abs/2017MNRAS.472..835D} {472, 835}

\bibitem[\protect\citeauthoryear{{Davis}, {Nusser}  \& {Willick}}{{Davis}
  et~al.}{1996}]{DavisNusserWillick1996}
{Davis} M.,  {Nusser} A.,   {Willick} J.~A.,  1996, \mn@doi [\apj]
  {10.1086/178124}, \href
  {https://ui.adsabs.harvard.edu/abs/1996ApJ...473...22D} {473, 22}

\bibitem[\protect\citeauthoryear{Davis et~al.,}{Davis et~al.}{2011}]{Davis2011}
Davis T.~M.,  et~al., 2011, \mn@doi [The Astrophysical Journal]
  {10.1088/0004-637x/741/1/67}, 741

\bibitem[\protect\citeauthoryear{Di~Valentino et~al.,}{Di~Valentino
  et~al.}{2021}]{DiValentino:2021izs}
Di~Valentino E.,  et~al., 2021, \mn@doi [Class. Quant. Grav.]
  {10.1088/1361-6382/ac086d}, 38, 153001

\bibitem[\protect\citeauthoryear{Feroz \& Hobson}{Feroz \&
  Hobson}{2008}]{feroz2008}
Feroz F.,  Hobson M.~P.,  2008, \mn@doi [Monthly Notices of the Royal
  Astronomical Society] {10.1111/j.1365-2966.2007.12353.x}, 384, 449

\bibitem[\protect\citeauthoryear{Feroz, Hobson  \& Bridges}{Feroz
  et~al.}{2009}]{feroz2009}
Feroz F.,  Hobson M.~P.,   Bridges M.,  2009, \mn@doi [Monthly Notices of the
  Royal Astronomical Society] {10.1111/j.1365-2966.2009.14548.x}, 398, 1601

\bibitem[\protect\citeauthoryear{{Feroz}, {Hobson}, {Cameron}  \&
  {Pettitt}}{{Feroz} et~al.}{2013}]{feroz2013}
{Feroz} F.,  {Hobson} M.~P.,  {Cameron} E.,   {Pettitt} A.~N.,  2013, arXiv
  e-prints, \href {http://adsabs.harvard.edu/abs/2013arXiv1306.2144F} {}

\bibitem[\protect\citeauthoryear{{Foley} et~al.,}{{Foley}
  et~al.}{2018}]{FoundationSNe1}
{Foley} R.~J.,  et~al., 2018, \mn@doi [\mnras] {10.1093/mnras/stx3136}, \href
  {https://ui.adsabs.harvard.edu/abs/2018MNRAS.475..193F} {475, 193}

\bibitem[\protect\citeauthoryear{Freedman}{Freedman}{2021}]{Freedman2021}
Freedman W.~L.,  2021, \mn@doi [The Astrophysical Journal]
  {10.3847/1538-4357/ac0e95}, 919, 16

\bibitem[\protect\citeauthoryear{{Frieman} et~al.,}{{Frieman}
  et~al.}{2008}]{sdss_frieman}
{Frieman} J.~A.,  et~al., 2008, \mn@doi [\aj] {10.1088/0004-6256/135/1/338},
  \href {https://ui.adsabs.harvard.edu/abs/2008AJ....135..338F} {135, 338}

\bibitem[\protect\citeauthoryear{{Gon{\c{c}}alves}, {Carvalho}, {Bengaly},
  {Carvalho}  \& {Alcaniz}}{{Gon{\c{c}}alves}
  et~al.}{2018}]{2018MNRAS.481.5270G}
{Gon{\c{c}}alves} R.~S.,  {Carvalho} G.~C.,  {Bengaly} C.~A.~P.,  {Carvalho}
  J.~C.,   {Alcaniz} J.~S.,  2018, \mn@doi [\mnras] {10.1093/mnras/sty2670},
  \href {https://ui.adsabs.harvard.edu/abs/2018MNRAS.481.5270G} {481, 5270}

\bibitem[\protect\citeauthoryear{{Gon{\c{c}}alves}, {Carvalho}, {Andrade},
  {Bengaly}, {Carvalho}  \& {Alcaniz}}{{Gon{\c{c}}alves}
  et~al.}{2020}]{2020arXiv201006635G}
{Gon{\c{c}}alves} R.~S.,  {Carvalho} G.~C.,  {Andrade} U.,  {Bengaly} C. A.~P.,
   {Carvalho} J.~C.,   {Alcaniz} J.,  2020, arXiv e-prints, \href
  {https://ui.adsabs.harvard.edu/abs/2020arXiv201006635G} {p. arXiv:2010.06635}

\bibitem[\protect\citeauthoryear{Gupta, Saini  \& Laskar}{Gupta
  et~al.}{2008}]{doi:10.1111/j.1365-2966.2008.13377.x}
Gupta S.,  Saini T.~D.,   Laskar T.,  2008, \mn@doi [Monthly Notices of the
  Royal Astronomical Society] {10.1111/j.1365-2966.2008.13377.x}, 388, 242

\bibitem[\protect\citeauthoryear{{Guy}, {Astier}, {Nobili}, {Regnault}  \&
  {Pain}}{{Guy} et~al.}{2005}]{guy2005}
{Guy} J.,  {Astier} P.,  {Nobili} S.,  {Regnault} N.,   {Pain} R.,  2005,
  \mn@doi [\aap] {10.1051/0004-6361:20053025}, \href
  {https://ui.adsabs.harvard.edu/#abs/2005A&A...443..781G} {443, 781}

\bibitem[\protect\citeauthoryear{{Guy} et~al.,}{{Guy} et~al.}{2007}]{guy2007}
{Guy} J.,  et~al., 2007, \mn@doi [\aap] {10.1051/0004-6361:20066930}, \href
  {https://ui.adsabs.harvard.edu/#abs/2007A&A...466...11G} {466, 11}

\bibitem[\protect\citeauthoryear{{Hamuy}, {Phillips}, {Maza}, {Suntzeff},
  {Schommer}  \& {Aviles}}{{Hamuy} et~al.}{1995}]{1995AJ....109....1H}
{Hamuy} M.,  {Phillips} M.~M.,  {Maza} J.,  {Suntzeff} N.~B.,  {Schommer}
  R.~A.,   {Aviles} R.,  1995, \mn@doi [\aj] {10.1086/117251}, \href
  {https://ui.adsabs.harvard.edu/abs/1995AJ....109....1H} {109, 1}

\bibitem[\protect\citeauthoryear{{Heneka}, {Marra}  \& {Amendola}}{{Heneka}
  et~al.}{2014}]{2014MNRAS.439.1855H}
{Heneka} C.,  {Marra} V.,   {Amendola} L.,  2014, \mn@doi [\mnras]
  {10.1093/mnras/stu066}, \href
  {https://ui.adsabs.harvard.edu/abs/2014MNRAS.439.1855H} {439, 1855}

\bibitem[\protect\citeauthoryear{{Hicken}, {Wood-Vasey}, {Blondin}, {Challis},
  {Jha}, {Kelly}, {Rest}  \& {Kirshner}}{{Hicken}
  et~al.}{2009}]{Constitution2009}
{Hicken} M.,  {Wood-Vasey} W.~M.,  {Blondin} S.,  {Challis} P.,  {Jha} S.,
  {Kelly} P.~L.,  {Rest} A.,   {Kirshner} R.~P.,  2009, \mn@doi [\apj]
  {10.1088/0004-637X/700/2/1097}, \href
  {https://ui.adsabs.harvard.edu/abs/2009ApJ...700.1097H} {700, 1097}

\bibitem[\protect\citeauthoryear{{Hinton} et~al.,}{{Hinton}
  et~al.}{2018}]{hinton2018}
{Hinton} S.~R.,  et~al., 2018, arXiv e-prints, \href
  {https://ui.adsabs.harvard.edu/#abs/2018arXiv181102381H} {p.
  arXiv:1811.02381}

\bibitem[\protect\citeauthoryear{{Hollinger} \& {Hudson}}{{Hollinger} \&
  {Hudson}}{2021}]{Hollinger:2021hwx}
{Hollinger} A.~M.,  {Hudson} M.~J.,  2021, \mn@doi [\mnras]
  {10.1093/mnras/staa4039}, \href
  {https://ui.adsabs.harvard.edu/abs/2021MNRAS.502.3723H} {502, 3723}

\bibitem[\protect\citeauthoryear{{Hudson}}{{Hudson}}{1993}]{Hudson1993}
{Hudson} M.~J.,  1993, \mn@doi [\mnras] {10.1093/mnras/265.1.43}, \href
  {https://ui.adsabs.harvard.edu/abs/1993MNRAS.265...43H} {265, 43}

\bibitem[\protect\citeauthoryear{{Huterer}}{{Huterer}}{2020}]{2020arXiv201005765H}
{Huterer} D.,  2020, arXiv e-prints, \href
  {https://ui.adsabs.harvard.edu/abs/2020arXiv201005765H} {p. arXiv:2010.05765}

\bibitem[\protect\citeauthoryear{{Hutsem\'ekers, D.}, {Braibant, L.},
  {Pelgrims, V.}  \& {Sluse, D.}}{{Hutsem\'ekers, D.} et~al.}{2014}]{refId0}
{Hutsem\'ekers, D.} {Braibant, L.} {Pelgrims, V.}  {Sluse, D.} 2014, \mn@doi
  [A\&A] {10.1051/0004-6361/201424631}, 572, A18

\bibitem[\protect\citeauthoryear{{Hutsem{\'e}kers}, {Cabanac}, {Lamy}  \&
  {Sluse}}{{Hutsem{\'e}kers} et~al.}{2005}]{hutsemekers2005}
{Hutsem{\'e}kers} D.,  {Cabanac} R.,  {Lamy} H.,   {Sluse} D.,  2005, \mn@doi
  [\aap] {10.1051/0004-6361:20053337}, \href
  {http://adsabs.harvard.edu/abs/2005A%26A...441..915H} {441, 915}

\bibitem[\protect\citeauthoryear{{Javanmardi}, {Porciani}, {Kroupa}  \&
  {Pflamm-Altenburg}}{{Javanmardi} et~al.}{2015}]{Javanmardi2015}
{Javanmardi} B.,  {Porciani} C.,  {Kroupa} P.,   {Pflamm-Altenburg} J.,  2015,
  \mn@doi [\apj] {10.1088/0004-637X/810/1/47}, \href
  {https://ui.adsabs.harvard.edu/abs/2015ApJ...810...47J} {810, 47}

\bibitem[\protect\citeauthoryear{Jim{\'e}nez, Salzano  \& Lazkoz}{Jim{\'e}nez
  et~al.}{2015}]{beltran2015}
Jim{\'e}nez J.~B.,  Salzano V.,   Lazkoz R.,  2015, \mn@doi [Physics Letters B]
  {https://doi.org/10.1016/j.physletb.2014.12.031}, 741, 168

\bibitem[\protect\citeauthoryear{{Jones} et~al.,}{{Jones}
  et~al.}{2019}]{FoundationSNe2}
{Jones} D.~O.,  et~al., 2019, \mn@doi [\apj] {10.3847/1538-4357/ab2bec}, \href
  {https://ui.adsabs.harvard.edu/abs/2019ApJ...881...19J} {881, 19}

\bibitem[\protect\citeauthoryear{{Kelly}}{{Kelly}}{2007}]{2007ApJ...665.1489K}
{Kelly} B.~C.,  2007, \mn@doi [\apj] {10.1086/519947}, \href
  {https://ui.adsabs.harvard.edu/abs/2007ApJ...665.1489K} {665, 1489}

\bibitem[\protect\citeauthoryear{{Kessler} \& {Scolnic}}{{Kessler} \&
  {Scolnic}}{2017}]{2017ApJ...836...56K}
{Kessler} R.,  {Scolnic} D.,  2017, \mn@doi [\apj]
  {10.3847/1538-4357/836/1/56}, \href
  {https://ui.adsabs.harvard.edu/abs/2017ApJ...836...56K} {836, 56}

\bibitem[\protect\citeauthoryear{Kessler et~al.,}{Kessler
  et~al.}{2009}]{Kessler2009}
Kessler R.,  et~al., 2009, \mn@doi [The Astrophysical Journal Supplement
  Series] {10.1088/0067-0049/185/1/32}, 185, 32

\bibitem[\protect\citeauthoryear{{Kolatt} \& {Lahav}}{{Kolatt} \&
  {Lahav}}{2001}]{2001MNRAS.323..859K}
{Kolatt} T.~S.,  {Lahav} O.,  2001, \mn@doi [\mnras]
  {10.1046/j.1365-8711.2001.04262.x}, \href
  {https://ui.adsabs.harvard.edu/abs/2001MNRAS.323..859K} {323, 859}

\bibitem[\protect\citeauthoryear{Kowalski et~al.,}{Kowalski
  et~al.}{2008}]{kowalski2008}
Kowalski M.,  et~al., 2008, The Astrophysical Journal, 686, 749

\bibitem[\protect\citeauthoryear{{LSST Science Collaboration} et~al.,}{{LSST
  Science Collaboration} et~al.}{2009}]{lsstsciencecollaboration2009lsst}
{LSST Science Collaboration} et~al., 2009, LSST Science Book, Version 2.0
  (\mn@eprint {arXiv} {0912.0201})

\bibitem[\protect\citeauthoryear{Lavaux \& Hudson}{Lavaux \&
  Hudson}{2011}]{10.1111/j.1365-2966.2011.19233.x}
Lavaux G.,  Hudson M.~J.,  2011, \mn@doi [Monthly Notices of the Royal
  Astronomical Society] {10.1111/j.1365-2966.2011.19233.x}, 416, 2840

\bibitem[\protect\citeauthoryear{{Lilow} \& {Nusser}}{{Lilow} \&
  {Nusser}}{2021}]{Lilow:2021omg}
{Lilow} R.,  {Nusser} A.,  2021, arXiv e-prints, \href
  {https://ui.adsabs.harvard.edu/abs/2021arXiv210207291L} {p. arXiv:2102.07291}

\bibitem[\protect\citeauthoryear{Lin, Wang, Chang  \& Li}{Lin
  et~al.}{2016a}]{lin2016}
Lin H.-N.,  Wang S.,  Chang Z.,   Li X.,  2016a, \mn@doi [Monthly Notices of
  the Royal Astronomical Society] {10.1093/mnras/stv2804}, 456, 1881

\bibitem[\protect\citeauthoryear{{Lin}, {Li}  \& {Chang}}{{Lin}
  et~al.}{2016b}]{2016MNRASLin2}
{Lin} H.-N.,  {Li} X.,   {Chang} Z.,  2016b, \mn@doi [\mnras]
  {10.1093/mnras/stw995}, \href
  {https://ui.adsabs.harvard.edu/abs/2016MNRAS.460..617L} {460, 617}

\bibitem[\protect\citeauthoryear{{Maartens}}{{Maartens}}{2011}]{maartens2011}
{Maartens} R.,  2011, \mn@doi [Philosophical Transactions of the Royal Society
  of London Series A] {10.1098/rsta.2011.0289}, \href
  {https://ui.adsabs.harvard.edu/abs/2011RSPTA.369.5115M} {369, 5115}

\bibitem[\protect\citeauthoryear{Mandel, Scolnic, Shariff, Foley  \&
  Kirshner}{Mandel et~al.}{2017}]{Mandel_2017}
Mandel K.~S.,  Scolnic D.~M.,  Shariff H.,  Foley R.~J.,   Kirshner R.~P.,
  2017, \mn@doi [The Astrophysical Journal] {10.3847/1538-4357/aa6038}, 842, 93

\bibitem[\protect\citeauthoryear{{March}, {Trotta}, {Berkes}, {Starkman}  \&
  {Vaudrevange}}{{March} et~al.}{2011}]{march2011}
{March} M.~C.,  {Trotta} R.,  {Berkes} P.,  {Starkman} G.~D.,   {Vaudrevange}
  P.~M.,  2011, \mn@doi [\mnras] {10.1111/j.1365-2966.2011.19584.x}, \href
  {http://adsabs.harvard.edu/abs/2011MNRAS.418.2308M} {418, 2308}

\bibitem[\protect\citeauthoryear{Mariano \& Perivolaropoulos}{Mariano \&
  Perivolaropoulos}{2012}]{mariano2012}
Mariano A.,  Perivolaropoulos L.,  2012, \mn@doi [Phys. Rev. D]
  {10.1103/PhysRevD.86.083517}, 86, 083517

\bibitem[\protect\citeauthoryear{Neill, Hudson  \& Conley}{Neill
  et~al.}{2007}]{RN224}
Neill J.~D.,  Hudson M.~J.,   Conley A.,  2007, \mn@doi [The Astrophysical
  Journal] {10.1086/518808}, 661, L123

\bibitem[\protect\citeauthoryear{Nielsen, Guffanti  \& Sarkar}{Nielsen
  et~al.}{2016}]{Nielsen16}
Nielsen J.~T.,  Guffanti A.,   Sarkar S.,  2016, \mn@doi [Scientific Reports]
  {10.1038/srep35596}, 6, 35596

\bibitem[\protect\citeauthoryear{Perlmutter et~al.,}{Perlmutter
  et~al.}{1997}]{Perlmutter1997}
Perlmutter S.,  et~al., 1997, \mn@doi [ApJ] {10.1086/304265}, 483, 565

\bibitem[\protect\citeauthoryear{{Perlmutter} et~al.,}{{Perlmutter}
  et~al.}{1999}]{perlmutter1999}
{Perlmutter} S.,  et~al., 1999, \mn@doi [\apj] {10.1086/307221}, \href
  {https://ui.adsabs.harvard.edu/abs/1999ApJ...517..565P} {517, 565}

\bibitem[\protect\citeauthoryear{{Phillips}}{{Phillips}}{1993}]{phillips1993}
{Phillips} M.~M.,  1993, \mn@doi [\apjl] {10.1086/186970}, \href
  {http://adsabs.harvard.edu/abs/1993ApJ...413L.105P} {413, L105}

\bibitem[\protect\citeauthoryear{Phillips, Lira, Suntzeff, Schommer, Hamuy  \&
  Maza}{Phillips et~al.}{1999}]{phillips1999}
Phillips M.~M.,  Lira P.,  Suntzeff N.~B.,  Schommer R.~A.,  Hamuy M.,   Maza
  J.,  1999, The Astronomical Journal, 118, 1766

\bibitem[\protect\citeauthoryear{{Pike} \& {Hudson}}{{Pike} \&
  {Hudson}}{2005}]{PikeHudson2005}
{Pike} R.~W.,  {Hudson} M.~J.,  2005, \mn@doi [\apj] {10.1086/497359}, \href
  {https://ui.adsabs.harvard.edu/abs/2005ApJ...635...11P} {635, 11}

\bibitem[\protect\citeauthoryear{{Planck Collaboration} et~al.,}{{Planck
  Collaboration} et~al.}{2016}]{planck2015_16}
{Planck Collaboration} et~al., 2016, \mn@doi [\aap]
  {10.1051/0004-6361/201526681}, \href
  {http://adsabs.harvard.edu/abs/2016A%26A...594A..16P} {594, A16}

\bibitem[\protect\citeauthoryear{{Planck Collaboration} et~al.,}{{Planck
  Collaboration} et~al.}{2020}]{collaboration2018planckI}
{Planck Collaboration} et~al., 2020, \mn@doi [\aap]
  {10.1051/0004-6361/201833880}, \href
  {https://ui.adsabs.harvard.edu/abs/2020A&A...641A...1P} {641, A1}

\bibitem[\protect\citeauthoryear{Pskovskii}{Pskovskii}{1977}]{Pskovskii1977}
Pskovskii Y.~P.,  1977, SvA, 21, 675

\bibitem[\protect\citeauthoryear{Pskovskii}{Pskovskii}{1984}]{Pskovskii1984}
Pskovskii Y.~P.,  1984, Soviet Astronomy, 28, 658

\bibitem[\protect\citeauthoryear{{Radburn-Smith}, {Lucey}  \&
  {Hudson}}{{Radburn-Smith} et~al.}{2004}]{Radburn-SmithLuceyHudson2004}
{Radburn-Smith} D.~J.,  {Lucey} J.~R.,   {Hudson} M.~J.,  2004, \mn@doi
  [\mnras] {10.1111/j.1365-2966.2004.08420.x}, \href
  {https://ui.adsabs.harvard.edu/abs/2004MNRAS.355.1378R} {355, 1378}

\bibitem[\protect\citeauthoryear{Riess, Press  \& Kirshner}{Riess
  et~al.}{1996}]{riess1996}
Riess A.~G.,  Press W.~H.,   Kirshner R.~P.,  1996, The Astrophysical Journal,
  473, 88

\bibitem[\protect\citeauthoryear{{Riess}, {Davis}, {Baker}  \&
  {Kirshner}}{{Riess} et~al.}{1997}]{RiessDavisBaker1997}
{Riess} A.~G.,  {Davis} M.,  {Baker} J.,   {Kirshner} R.~P.,  1997, \mn@doi
  [\apjl] {10.1086/310917}, \href
  {https://ui.adsabs.harvard.edu/abs/1997ApJ...488L...1R} {488, L1}

\bibitem[\protect\citeauthoryear{{Riess} et~al.,}{{Riess}
  et~al.}{1998}]{riess1998}
{Riess} A.~G.,  et~al., 1998, \mn@doi [\aj] {10.1086/300499}, \href
  {https://ui.adsabs.harvard.edu/abs/1998AJ....116.1009R} {116, 1009}

\bibitem[\protect\citeauthoryear{{Riess} et~al.,}{{Riess}
  et~al.}{2007}]{2007ApJ...659...98R}
{Riess} A.~G.,  et~al., 2007, \mn@doi [\apj] {10.1086/510378}, \href
  {https://ui.adsabs.harvard.edu/abs/2007ApJ...659...98R} {659, 98}

\bibitem[\protect\citeauthoryear{Riess et~al.,}{Riess et~al.}{2021}]{RN1573}
Riess A.~G.,  et~al., 2021, A Comprehensive Measurement of the Local Value of
  the Hubble Constant with 1 km/s/Mpc Uncertainty from the Hubble Space
  Telescope and the SH0ES Team, \url
  {https://ui.adsabs.harvard.edu/abs/2021arXiv211204510R}

\bibitem[\protect\citeauthoryear{Rubin \& Hayden}{Rubin \& Hayden}{2016}]{RH16}
Rubin D.,  Hayden B.,  2016, \mn@doi [The Astrophysical Journal]
  {10.3847/2041-8213/833/2/l30}, 833, L30

\bibitem[\protect\citeauthoryear{Rubin \& Heitlauf}{Rubin \&
  Heitlauf}{2020}]{Rubin_2020}
Rubin D.,  Heitlauf J.,  2020, \mn@doi [The Astrophysical Journal]
  {10.3847/1538-4357/ab7a16}, 894, 68

\bibitem[\protect\citeauthoryear{Rubin et~al.,}{Rubin et~al.}{2015}]{rubin2015}
Rubin D.,  et~al., 2015, \mn@doi [The Astrophysical Journal]
  {10.1088/0004-637x/813/2/137}, 813, 137

\bibitem[\protect\citeauthoryear{Rust}{Rust}{1974}]{Rust1974}
Rust B.~W.,  1974, PhD thesis, Oak Ridge National Lab., TN.

\bibitem[\protect\citeauthoryear{Said, Colless, Magoulas, Lucey  \&
  Hudson}{Said et~al.}{2020}]{Said:2020epb}
Said K.,  Colless M.,  Magoulas C.,  Lucey J.~R.,   Hudson M.~J.,  2020,
  \mn@doi [Mon. Not. Roy. Astron. Soc.] {10.1093/mnras/staa2032}, 497, 1275

\bibitem[\protect\citeauthoryear{{Sako} et~al.,}{{Sako}
  et~al.}{2018}]{sdss_sako}
{Sako} M.,  et~al., 2018, \mn@doi [\pasp] {10.1088/1538-3873/aab4e0}, \href
  {https://ui.adsabs.harvard.edu/abs/2018PASP..130f4002S} {130, 064002}

\bibitem[\protect\citeauthoryear{{Sarkar}, {Pandey}  \& {Khatri}}{{Sarkar}
  et~al.}{2019}]{sarkar2019}
{Sarkar} S.,  {Pandey} B.,   {Khatri} R.,  2019, \mn@doi [\mnras]
  {10.1093/mnras/sty3272}, \href
  {https://ui.adsabs.harvard.edu/abs/2019MNRAS.483.2453S} {483, 2453}

\bibitem[\protect\citeauthoryear{{Schwarz} \& {Weinhorst}}{{Schwarz} \&
  {Weinhorst}}{2007}]{2007AA...474..717S}
{Schwarz} D.~J.,  {Weinhorst} B.,  2007, \mn@doi [\aap]
  {10.1051/0004-6361:20077998}, \href
  {https://ui.adsabs.harvard.edu/abs/2007A&A...474..717S} {474, 717}

\bibitem[\protect\citeauthoryear{{Schwarz}, {Copi}, {Huterer}  \&
  {Starkman}}{{Schwarz} et~al.}{2016}]{schwarz2016}
{Schwarz} D.~J.,  {Copi} C.~J.,  {Huterer} D.,   {Starkman} G.~D.,  2016,
  \mn@doi [Classical and Quantum Gravity] {10.1088/0264-9381/33/18/184001},
  \href {https://ui.adsabs.harvard.edu/abs/2016CQGra..33r4001S} {33, 184001}

\bibitem[\protect\citeauthoryear{{Scolnic} et~al.,}{{Scolnic}
  et~al.}{2018a}]{scolnic2018}
{Scolnic} D.~M.,  et~al., 2018a, \mn@doi [\apj] {10.3847/1538-4357/aab9bb},
  \href {http://adsabs.harvard.edu/abs/2018ApJ...859..101S} {859, 101}

\bibitem[\protect\citeauthoryear{Scolnic et~al.}{Scolnic
  et~al.}{2018b}]{Scolnic:2017caz}
Scolnic D.,  et~al., 2018b, \mn@doi [Astrophys. J.] {10.3847/1538-4357/aab9bb},
  859, 101

\bibitem[\protect\citeauthoryear{{Secrest}, {von Hausegger}, {Rameez},
  {Mohayaee}, {Sarkar}  \& {Colin}}{{Secrest}
  et~al.}{2020}]{2020arXiv200914826S}
{Secrest} N.,  {von Hausegger} S.,  {Rameez} M.,  {Mohayaee} R.,  {Sarkar} S.,
   {Colin} J.,  2020, arXiv e-prints, \href
  {https://ui.adsabs.harvard.edu/abs/2020arXiv200914826S} {p. arXiv:2009.14826}

\bibitem[\protect\citeauthoryear{{Shariff}, {Jiao}, {Trotta}  \& {van
  Dyk}}{{Shariff} et~al.}{2016}]{shariff2016}
{Shariff} H.,  {Jiao} X.,  {Trotta} R.,   {van Dyk} D.~A.,  2016, \mn@doi
  [\apj] {10.3847/0004-637X/827/1/1}, \href
  {http://adsabs.harvard.edu/abs/2016ApJ...827....1S} {827, 1}

\bibitem[\protect\citeauthoryear{Smith et~al.,}{Smith et~al.}{2020}]{smith2020}
Smith M.,  et~al., 2020, \mn@doi [Monthly Notices of the Royal Astronomical
  Society] {10.1093/mnras/staa946}, 494, 4426

\bibitem[\protect\citeauthoryear{{Soltis}, {Farahi}, {Huterer}  \&
  {Liberato}}{{Soltis} et~al.}{2019}]{2019PhRvL.122i1301S}
{Soltis} J.,  {Farahi} A.,  {Huterer} D.,   {Liberato} C.~M.,  2019, \mn@doi
  [\prl] {10.1103/PhysRevLett.122.091301}, \href
  {https://ui.adsabs.harvard.edu/abs/2019PhRvL.122i1301S} {122, 091301}

\bibitem[\protect\citeauthoryear{{Stahl}, {de Jaeger}, {Boruah}, {Zheng},
  {Filippenko}  \& {Hudson}}{{Stahl} et~al.}{2021}]{StahldeJaegerBoruah2021}
{Stahl} B.~E.,  {de Jaeger} T.,  {Boruah} S.~S.,  {Zheng} W.,  {Filippenko}
  A.~V.,   {Hudson} M.~J.,  2021, \mn@doi [\mnras] {10.1093/mnras/stab1446},
  \href {https://ui.adsabs.harvard.edu/abs/2021MNRAS.tmp.1412S} {}

\bibitem[\protect\citeauthoryear{Strauss \& Willick}{Strauss \&
  Willick}{1995}]{Strauss1995}
Strauss M.~A.,  Willick J.~A.,  1995, \mn@doi [Physics Reports]
  {https://doi.org/10.1016/0370-1573(95)00013-7}, 261, 271

\bibitem[\protect\citeauthoryear{Sullivan et~al.,}{Sullivan
  et~al.}{2010}]{sullivan2010}
Sullivan M.,  et~al., 2010, \mn@doi [Monthly Notices of the Royal Astronomical
  Society] {10.1111/j.1365-2966.2010.16731.x}, 406, 782

\bibitem[\protect\citeauthoryear{{Sullivan} et~al.,}{{Sullivan}
  et~al.}{2011}]{snls_sullivan}
{Sullivan} M.,  et~al., 2011, \mn@doi [\apj] {10.1088/0004-637X/737/2/102},
  \href {https://ui.adsabs.harvard.edu/abs/2011ApJ...737..102S} {737, 102}

\bibitem[\protect\citeauthoryear{{Sun} \& {Wang}}{{Sun} \&
  {Wang}}{2018a}]{sun2018}
{Sun} Z.~Q.,  {Wang} F.~Y.,  2018a, preprint, \href
  {http://adsabs.harvard.edu/abs/2018arXiv180405191S} {} (\mn@eprint {arXiv}
  {1804.05191})

\bibitem[\protect\citeauthoryear{{Sun} \& {Wang}}{{Sun} \&
  {Wang}}{2018b}]{2018MNRAS.478.5153S}
{Sun} Z.~Q.,  {Wang} F.~Y.,  2018b, \mn@doi [\mnras] {10.1093/mnras/sty1391},
  \href {https://ui.adsabs.harvard.edu/abs/2018MNRAS.478.5153S} {478, 5153}

\bibitem[\protect\citeauthoryear{Sun \& Wang}{Sun \& Wang}{2018c}]{Sun_2018}
Sun Z.~Q.,  Wang F.~Y.,  2018c, \mn@doi [Monthly Notices of the Royal
  Astronomical Society] {10.1093/mnras/sty1391}, 478, 5153–5158

\bibitem[\protect\citeauthoryear{{Sun} \& {Wang}}{{Sun} \&
  {Wang}}{2019}]{2019EPJC...79..783S}
{Sun} Z.~Q.,  {Wang} F.~Y.,  2019, \mn@doi [European Physical Journal C]
  {10.1140/epjc/s10052-019-7293-3}, \href
  {https://ui.adsabs.harvard.edu/abs/2019EPJC...79..783S} {79, 783}

\bibitem[\protect\citeauthoryear{{Thorp}, {Mandel}, {Jones}, {Ward}  \&
  {Narayan}}{{Thorp} et~al.}{2021}]{2021arXiv210205678T}
{Thorp} S.,  {Mandel} K.~S.,  {Jones} D.~O.,  {Ward} S.~M.,   {Narayan} G.,
  2021, arXiv e-prints, \href
  {https://ui.adsabs.harvard.edu/abs/2021arXiv210205678T} {p. arXiv:2102.05678}

\bibitem[\protect\citeauthoryear{{Tripp}}{{Tripp}}{1998}]{Tripp1998}
{Tripp} R.,  1998, \aap, \href
  {https://ui.adsabs.harvard.edu/abs/1998A&A...331..815T} {331, 815}

\bibitem[\protect\citeauthoryear{{Trotta}}{{Trotta}}{2008}]{2008ConPh..49...71T}
{Trotta} R.,  2008, \mn@doi [Contemporary Physics] {10.1080/00107510802066753},
  \href {https://ui.adsabs.harvard.edu/abs/2008ConPh..49...71T} {49, 71}

\bibitem[\protect\citeauthoryear{{Turnbull}, {Hudson}, {Feldman}, {Hicken},
  {Kirshner}  \& {Watkins}}{{Turnbull}
  et~al.}{2012}]{TurnbullHudsonFeldman2012}
{Turnbull} S.~J.,  {Hudson} M.~J.,  {Feldman} H.~A.,  {Hicken} M.,  {Kirshner}
  R.~P.,   {Watkins} R.,  2012, \mn@doi [\mnras]
  {10.1111/j.1365-2966.2011.20050.x}, \href
  {https://ui.adsabs.harvard.edu/abs/2012MNRAS.420..447T} {420, 447}

\bibitem[\protect\citeauthoryear{Visser}{Visser}{2004}]{visser2004}
Visser M.,  2004, \mn@doi [Classical and Quantum Gravity]
  {10.1088/0264-9381/21/11/006}, 21, 2603–2615

\bibitem[\protect\citeauthoryear{Wang \& Wang}{Wang \& Wang}{2014}]{wang2014}
Wang J.~S.,  Wang F.~Y.,  2014, \mn@doi [Monthly Notices of the Royal
  Astronomical Society] {10.1093/mnras/stu1279}, 443, 1680

\bibitem[\protect\citeauthoryear{Yang, Wang  \& Chu}{Yang
  et~al.}{2014}]{yang2014}
Yang X.,  Wang F.~Y.,   Chu Z.,  2014, \mn@doi [Monthly Notices of the Royal
  Astronomical Society] {10.1093/mnras/stt2015}, 437, 1840

\bibitem[\protect\citeauthoryear{{Zhao}, {Zhou}  \& {Chang}}{{Zhao}
  et~al.}{2019}]{2019MNRAS.486.5679Z}
{Zhao} D.,  {Zhou} Y.,   {Chang} Z.,  2019, \mn@doi [\mnras]
  {10.1093/mnras/stz1259}, \href
  {https://ui.adsabs.harvard.edu/abs/2019MNRAS.486.5679Z} {486, 5679}

\bibitem[\protect\citeauthoryear{{Zheng} et~al.,}{{Zheng}
  et~al.}{2008}]{2008AJ....135.1766Z}
{Zheng} C.,  et~al., 2008, \mn@doi [\aj] {10.1088/0004-6256/135/5/1766}, \href
  {https://ui.adsabs.harvard.edu/abs/2008AJ....135.1766Z} {135, 1766}

\makeatother
\end{thebibliography}

%\begin{thebibliography}{99}
%\end{thebibliography}

%%%%%%%%%%%%%%%%%%%%%%%%%%%%%%%%%%%%%%%%%%%%%%%%%%

%%%%%%%%%%%%%%%%% APPENDICES %%%%%%%%%%%%%%%%%%%%%

\appendix

\section{Derivation and Test of Method of Moments} 
\label{app:moments}
We present here the derivation of the the first and second moment of the moments generating function. We also demonstrate that our method of moments correctly recovers the selection function from simulations and that inference from replica of the data under the model is unbiased. 

Consider the distribution of the random variable $C$, denoting the observed colour within a single survey and redshift bin $sj$. From Eqs.~\eqref{eq:moments_function} and \eqref{eq:fc}, we wish to compute the moment generating function,
\begin{align}
M_{C}(t) &= \int_{-\infty}^{\infty}e^{t \ch}f_C(\ch)d\ch \\
&= \frac{1}{p(I=1|\Psi, \Theta)} \times \nonumber\\
&\int_{-\infty}^{\infty}e^{t\ch}.\frac{1}{\sqrt{2\pi\sigma^2}}e^{-\frac{1
    }{2\sigma^{2}}(\ch - c_\star)^2}.\Phi\left(\frac{\cobs - \ch}{\sigmaobs}\right)d\ch.
\end{align}
where $\sigma^2 \equiv  R_c^2+\sigma_{\ch}^2$, and $\sigma_\ch$ is the average measurement noise for colour observations (which we approximate as being the same for all data points in a given survey and redshift bin).

The above can be recast as: 
\begin{align} 
M_{C}(t) &= \frac{e^{c_\star t + \frac{1}{2}\sigma^2 t^2}}{p(I=1|\Psi, \Theta)} \times \nonumber\\ &\int_{-\infty}^{\infty}\frac{1}{\sqrt{2\pi\sigma^2}}e^{-\frac{1
    }{2\sigma^{2}}(\ch - (c_\star + \sigma^2 t))^2}\Phi\left(\frac{\cobs - \ch}{\sigmaobs}\right) d\ch\\
&= \frac{e^{c_\star t + \frac{1}{2}\sigma^2 t^2}}{p(I=1|\Psi, \Theta)} \int_{-\infty}^{\infty}\norm_{\ch}(c_\star + \sigma^2 t,\sigma^2).\Phi\left(\frac{\cobs - \ch}{\sigmaobs}\right)d\ch \\
& = \frac{e^{c_{\star}t + \frac{1}{2}\sigma^{2} t^{2}}}{p(I=1|\Psi, \Theta)}\Phi \left(\frac{\cobs - (c_{\star} + \sigma^{2} t)}{\sqrt{\sigma^{2} + {\sigmaobs}^{2}}}\right). \label{eq:moments_function_result}
\end{align}
%This is a convenient form to apply the relation $\int_{-\infty}^{\infty}\Phi\left(\frac{\mu-x}{\sigma}\right) \norm_x(\nu, \tau^2) dx =  \Phi\left(\frac{\mu-\nu}{\sqrt{\sigma^2+\tau^2}}\right)$ which leads us to

We now compute the first and second moments, set $h(t) \equiv e^{c_{\star} t + \frac{1}{2}\sigma^{2} t^{2}}$ and $g(t) = \frac{\cobs - (c_{\star} + \sigma^{2} t)}{\sqrt{\sigma^{2} + {\sigmaobs}^{2}}}$. A dash ($'$) symbol indicates a derivative with respect to $t$. Hence:

\begin{align}
\frac{dM_{C}(t)}{dt}{\Big \vert}_{t=0} & = \frac{1}{p(I=1|\Psi, \Theta)}\frac{d}{dt}\left[ h(t)\Phi(g(t)) \right]_{t=0} \nonumber\\ 
&= \frac{1}{p(I=1|\Psi, \Theta)} \left[h'(t)\Phi(g(t)) + h(t)\Phi'(g(t))\right]_{t=0}
\end{align}
and 
\begin{align}
\frac{d^2M_{C(t)}}{dt^2}{\Big \vert}_{t=0} &= \frac{1}{p(I_i=1|\Psi, \Theta)} \times \nonumber\\ 
&\left[h''(t)\Phi(g(t)) + 2h'(t)\Phi'((g(t)) +  h(t)\Phi''(g(t))\right]_{t=0}
\end{align}

We derive each of the terms $h(t),\, h'(t),\, h''(t),\, \Phi(g(t)),\,\Phi'(g(t))$ and $\Phi''(g(t)$ evaluated at $t=0$:
\begin{align}
h(0) & = 1. \\ 
h'(0) & = h(t)(c_\star + \sigma^2 t)|_{t=0} = c_\star . \\
h''(0) & = h(t)\sigma^2 + h'(t)(c_\star + \sigma^2t)|_{t=0} = \sigma^2 + c_\star^2.
\end{align}
To determine $\Phi'(g(t))$ and $\Phi''(g(t))$ we use the Leibniz rule for differentiating under an integral (the CDF). As a reminder, our CDF is from the integral of $\norm_{x}(0,1)$ from $-\infty$ up to $g(t)$ with $\cobs$ and $\sigmaobs$ used to control the width as opposed to the normal distribution hyperparameters. This gives:
\begin{align}
\Phi(g(t=0)) & = \int_{-\infty}^{g(0)} \frac{1}{\sqrt{2\pi}}.e^{-\frac{1}{2}x^2}dx \\
\Phi'(g(t=0))| &= \frac{d}{dt}\left[ \int_{-\infty}^{g(t)} \frac{1}{\sqrt{2\pi}}.e^{-\frac{1}{2}x^2}dx\right]_{t=0} \\
&= \frac{1}{\sqrt{2\pi}}.e^{-\frac{1}{2}g(t)^2}. \gprime(t)|_{t=0} \\
&= -\frac{\sigma^2}{\sqrt{2\pi}\sqrt{\sigma^2 +{\sigmaobs}^2}} e^{-\frac{1}{2} \frac{(\cobs -c_\star)^2}{\sigma^2 + {\sigmaobs}^2} }
\end{align}
where $g(0) = \frac{\cobs - c_\star}{\sqrt{\sigma^2 + {\sigmaobs}^2}}$ and $\gprime(t)|_{t=0} = -\frac{\sigma^2}{\sqrt{\sigma^2 + {\sigmaobs}^2}}|_{t=0} = -\frac{\sigma^2}{\sqrt{\sigma^2 +{\sigmaobs}^2}}$.
Finally,
\begin{align}
\Phi''(t=0) &= \frac{1}{\sqrt{2\pi}}.e^{-\frac{1}{2}g(t)^2}.-g(t).\gprime^2(t) + \frac{1}{\sqrt{2\pi}}.e^{-\frac{1}{2}g(t)^2}\gprime'(t)|_{t=0} \\
&=\frac{1}{\sqrt{2\pi}}.e^{-\frac{1}{2}g(t)^2} \left( \gprime'(t) - g(t)\gprime^2(t)\right){\Big \vert}_{t=0}.
\end{align}
Given that, $\gprime'(t)|_{t=0} =0$, this reduces to
\be
\Phi''(g(t))|_{t=0} =-\frac{1}{\sqrt{2\pi}}.e^{-\frac{1}{2}g(0)^2} \left(g(0)\gprime^2(0)\right).
\ee
This leads to the first and second moments:

\begin{align}
\frac{dM_{C}(t)}{dt}{\Big \vert}_{t=0} &=  \frac{1}{p(I=1|\Psi, \Theta)}\left(c_\star \Phi(g(0)) -\frac{\sigma^2}{\sqrt{2\pi}\sqrt{\sigma^2 +{\sigmaobs}^2}} e^{-\frac{1}{2} \frac{(\cobs -c_\star)^2}{\sigma^2 + {\sigmaobs}^2} }\right), \label{eq:moments_mean}\\
\frac{d^{2}M_{C}(t)}{dt^{2}}{\Big \vert}_{t=0} &= \frac{1}{p(I=1|\Psi, \Theta)}\left((\sigma^2 + c_{\star}^{2})\Phi(g(0)) -\frac{2c_\star\sigma^2}{\sqrt{2\pi}\sqrt{\sigma^2 +{\sigmaobs}^2}} e^{-\frac{1}{2} \frac{(\cobs -c_\star)^2}{\sigma^2 + {\sigmaobs}^2} }   -\frac{\sigma^4}{\sqrt{2\pi}}e^{-\frac{1}{2}\left(\frac{(\cobs -c_\star)^2}{\sigma^2 + {\sigmaobs}^2}\right)}\left(\frac{\cobs - c_\star}{(\sigma^2 + {\sigmaobs}^2)^{3/2}}\right), \right) \label{eq:moments_var}
\end{align}
where the normalization constant is given by Eq.~\eqref{eq:correction_factor}. 

\begin{figure*}
    \centering
\includegraphics[width=2\columnwidth]{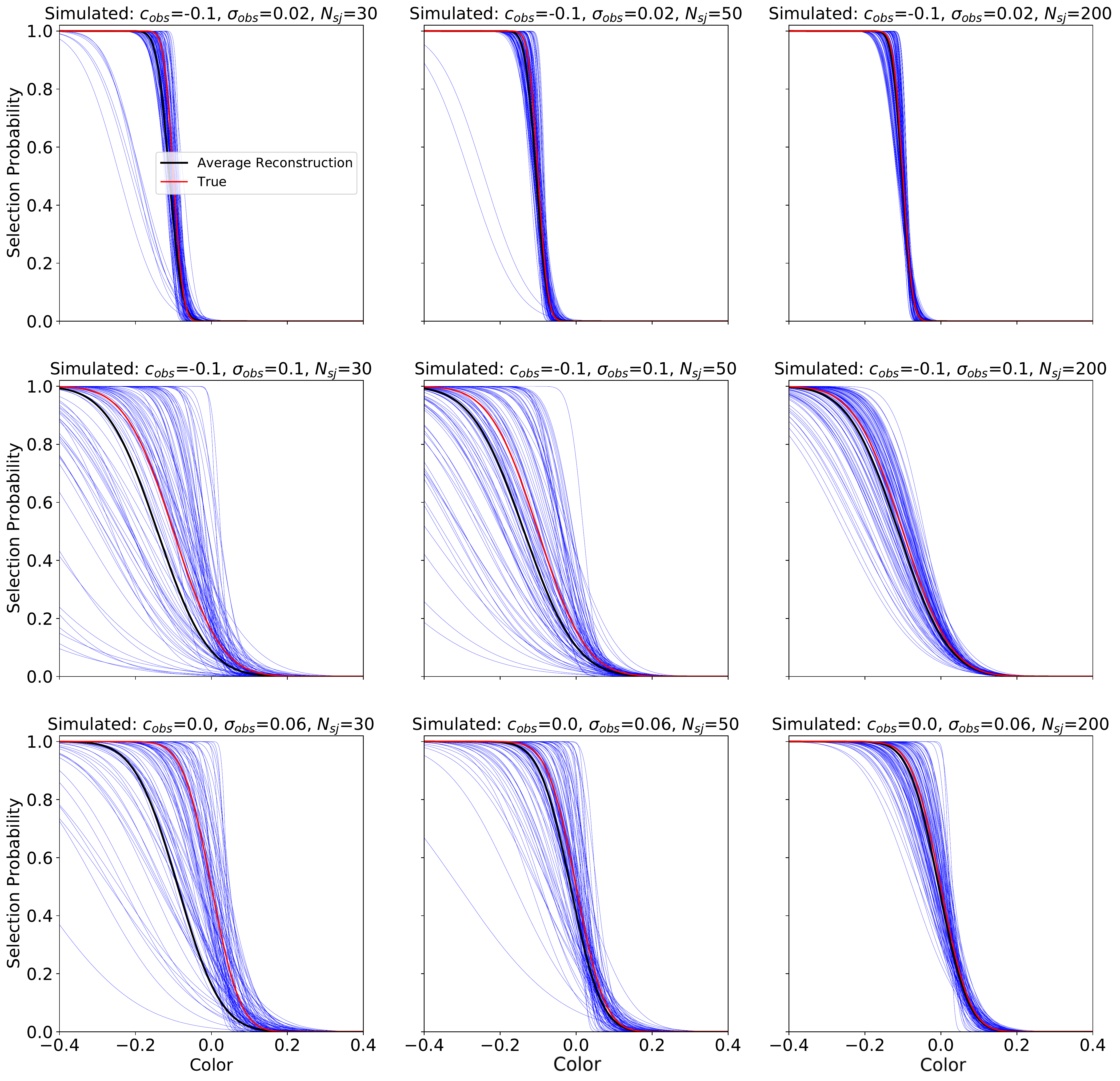}
   \caption{Reconstructions of the selection function from the two first moments of the moment generating function from simulated data: we show three representative choices for the selection function parameters, $\cobs, \sigmaobs$ (columns) and three different sample sizes (\SNe\ per bin, $N_{sj}=30,50, 100$, top to bottom in rows). Blue lines are the individual reconstruction from each of $N_{\rm sim} = 100$ simulations; dashed green is the mean reconstruction averaged over realizations, and solid red is the true selection function.     \label{fig:test_reconstructions}}
\end{figure*}

We tested our method of moments to reconstruct the selection function on a suite of simulations, with $c_\star=0.0, R_c=0.1$, and three different choices of selection function parameters: $\{ \cobs, \sigmaobs\} = \{ [-0,1, 0.02], [-0.1, 0.10], [0.0, 0.06] \}$, chosen the span the parameter space of interest in our application. We show the results of the reconstructed selection function for $N_{sj} = 30, 50, 200$ (from top to bottom) in Fig.~\ref{fig:test_reconstructions}. The results show that the reconstruction, when averaged over realizations, is extremely close to the underlying true selection function, thus validating the method. 

\begin{figure*}
\caption{Posterior marginal distributions on simulated data with colour-based selection effects, and a simulated dipole, averaged over $N=100$ data realizations. The posterior includes a correction for colour-based selection effects according to our method. Vertical lines give the true value of the parameters. \label{fig:reconstruction_w_selectioneffects}}
\includegraphics[width =2\columnwidth]{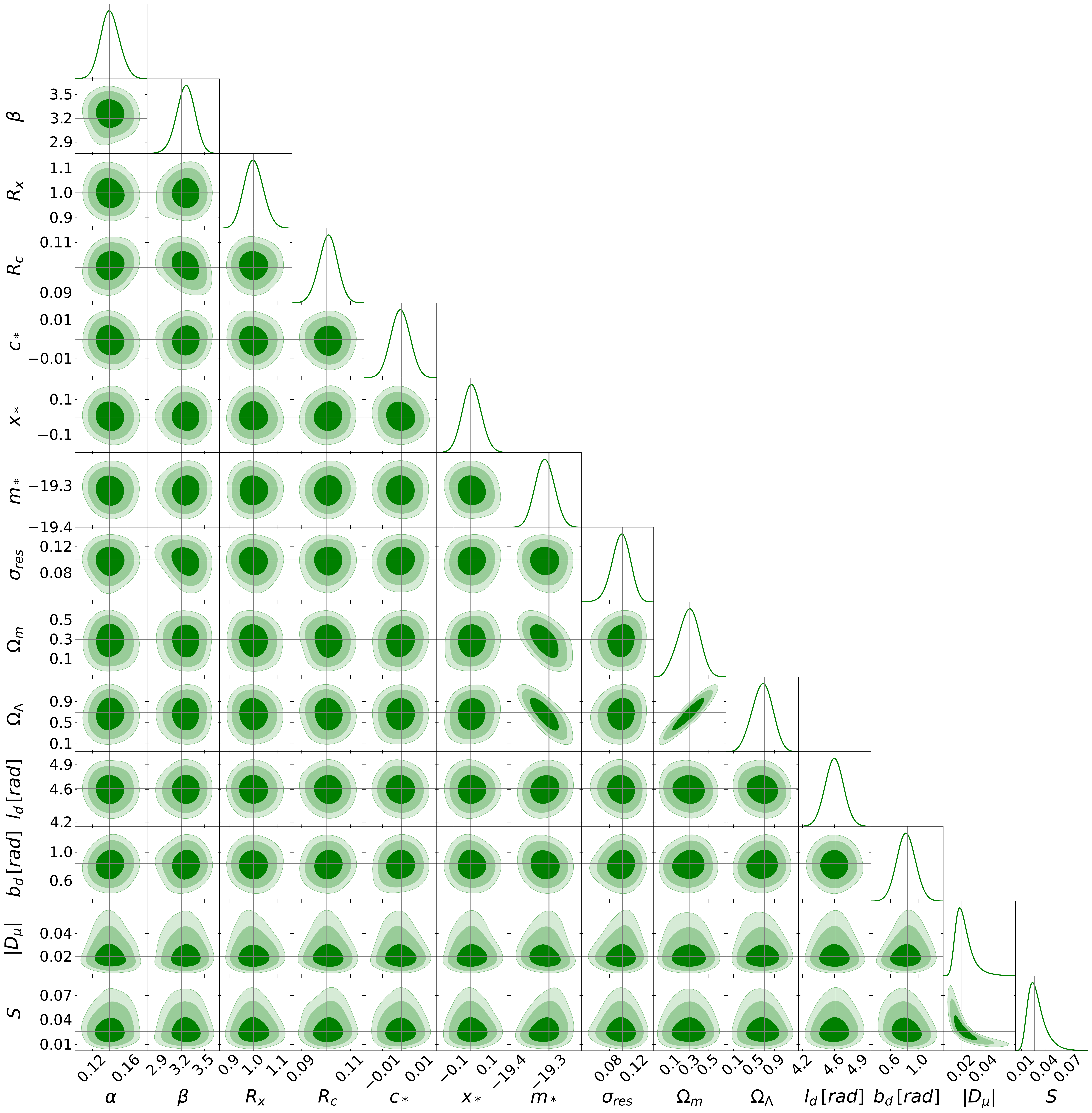}
\end{figure*}

We tested parameter inference in the presence of residual colour-based selection effects data simulated according to the method presented in section \ref{sec:simulating_data}, with colour-based selection effects as described in section~\ref{sec:colour_selection}, with selection function parameters for each survey being:

SDSS = \{(-0.5, 3.4), (-0.5, 0.57), (-0.35, 0.29), (0.20, 0.20), (0.20, 0.20) \},

SNLS = \{(0.20, 0.20), (-0.50, 1.17), (0.14, 0.17), (-0.06, 0.13), (-0.18, 0.14)\},

Low-z = \{(-0.50, 2.95), (-0.50, 4.47), (0.09, 0.01), (0.017, 0.01)\},

HST = \{(-0.01, 0.12)\},
where each tuple gives the values of $(\cobs,\sigmaobs)$ in order from lowest redshift bin to highest within each survey. In the reconstruction, we estimate the selection function parameters as described above, and present 1- and 2-D marginal posteriors on all parameters in Fig.~\ref{fig:reconstruction_w_selectioneffects}. The posterior distributions have been averaged over $N=100$ replicas. We observe that the posterior for all of the parameters has a mode very close to the true value, thus validating our methodology.  

%%%%%%%%%%%%%%%%%%%%%%%%%%%%%%%%%%%%%%%%%%%%%%%%%%

% Don't change these lines
\bsp	% typesetting comment
\label{lastpage}
\end{document}